\documentclass[%
 reprint,
 amsmath,amssymb,
 aps,
 prd,
]{revtex4-1}

\usepackage{graphicx}
\usepackage{dcolumn}
\usepackage{bm}

\usepackage{units}
\usepackage{xspace}
\usepackage{subfig}
\usepackage{siunitx}

\pdfoutput=1

\def\EeV{\ifmmode {\mathrm{Ee\kern -0.07em V}}\else
                   \textrm{Ee\kern -0.07em V}\fi}
\def\TeV{\ifmmode {\mathrm{Te\kern -0.07em V}}\else
                   \textrm{Te\kern -0.07em V}\fi}
\def\eV{\ifmmode {\mathrm{\ e\kern -0.07em V}}\else
                   \textrm{e\kern -0.07em V}\fi}                   
\def\gcm{\ensuremath{\mathrm{g/cm}^2}\xspace}
\def\sigmaXmax{\ensuremath{\sigma(X_\text{max})}\xspace}

\newcommand{\gcmdec}{\ifmmode {\gcm/\mathrm{decade}}\else
                     {\gcm/decade}\fi\xspace}%
\newcommand{\depth}[1]{\unit[#1]{\gcm}}
\newcommand{\energy}[1]{\ensuremath{10^{#1}\,}\eV}

\newcommand{\bea}{\begin{linenomath}\begin{eqnarray}}
\newcommand{\eea}{\end{eqnarray}\end{linenomath}}
\newcommand{\bean}{\begin{linenomath}\begin{eqnarray*}}
\newcommand{\eean}{\end{eqnarray*}\end{linenomath}}
\newcommand{\be}{\begin{linenomath}\begin{equation}}
\newcommand{\ee}{\end{equation}\end{linenomath}}
\newcommand{\ben}{\begin{linenomath}\begin{equation*}}
\newcommand{\een}{\end{equation*}\end{linenomath}}

\newcommand{\bem}{\begin{linenomath}\begin{pmatrix}}
\newcommand{\eem}{\end{pmatrix}\end{linenomath}}

\newcommand{\sib}{{Sibyll2.3}\xspace}
\newcommand{\epos}{{Epos-LHC}\xspace}
\newcommand{\qgs}{{QGSJetII-04}\xspace}
\newcommand{\conex}{{CONEX v4r37}\xspace}

\def\Xmax{\ensuremath{X_\mathrm{max}}\xspace}
\def\bfXmax{\ensuremath{\bf{X_\mathrm{max}\;}}}
\def\meanXmax{\ensuremath{\left\langle \Xmax \negthickspace \; \right\rangle}\xspace }

\newcommand{\lnA}{\ensuremath{\ln A}\;}

\newcommand{\meanlnA}{\ensuremath {\langle \ln A \rangle\;}}

\def\tnorm{\ensuremath{t_{0_\mathrm{norm}}}\xspace}
\def\sigmanorm{\ensuremath{\sigma_{\mathrm{norm}}}\xspace}
\def\lambdanorm{\ensuremath{\lambda_{\mathrm{norm}}}\xspace}
\newcommand{\fig}[1]{Fig.~\ref{#1}}
\newcommand{\figs}[2]{Figs.~\ref{#1} and \ref{#2}}
\newcommand{\figsThree}[3]{Figs.~\ref{#1}, \ref{#2} and \ref{#3}}

\newcommand{\tab}[1]{Table~\ref{#1}}

\graphicspath{{pics/}}

\begin{document}


\title{Extracting a less model dependent cosmic ray composition from \bfXmax distributions}

\author{Simon Blaess}
\author{Jose A. Bellido}%
 \author{Bruce R. Dawson}%
\affiliation{%
 Department of Physics, University of Adelaide, Adelaide, Australia
}%


%
%

\begin{abstract}
At higher energies the uncertainty in the estimated cosmic ray mass composition, extracted from the observed distributions of the depth of shower maximum $\text{X}_{\text{max}}$, is dominated by uncertainties in the hadronic interaction models. Thus, the estimated composition depends strongly on the particular model used for its interpretation. To reduce this model dependency in the interpretation of the mass composition, we have developed a novel approach which allows the adjustment of the normalisation levels of the proton \meanXmax and \sigmaXmax guided by real observations of \Xmax distributions. In this paper we describe the details of this approach and present a study of its performance and its limitations. Using this approach we extracted cosmic ray mass composition information from the published Pierre Auger \Xmax distributions.  We have obtained a consistent mass composition interpretation for \epos, \qgs and \sib. Our fits suggest a composition consisting of predominantly iron. Below \energy{18.8}, the small proportions of proton, helium and nitrogen vary. Above \energy{18.8}, there is little proton or helium, and with increasing energy the nitrogen component gradually gives way to the growing iron component, which dominates at the highest energies. The fits suggest that the normalisation level for proton \meanXmax is much deeper than the initial predictions of the hadronic interaction models. The fitted normalisation level for proton \sigmaXmax is also greater than the model predictions. When fixing the expected normalisation of \sigmaXmax to that suggested by the \qgs model, a slightly larger fraction of protons is obtained. These results remain sensitive to the other model parameters that we keep fixed, such as the elongation rate and the \meanXmax separation between p and Fe.


\end{abstract}

\maketitle

 \section{Introduction}
 A common parameter used to extract mass composition information is \Xmax, the atmospheric depth in \gcm from the top of the atmosphere where the longitudinal development of an air shower reaches the maximum number of particles or the maximum of the energy deposited in the atmosphere. Different cosmic ray primaries propagate through the atmosphere differently, resulting in different observed distributions of \Xmax~\cite{1977ICRC....8..353G}. Due to statistical variability in the interaction between cosmic rays of a specific primary mass and the atmosphere, a cosmic ray's primary mass cannot be determined on an event by event basis by examining \Xmax. Instead we study the \Xmax distribution of cosmic rays of similar energy to infer the mass composition distribution of the events. Differences in the mode, width and tail of the \Xmax distribution provide information on the mass composition distribution of the events and on the hadronic interaction properties~\cite{Kampert:2012mx,Collaboration:2012wt}.

\begin{figure}[!htb]
\centering
  \vspace{0.6cm}
    \includegraphics[trim={0 1.3cm 0 0}, width=0.4\textwidth]{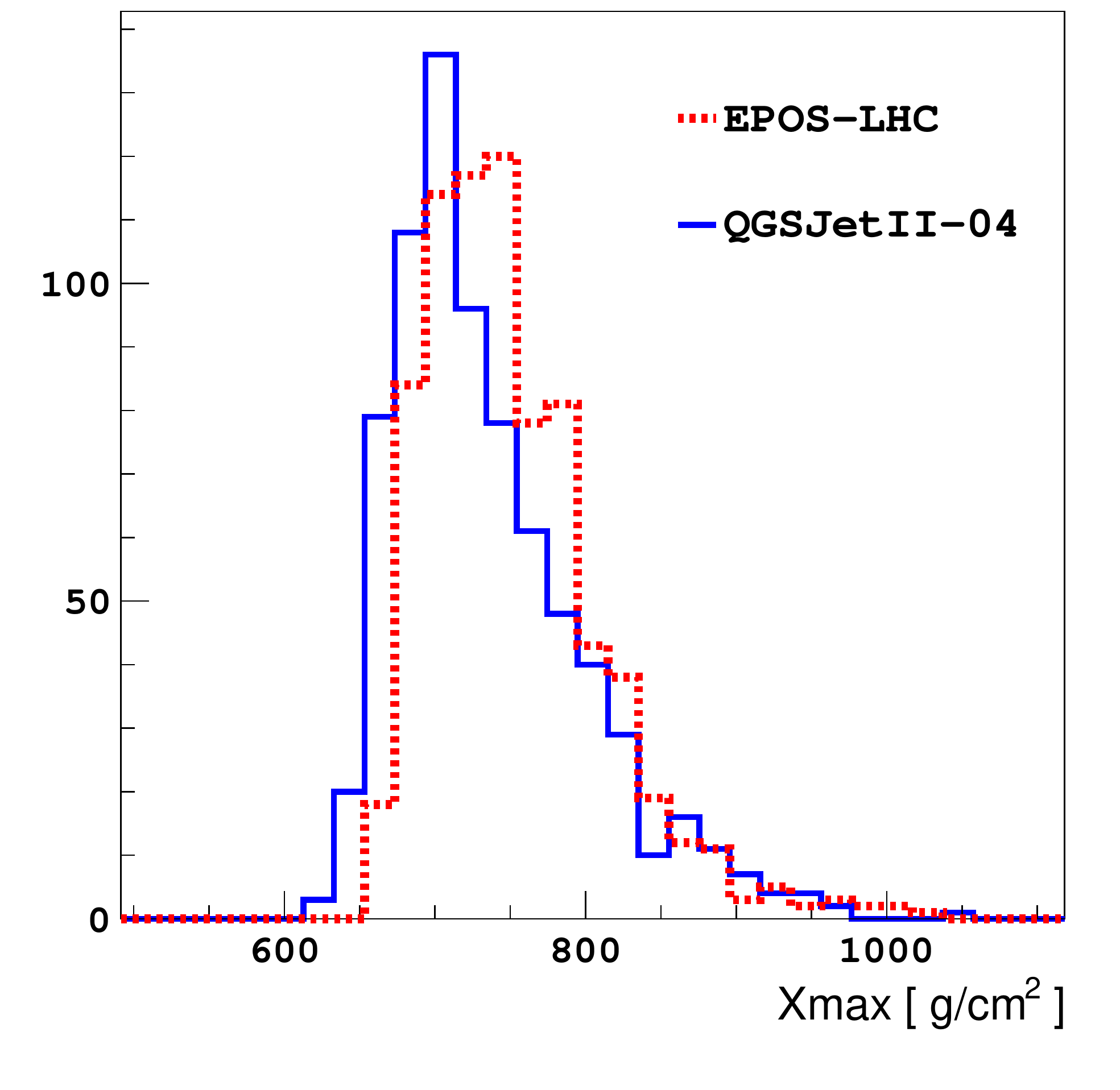}
      \caption{An \Xmax distribution of 750 \epos simulated proton events (red), and separately 750 \qgs simulated protons events (blue), of energy \energy{18}.}      
\label{fig:conex_proof_p_EPOS_QGSJET}
\end{figure}  

\fig{fig:conex_proof_p_EPOS_QGSJET} shows the \Xmax distribution resulting from the \conex simulation of 750 proton events according to the \epos model, and separately 750 proton events according to the \qgs model, of energy \energy{18}. The figure illustrates the differences in the \Xmax distribution predicted by different hadronic interaction models. Most noticeable is the difference in the modes of the distributions, but there are also marginal differences in the width and tails of the distributions. These differences between the hadronic interaction models change with energy to some degree. Although the dissimilarity between these predicted distributions may appear minor, applying a parameterisation based on these different predictions to data can have a considerable impact on the mass composition inferred. Consequently, typical mass composition studies of \Xmax are strongly dependent on the hadronic interaction model assumed. 

The algorithm \conex~\cite{Bergmann:2006yz,Pierog:2004re}, along with the hadronic interaction packages \epos \cite{Pierog:2013ria}, \qgs \cite{Ostapchenko:2010vb} and \sib \cite{sib}, were used to simulate air showers to obtain \Xmax distributions according to each of these models. We have developed a parameterisation for des\-cri\-bing these expected \Xmax distributions for cosmic rays of some energy and mass. Our parameterisation of the \Xmax distributions can then be used to fit observed \Xmax distributions, to extract primary mass information (composition fractions) from each energy bin. By  in\-clu\-ding some of the coefficients of our \Xmax parameterisation in the fit, mass composition results are obtained which are somewhat independent of the hadronic interaction model assumed. 

Assuming the \epos, \qgs or \sib hadronic models, the Auger \Xmax distributions can be well reproduced assuming a composition of at least four components consisting of proton, Helium, Nitrogen and Iron~\cite{Aab:2014kda,Aab:2014aea,AugerCombinedFit}. Therefore, in this work we have used mock data sets to evaluate the performance of our method for retrieving the true relative amounts of p, He, N, Fe (composition fractions). The results of applying this method to interpret the published Auger \Xmax distributions in~\cite{Aab:2014kda} in terms of the mass composition of cosmic rays are presented.

\section{\label{sec:param} Parameterisation of \bfXmax distributions}
An \Xmax distribution of some primary energy and mass can be modelled as the convolution of a Gaussian with an exponential \cite{Peixoto:2013tu}. Three shape parameters $(t_{0},\sigma,\lambda)$ define the \Xmax distribution:
\begin{equation} \label{eq:Xmaxbasic}
  \frac{dN}{d\text{X}_{\text{max}}}(t)\ = \frac{1}{2\lambda} \exp\left({\frac{t_{0} -t}{\lambda} + \frac{\sigma^2}{2\lambda^2}}\right)Erfc\left(\frac{t_{0}-t+\frac{\sigma^2}{\lambda}}{\sigma\sqrt{2}}\right)
\end{equation}
where $t_{0}$ defines the mode of the Gaussian component, $\sigma$ defines the width of the Gaussian component and $\lambda$ defines the exponential tail of the \Xmax distribution, and $t$ is the \Xmax bin. The mode and spread of the distribution defined in Equation~\eqref{eq:Xmaxbasic} is sensitive to $t_{0}$ and $\sigma$ respectively.

\begin{figure}
 \includegraphics[width=0.48\textwidth]{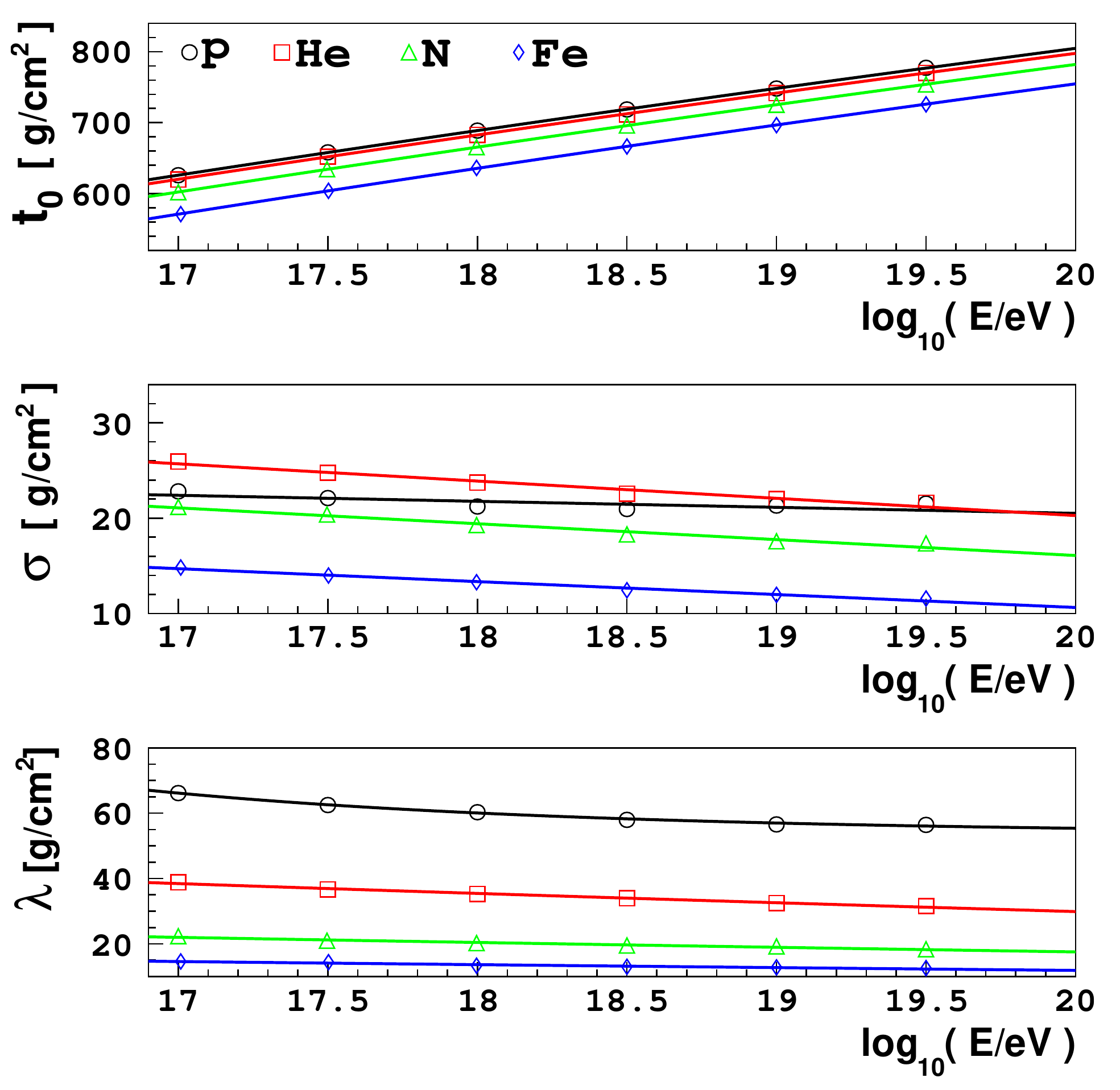}
 \caption{Fits to the shape parameter as a function of energy according to the \epos model.}
 \label{fig:conex_epos}
\end{figure}

We fit Equation~\eqref{eq:Xmaxbasic} to the \Xmax distributions from \conex simulations of cosmic rays of a particular primary energy, mass (either proton, Helium, Nitrogen or Iron primaries) and hadronic interaction model, obtaining the values of $t_{0}$, $\sigma$ and $\lambda$ for that distribution (see Appendix \ref{AppA}). The fit results as a function of energy are displayed in \figsThree{fig:conex_epos}{fig:conex_qgs}{fig:conex_sib}. The solid lines are fits to the shape parameters ( $t_{0}$, $\sigma$ and $\lambda$ ) as a function of energy. The functions fitted are defined as follows:
\begin{equation} \begin{split} \label{eq:Xmaxbasicshape} 
t_{0}(E) &= \tnorm + B\cdot \log_{10}\left(\frac{\log_{10}E}{\log_{10}E_0}\right),
\\ 
\sigma(E) &=  \sigmanorm + C\cdot \log_{10}\left(\frac{E}{E_0}\right), 
\\ 
\lambda(E) &= \lambdanorm - K + K \cdot \left(\frac{\log_{10}E}{\log_{10}E_0}\right)^{\frac{L}{\ln10}} \;  ,
\end{split}\end{equation}
where E is the energy in eV and $E_0 = \energy{18.24} $, the energy at which we choose to normalise the equations. This energy corresponds to the energy at which Auger has measured $\lambda$ for a proton dominated composition~\cite{Collaboration:2012wt}. This means that $ \lambdanorm $ for proton can be directly compared with  $\Lambda_\eta$, the exponential tail measured by Auger, which is shown in Equation~\eqref{eq:AugerLambda}. We even considered adopting  $\Lambda_\eta$ as the value for \lambdanorm, but this could potentially break self consistency in the models.
\begin{equation} \label{eq:AugerLambda} 
\Lambda_\eta = [55.8 \pm 2.3(stat) \pm 1.6(sys)]\; \gcm
\end{equation}
The coefficients in Equation~\eqref{eq:Xmaxbasicshape} are specified in Appendix \ref{AppB} for each mass component and hadronic model. 

\begin{figure}
 \includegraphics[width=0.48\textwidth]{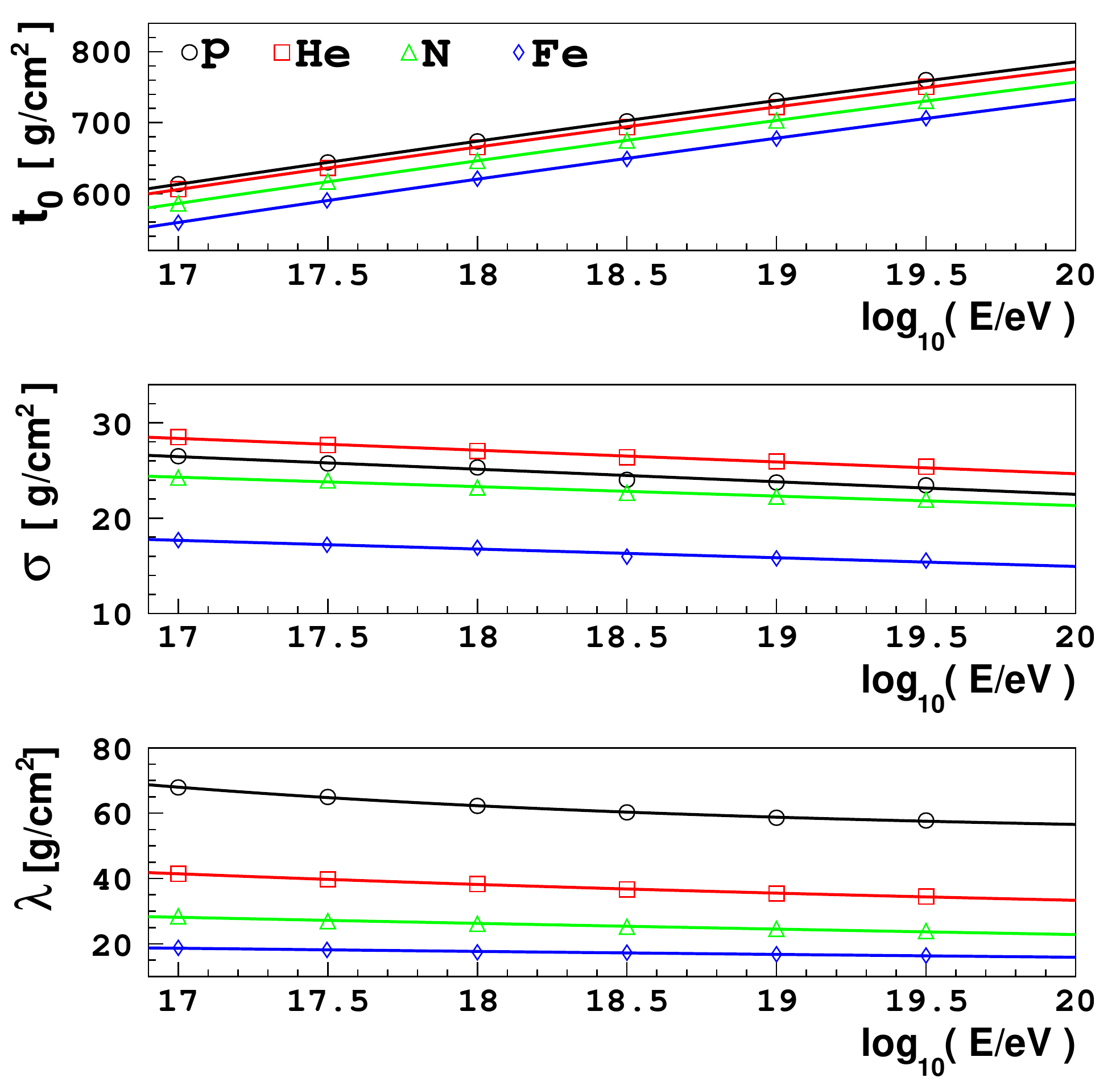}
 \caption{Fits to the shape parameter as a function of energy according to the \qgs model.}
 \label{fig:conex_qgs}
\end{figure}

\begin{figure}
 \includegraphics[width=0.48\textwidth]{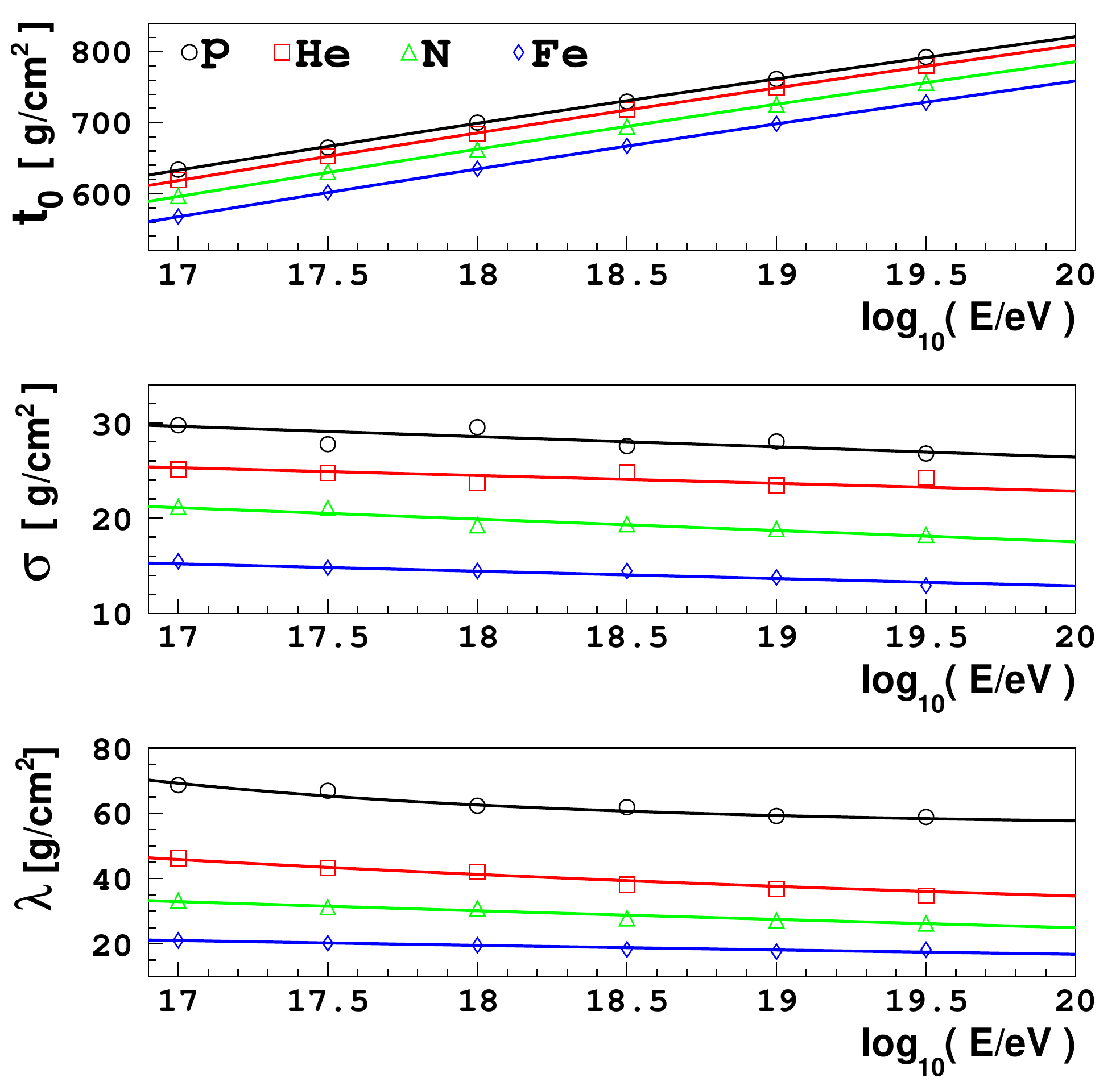}
 \caption{Fits to the shape parameter as a function of energy according to the \sib model.}
 \label{fig:conex_sib}
\end{figure}

The functions of Equation~\eqref{eq:Xmaxbasicshape} consist of two parts, the first part defining the value of a shape parameter at the normalisation energy, and the second part defining the change in the shape parameter as a function of energy. For example, for protons \tnorm would be the value of $t_0$ for protons at \energy{18.24}, and similarly \sigmanorm would be the value of $\sigma$ at \energy{18.24}.

\subsection{\label{sec.reso_acc} Accounting for the detector resolution and acceptance}

The expected \Xmax distributions are affected by the detector resolution and the detector acceptance. The Pierre Auger \Xmax publication~\cite{Aab:2014kda} provides parametrisations for the average detector \Xmax resolution as a function of energy ($Res(E)$) and the detector acceptance as a function of \Xmax for each energy bin, $Acc(E,t)$, where  $t$ is the \Xmax bin as in  Equation~\eqref{eq:Xmaxbasic}.

The detector \Xmax resolution is accounted for by adding it in quadrature with the corresponding $\sigma(E)$, to provide the total expected value of $\sigma(E)_{tot}$ for some primary:
\begin{equation} \label{eq:AcoountReso}
  \sigma(E)_{tot} = \sqrt{\sigma(E)^2 + Res(E)^2}
\end{equation}

We can combine Equations~\eqref{eq:Xmaxbasic}, \eqref{eq:Xmaxbasicshape}, \eqref{eq:AcoountReso} and the detector acceptance $Acc(E,t)$ to obtain the expected \Xmax distribution for cosmic rays of a mixture of primary masses in a particular energy bin according to a hadronic interaction model:

\begin{equation}
  \begin{split}\label{eq:Xmaxfinal}
    \frac{dN}{d\text{X}_{\text{max}}}(E,t)\bigg|_{\text{total}} &=  \\
    N(E)Acc(E,t)&\sum_{i=p,He,N,Fe}f_i(E)\:\frac{dN}{d\text{X}_{\text{max}}}(E,t)\bigg|_i
  \end{split}
\end{equation}
where $f_p(E)$, $f_{He}(E)$, $f_{N}(E)$ and $f_{Fe}(E)$ are the fractions of proton, Helium, Nitrogen and Iron events respectively, and $N(E)$ is the total number of events. The fractions $f_p$, $f_{He}$, $f_{N}$ and $f_{Fe}$ are all correlated. Furthermore, the range of allowed values is not always $[0,1]$. This range changes depending on the values of the other fractions. For example, if $f_p$ were $0.9$, the allowed range for any of the other fractions would be $[0,0.1]$. In order to avoid changing the fraction limits in an iterative way, we have expressed the fractions $f_p$, $f_{He}$, $f_{N}$ and $f_{Fe}$ in terms of $\eta_1$, $\eta_2$ and $\eta_3$ as follows:
\begin{align}
  \label{eq:massparameters} 
  f_p(E) &= \eta_1  \nonumber \\
  f_{He}(E) &= (1-\eta_1)\eta_2  \nonumber \\
  f_N(E) &= (1-\eta_1)(1-\eta_2)\eta_3  \nonumber \\
  f_{Fe}(E) &= 1 - f_p(E) - f_{He}(E) - f_{N}(E)
\end{align}
Therefore, each energy bin has a set of $\eta_1$, $\eta_2$ and $\eta_3$ which defines the mass fractions of that energy bin. The allowed range for $\eta_1$, $\eta_2$ and $\eta_3$ is always $[0,1]$, consequently the mass fractions are constrained to values between 0 and 1 whilst the sum of the mass fractions equals 1. So, in practice we fit $\eta_1$, $\eta_2$ and $\eta_3$ to determine the corresponding fractions ($f_p$, $f_{He}$, $f_{N}$, $f_{Fe}$).

\begin{figure*}
\centering \includegraphics[width=0.98\textwidth]{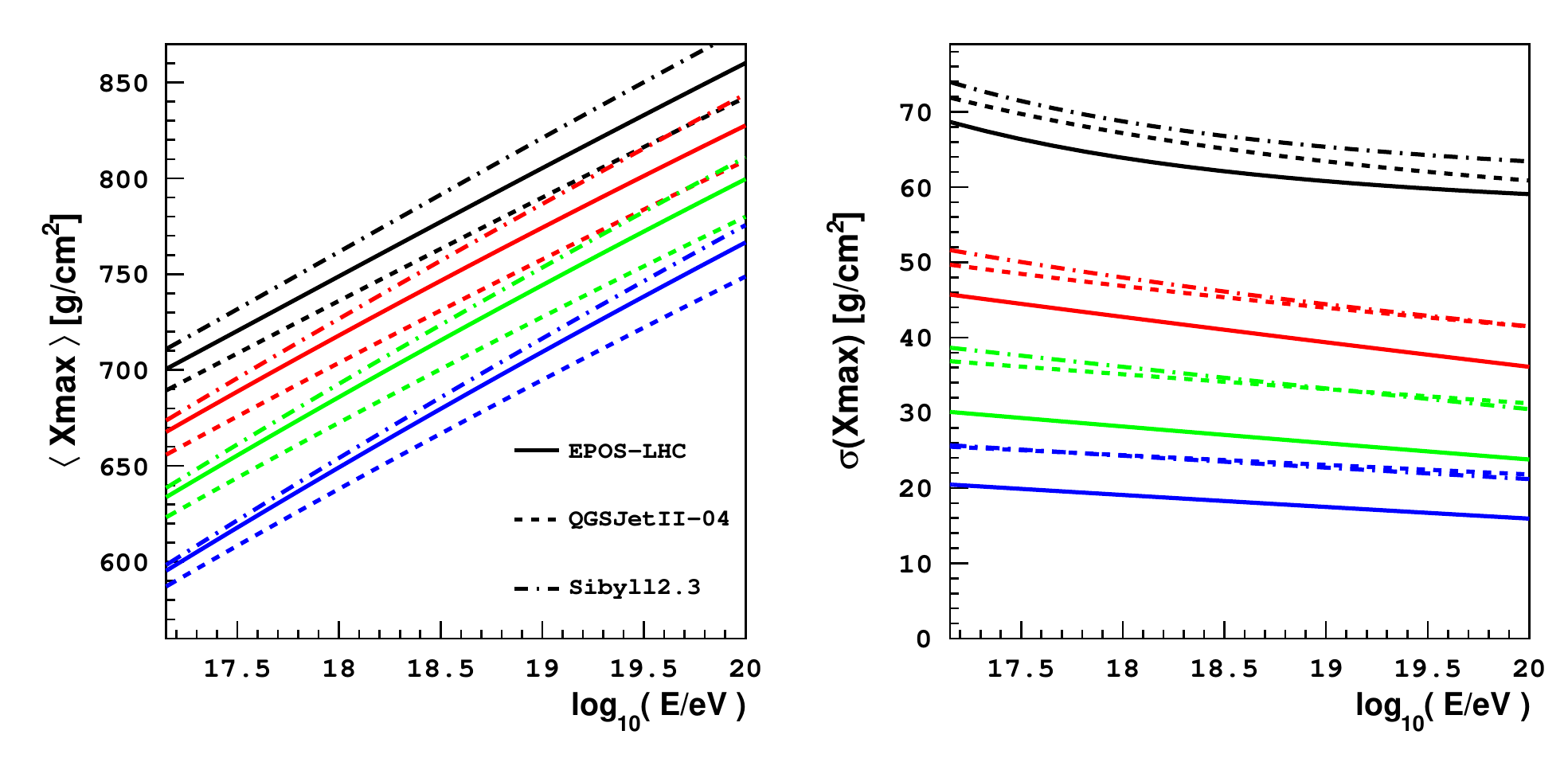}
\caption{The \meanXmax and \sigmaXmax predictions of the \epos, \qgs and \sib \Xmax parameterisations for proton (black), helium (red), nitrogen (green) and iron (blue).}
\label{fig:predicted_Xmax_RMS}
\end{figure*}

 \fig{fig:predicted_Xmax_RMS} displays the \meanXmax and \sigmaXmax predictions of the three parameterisations for each primary. The predicted \meanXmax separation of each adjacent mass component (eg. proton vs. helium, helium vs. nitrogen) within a parameterisation is approximately \SI{30}{\gcm} to \SI{40}{\gcm}. The predicted \sigmaXmax of the primaries is much larger for the \qgs and \sib parameterisations than the \epos parameterisation.

\subsection{\label{sec:validation}Validation of the parameterisation}

\begin{figure}
\centering
  \includegraphics[width=0.48\textwidth]{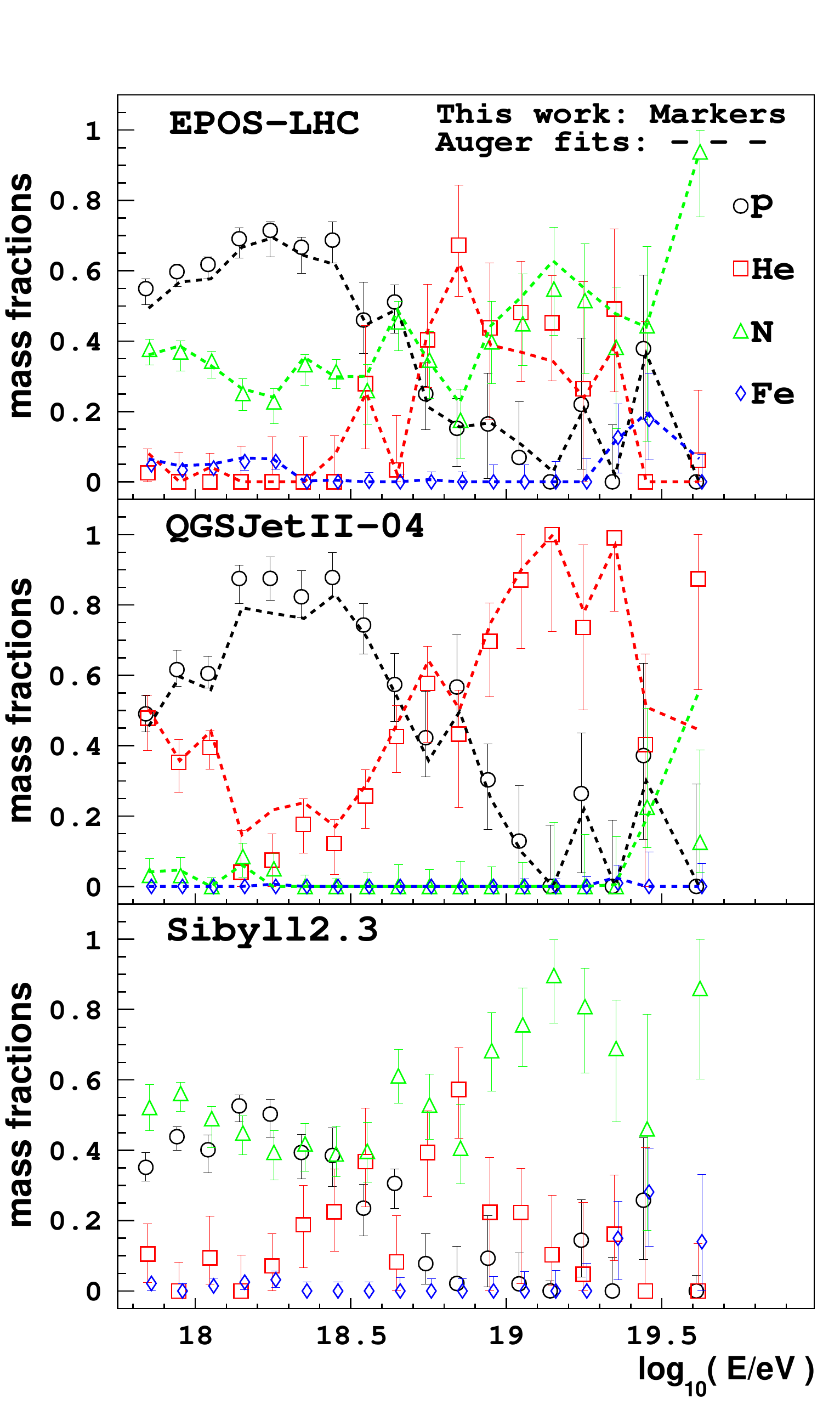}%
\caption{Fitting only the mass fractions of our parameterisations to FD \Xmax data measured by the Pierre Auger Observatory. The error bars represent the statistical error of the fits. Included is the mass composition results for each hadronic model from the Pierre Auger Observatory analysis (labelled `Auger fits'). \cite{Aab:2014aea}.}
\label{fig:realdata_comp_0_all}
\end{figure}

\fig{fig:realdata_comp_0_all} displays the mass composition results of fitting the mass fractions using our \epos, \qgs or \sib \Xmax parameterisations and the \Xmax data measured by the Pierre Auger Observatory fluorescence detector (FD) \cite{Aab:2014kda}. The fits took into account the detector resolution and acceptance. The mass composition obtained using our \Xmax parameterisations are consistent with the Auger analysis of the 2014 FD \Xmax data set \cite{Aab:2014aea}, where \Xmax distribution templates from hadronic interaction models were compared to the data. The compatibility of our results with the 2014 Auger analysis validates the accuracy of our \Xmax parameterisations.

\section{Method}\label{sec.method}
The parameters of Equation~\eqref{eq:Xmaxfinal} are fitted to energy binned \Xmax distributions. The coefficients of Equation~\eqref{eq:Xmaxbasicshape} shown in Appendix \ref{AppB} were obtained with a global fit which included all energy bins.

When fitting (the \Xmax distribution data) for the mass fraction parameters using our \epos, \qgs or \sib parameterisation with the coefficients fixed (as in \fig{fig:realdata_comp_0_all}), the resulting mass composition reflects the characteristics of the corresponding hadronic model. Therefore, the estimated composition depends on which hadronic model is used. Additionally, the mass composition fitted to each energy bin is independent of the mass composition fitted to other energy bins.  However, by including some of the coefficients shown in Appendix \ref{AppB}  in the fit, in addition to the mass composition fractions, the mass composition obtained has a reduced dependence on the hadronic interaction model assumed. In this alternative case the mass composition fitted at each energy bin has some dependence with the fits at other energy bins. This is because the fitted coefficients (from the \Xmax parameterisation) are fitted using all energy bins, while in the first case these coefficients were fixed.   

\begin{figure}
 \includegraphics[width=0.48\textwidth]{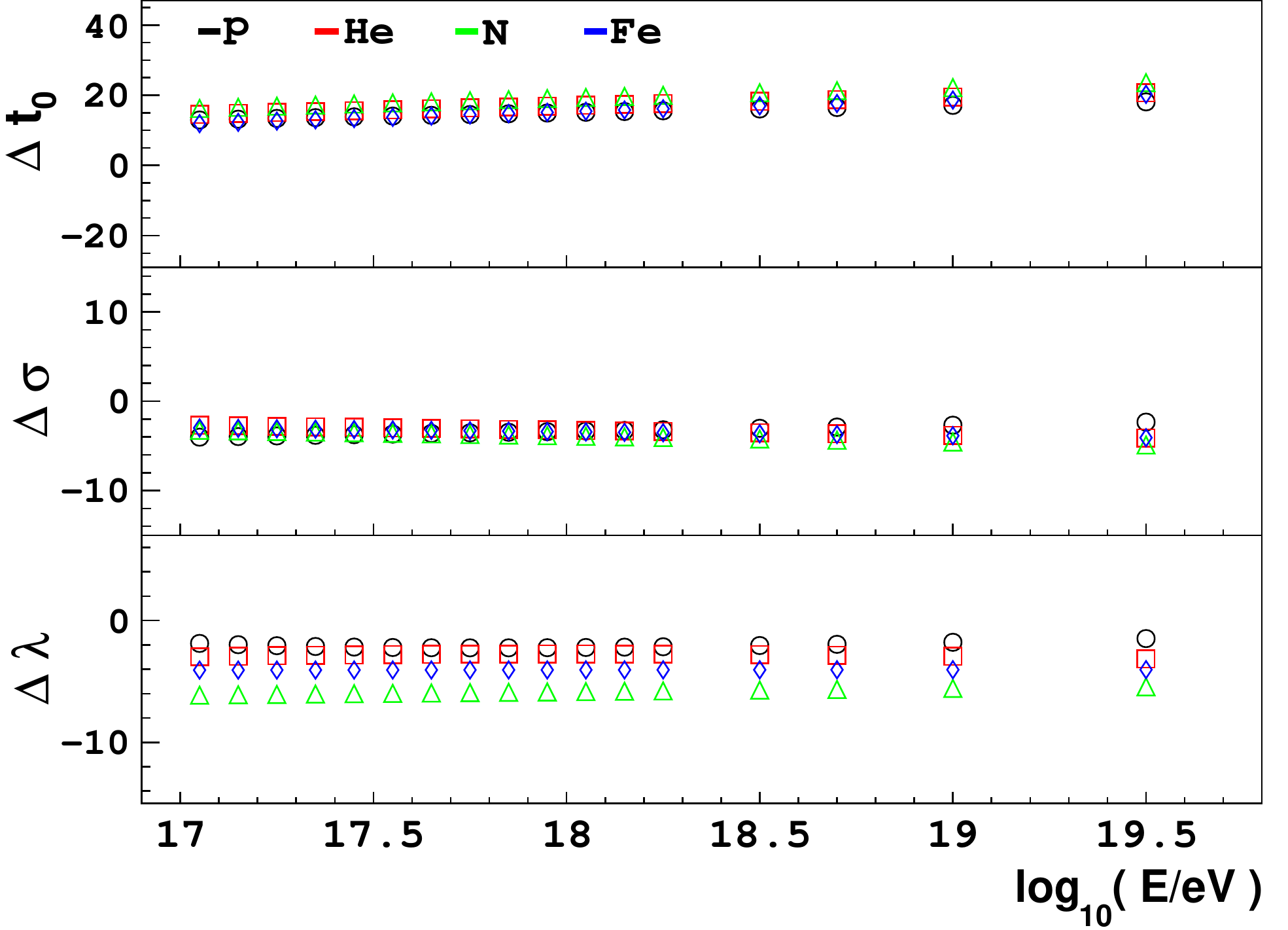}
 \caption{\epos shape parameter value minus \qgs shape parameter value for some mass and energy.}
 \label{fig:model_diff_comparison_eq}
\end{figure}

      \begin{figure}
    \includegraphics[width=0.48\textwidth]{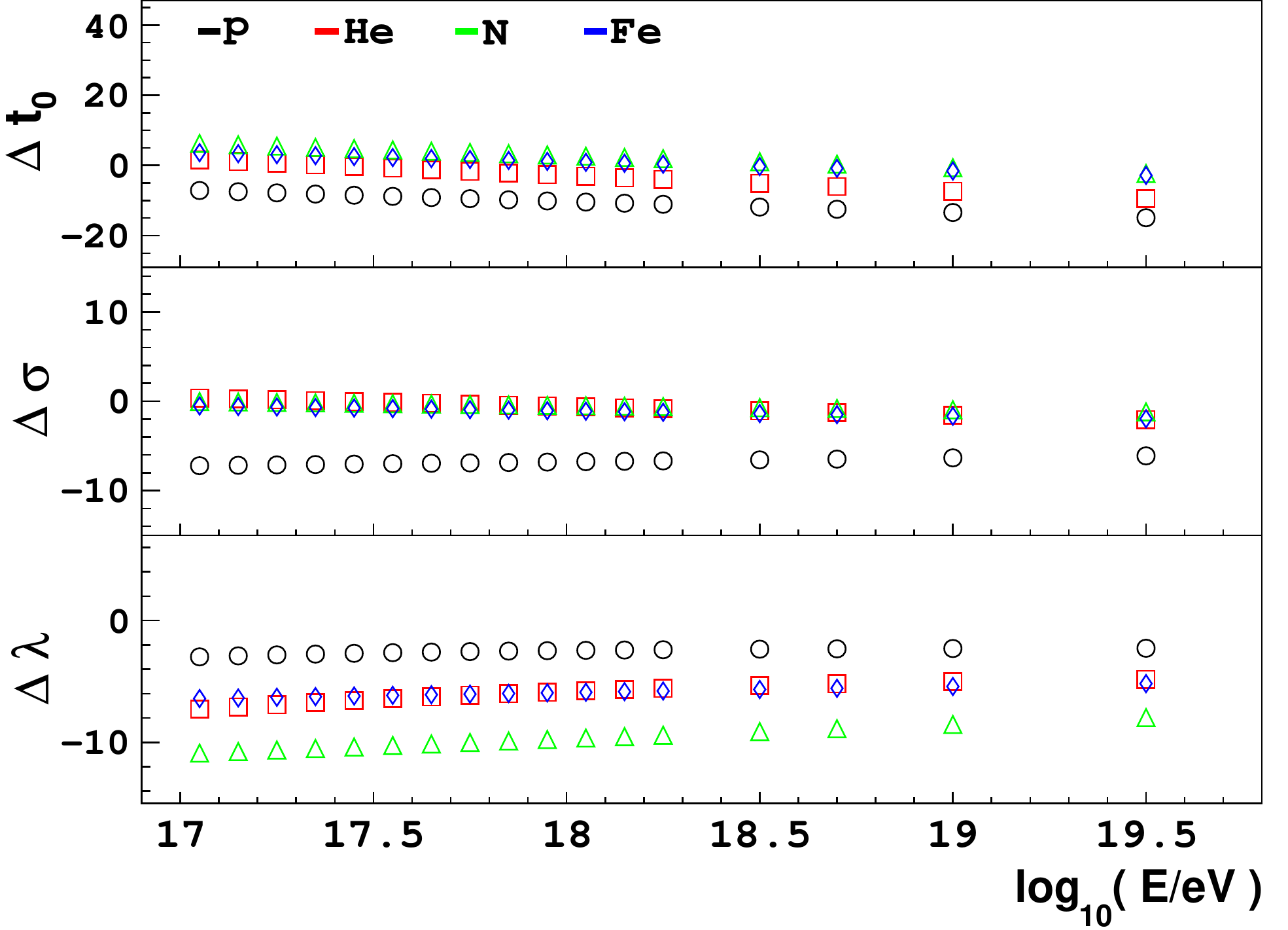}%
    \caption{\epos shape parameter value minus \sib shape parameter value for some mass and energy.}
     \label{fig:model_diff_comparison_es}
    \end{figure}  

     \begin{figure}
    \includegraphics[width=0.48\textwidth]{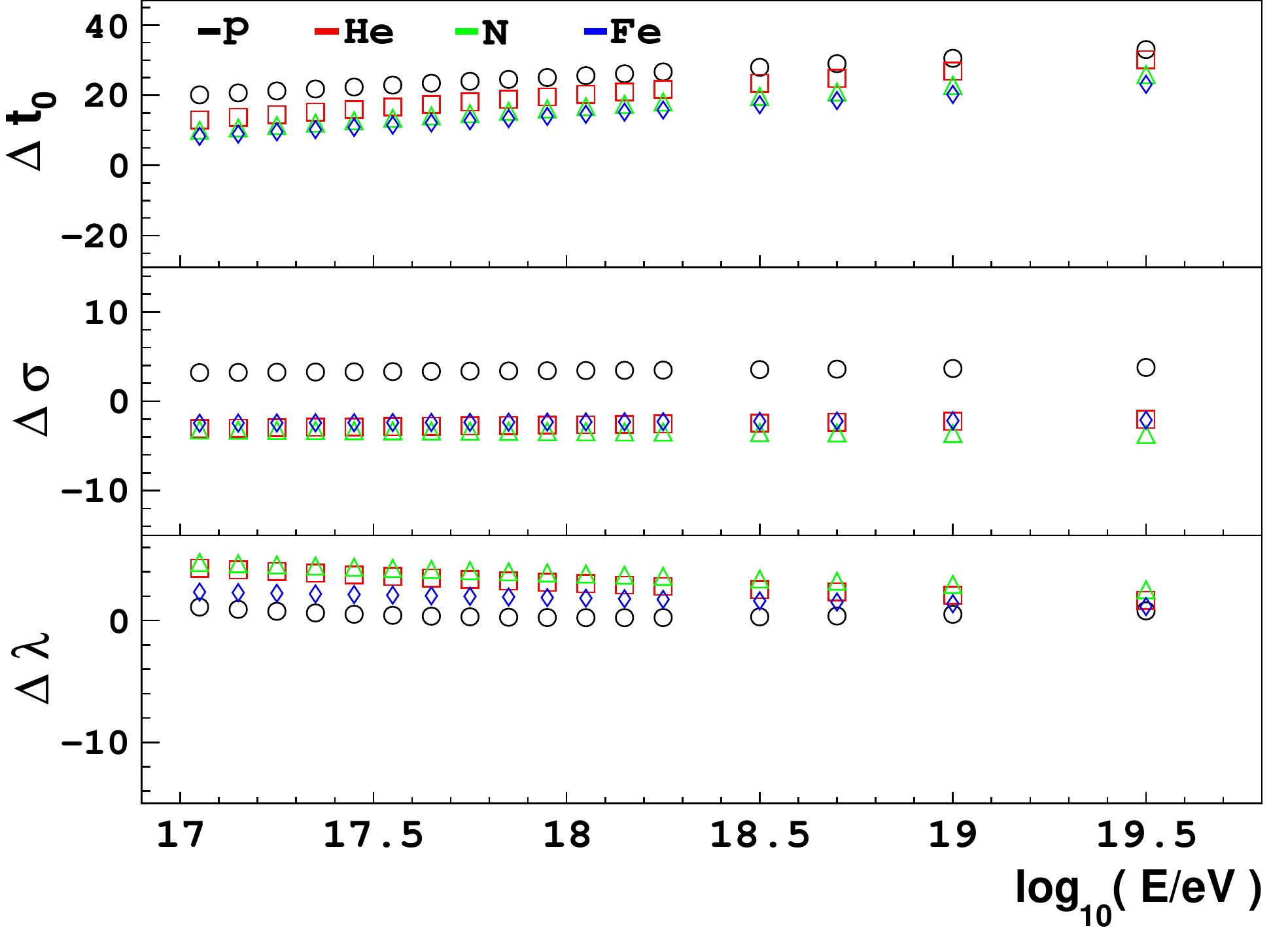}%
    \caption{\sib shape parameter value minus \qgs shape parameter value for some mass and energy.}
     \label{fig:model_diff_comparison_sq}
    \end{figure}  
In principle, if we were able to use the Auger \Xmax data to perform a global fit of the mass composition and all of the coefficients from Equation~\eqref{eq:Xmaxbasicshape}, the resulting composition would be independent of the hadronic models, depending only on the assumed functional forms of the equations. However, the degeneracy between the fitted mass fractions and the coefficients makes it impossible to unambiguously constrain all of these parameters (i.e. the solution would be degenerate). Therefore, we need to identify which coefficients are most relevant for interpreting the mass composition, and evaluate whether we can unambiguously fit these coefficients and the mass composition. One way to identify which coefficients to include in a global fit is to compare the values of  $t_{0}$, $\sigma$ and $\lambda$ between different models. This comparison will identify the parameters that are well or poorly constrained by our current knowledge of the high energy hadronic interaction physics.    

\figsThree{fig:model_diff_comparison_eq}{fig:model_diff_comparison_es}{fig:model_diff_comparison_sq} illustrates the  $t_{0}$, $\sigma$ and $\lambda$  difference between the \epos, \qgs and \sib parameterisations at some energy and mass. The differences as a function of energy are relatively small. For example, the slope of $\Delta t_{0}$ as a function of energy is less than $\sim5\;\gcm$ /energy-decade, which is small compared with an elongation rate of $60\;\gcm$ /energy-decade.  We have also verified that the separation between  different primaries in  the $t_{0}$, $\sigma$ and $\lambda$ space is similar for the three tested models. The main differences between our  \epos, \qgs and \sib \Xmax parameterisations are the normalisation of $t_0$ and $\sigma$. The difference in the normalization of $\lambda$ is not negligible, but it has little impact on the mass composition interpretation. Therefore, when including \tnorm and \sigmanorm in the global fit, we should obtain a similar interpretation of the mass composition with either the \epos, \qgs or \sib \Xmax distribution parameterisation. We choose to fit  \tnorm and \sigmanorm  in the following way:

\begin{itemize}

  \item \tnorm is fitted such that the absolute values of \tnorm for each primary change by the same amount. Therefore, the difference in \tnorm between primaries is conserved. 

\item \sigmanorm is fitted such that the ratio of $\sigma$ between primaries remains similar to the initial ratio over the energy range (differences in $C$ between primaries prevents the exact conservation of the initial ratio). Therefore, if \sigmanorm for protons changes by $\Delta$, \sigmanorm for other primaries will change by $\Delta$ multiplied by the initial average ratio of $\sigma$ between that primary and proton.

\end{itemize}

  Fitting \tnorm and \sigmanorm in this way assumes the hadronic models are correctly predicting the separation in $t_0$ between different species, and the ratio of $\sigma$ between different species, over the fitted energy range. 

  In Equation~\eqref{eq:Xmaxbasicshape}, the values of the shape parameters for Helium, Nitrogen and Iron can be expressed in terms of the corresponding values for protons, therefore fitting \tnorm and \sigmanorm in the way described above can be implemented by simply fitting \tnorm and \sigmanorm for protons.

 In order to avoid unphysical fit results, we constrain the possible fitted values for \tnorm and \sigmanorm.  These constraints are significantly wider than the separation between the \epos, \qgs and \sib \Xmax parameterisation predictions for these coefficients. The predicted value of \tnorm for protons according to \epos is $\sim \SI{703}{\gcm}$, according to \qgs is $\sim \SI{688}{\gcm}$, and according to \sib is $\sim \SI{714}{\gcm}$. The minimum and maximum limits of \tnorm for protons are set to \depth{670} and \depth{765} respectively. The predicted value of \sigmanorm for protons according to \epos, \qgs and \sib is $\sim$ \depth{22}, $\sim$ \depth{25} and $\sim$ \depth{28}  respectively. The minimum and maximum limits of \sigmanorm for protons are set to \depth{5} and \depth{55} respectively. 

With a suitable shift in \tnorm and \sigmanorm, many primary mixtures which produce a fairly smooth total distribution can be fitted well with a single dominant distribution, instead of a sum of distributions. On the other hand, a distribution dominated by a single primary can be well fitted by a balanced mixture of distributions when \tnorm and \sigmanorm are shifted appropriately. It is common that \Xmax distributions can be fitted with a value of \tnorm for protons much larger than the true \tnorm of the distributions, which results in the primary mass of the events being overestimated (i.e. biased towards heavier masses). Therefore, it is important that appropriate shape coefficient limits are chosen.


We have evaluated the performance of fitting \tnorm and \sigmanorm in addition to the mass fractions using simulated \Xmax distributions of a known composition (see details in Sec. \ref{sec.performance}). Provided there is enough dispersion of masses in the data, it is possible to fit with good accuracy, \tnorm, \sigmanorm and the corresponding abundance (fractions) of p, He, N and Fe.  An important achievement from including \tnorm and \sigmanorm in the fit is that the mass composition interpretation becomes consistent whether using the predicted \epos, \qgs or \sib parameterisation.

The requirement of a large dispersion of masses is evaluated over the entire energy range. For example, a data set consisting of a pure proton composition at higher energies can be fitted, provided that at lower energies we have populations consisting of other primaries. If the statistics or mass dispersion were not large enough, there would be some degeneracy in the fit between the mass fractions and \tnorm and \sigmanorm. A greater change in the mass composition with energy improves the accuracy of the fit.

 Apart from the dispersion of masses in the data, the performance of the fit depends on the intrinsic values for $\sigma$ of the data. This is nature's width for the \Xmax distribution of the different primaries. The separation of the distribution modes between primaries remains unchanged in the fit, therefore primary \Xmax distributions of larger width will increase the \Xmax distribution overlap of adjacent primaries, resulting in the fit of \tnorm, \sigmanorm and the mass composition becoming more uncertain.

We have also evaluated the performance of fitting \tnorm, $B$, and \sigmanorm in addition to the mass fractions, where $B$ defined in Equation~\eqref{eq:Xmaxbasicshape} describes the change in $t_0$ with energy. As the predicted mass composition is particularly sensitive to the predicted values of $t_0$, $B$ is a powerful coefficient which can significantly affect the fitted mass composition. We fit $B$ such that for each primary the value of $B$ changes by the same amount from the initial predicted value, thus the initial predicted differences among primaries in the rate of change of $t_0$ with energy are conserved (identical to how \tnorm is fitted). Our \Xmax parameterisations have similar values for $B$, therefore we do not expect fits of $B$ to yield results significantly different from the initial prediction of $B$ when we are fitting \epos, \qgs or \sib simulated \Xmax data. However, if the values of $B$ predicted by our parameterisations are significantly incorrect for the data being fitted, considerable systematics would be introduced to the reconstructed mass composition if $B$ remains fixed.

Data sets that can be fitted with \tnorm and \sigmanorm may not be accurately fitted when $B$ is included in the fit, as fitting extra coefficients increases the degeneracy between the fitted variables. Fitting these three coefficients accurately requires a greater spread of primaries and/or statistics than fitting just \tnorm and \sigmanorm. The predicted value of $B$ for protons according to \epos, \qgs and \sib is $\sim$ \depth{2533}, $\sim$ \depth{2445} and $\sim$ \depth{2666} respectively. With \tnorm normalised at \energy{18.24}, a change in $B$ of \depth{350} corresponds to a change in $t_0$ at \energy{19.5} of $\sim$\depth{10}. The  fitting range limits of $B$ for protons is \depth{1000} to \depth{4000}. 

We have also considered constraining $t_0$ at \energy{14}, where the hadronic models are more reliable, and fitting $B$ and \sigmanorm. Fitting $B$ in this way can also provide a consistent mass fraction result between the \epos, \qgs and \sib parameterisation fits of simulated \Xmax data, as the $t_0$ prediction of the fitted energy range adjusts in a way that is similar to the \tnorm fit, with the added advantage that unlike the \tnorm fit, the resulting fitted parameterisation of $t_0$ is consistent with the hadronic model predictions at lower energies. We have found that over the energy range of interest (\energy{17.8} to \energy{20}), fitting \tnorm and \sigmanorm results in a more accurate mass composition reconstruction compared to fitting $B$ and \sigmanorm. This is because there is less degeneracy between the fitted mass fractions and shape parameters when fitting \tnorm and \sigmanorm. Additionally, a $t_0$ parameterisation constrained at \energy{18.24} describes the energy range of interest better than a $t_0$ parameterisation extrapolated from \energy{14}. If a wider energy range was being fitted, then a \tnorm and \sigmanorm fit would be less accurate, because the $t_0$ and $\sigma$ parameterisations of different models do not adequately align over a wider energy range by only adjusting their normalisations. It is also important to recognise that this fit of $B$ is restricted, as we are fixing how $t_0$ changes with energy, and only fitting the rate of change of the $\log_{10}\left(\frac{\log_{10}E}{\log_{10}E_0}\right)$ factor. To properly fit the slope of $t_0$ with energy would require the fit of a third $t_0$ parameter (for example, fitting $B$ and $x$ in $B\cdot\log_{10}\left(\frac{\log_{10}E}{\log_{10}E_0}\right)^x$, where $x$ currently equals 1). 

We have evaluated the effect of different \Xmax bin sizes and energy bin sizes on the performance of the fit. When fitting only the mass fractions, \depth{1} \Xmax binning gives marginally more accurate results than \depth{20} \Xmax binning (\depth{20} is the \Xmax bin size of the Auger \Xmax distributions published in~\cite{Aab:2014kda}). The absolute improvement in the fitted mass fractions is no greater than $3\%$ in an energy bin. However, when fitting \tnorm and \sigmanorm in addition to the mass fractions, using a small \Xmax binning is more important, otherwise the chosen center of the \Xmax bins may significantly affect the fitted results, especially if the statistics are not large.  The predicted separation between different primaries in \tnorm and \sigmanorm can be very small. For example, our \epos parameterisation predicts the difference in \tnorm between proton and helium is only $\sim \depth{6}$.
Therefore, a \depth{20} \Xmax binning (as published in \cite{Aab:2014kda}) can be too coarse, and can shift the apparent \meanXmax of the distribution, which affects the fit of \tnorm. 

Due to similar reasons, the energy bin size is also important. Energy binning that is too large can result in data from the same primary mass, but on opposite extremes of the energy bin, being evaluated as data from different primaries. This is because the separation between the predicted \Xmax distributions of different primaries is small compared to the shift in these \Xmax distributions with energy. We find that an energy binning of $0.1$ in $\log_{10}(E/\text{eV})$ is reasonable. 

\section{Performance}\label{sec.performance}

Using \conex, 100 \Xmax data sets were generated according to both the \epos and \qgs hadronic interaction models for a number of different mass compositions.  The data consists of 17 energy bins, of which there are 13 energy bins of a width of $0.1$ in $\log_{10}(E/\text{eV})$ between \energy{17} and \energy{18.3}, and 4 fixed energy bins at \energy{18.5}, \energy{18.7}, \energy{19} and \energy{19.5}. Each energy bin contains approximately 750 events. The binning of the simulated \Xmax distributions is \depth{1}.

\begin{figure}[htb!]
  \includegraphics[width = 0.48\textwidth]{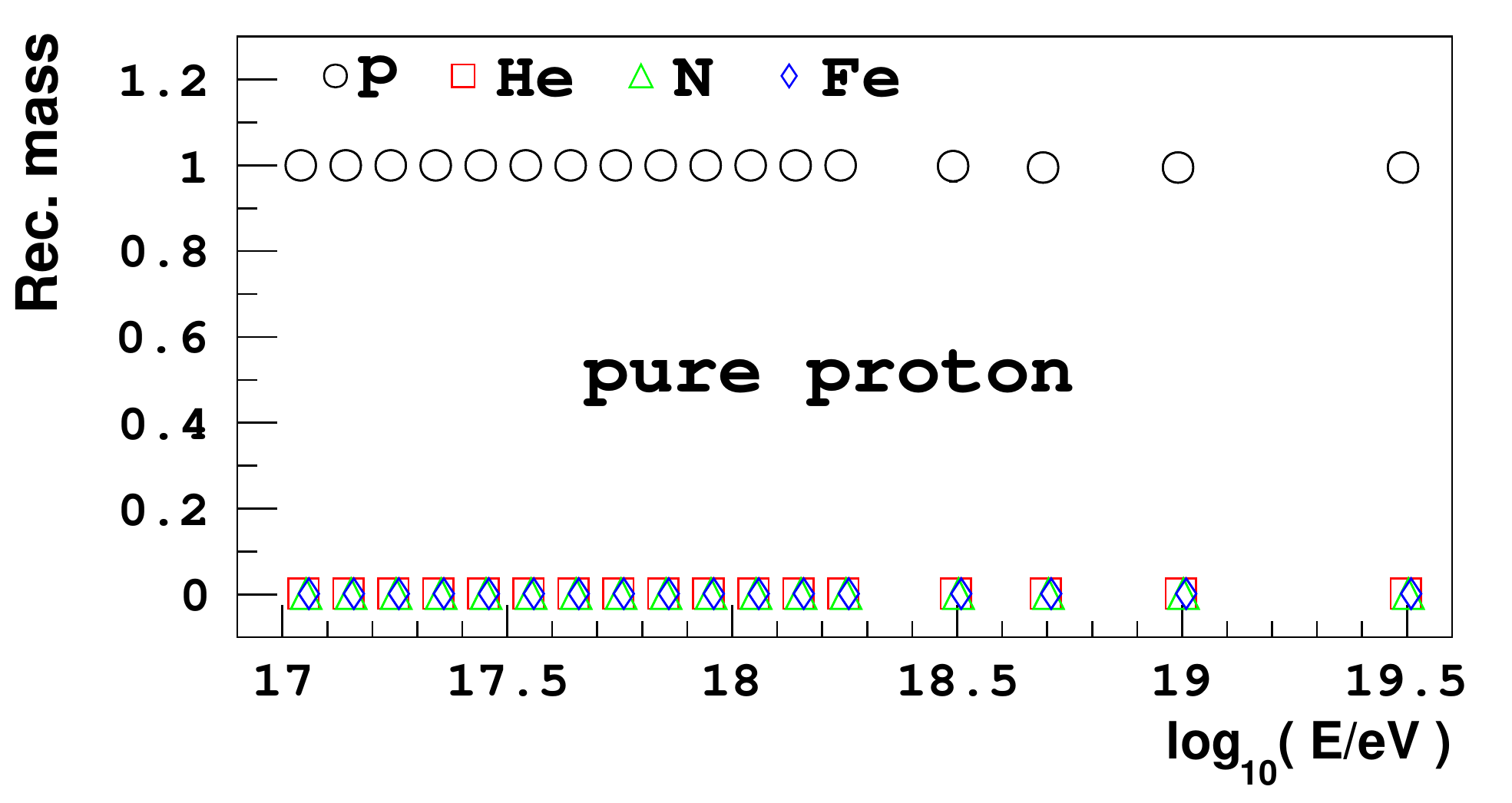}
  \caption{ Fitting only the mass fractions to mock data sets of \Xmax distributions.  The data sets have been generated using the \epos model and assuming a {\bf{proton}} primary composition over the whole energy range. The composition fits were performed using our \Xmax parameterisations for the \epos model predictions. `Rec. mass' refers to the mass fractions fitted to the data.}
  \label{fig:EPOSLHC_pureP}
\end{figure}

\begin{figure}[htb!]
  \includegraphics[width = 0.48\textwidth]{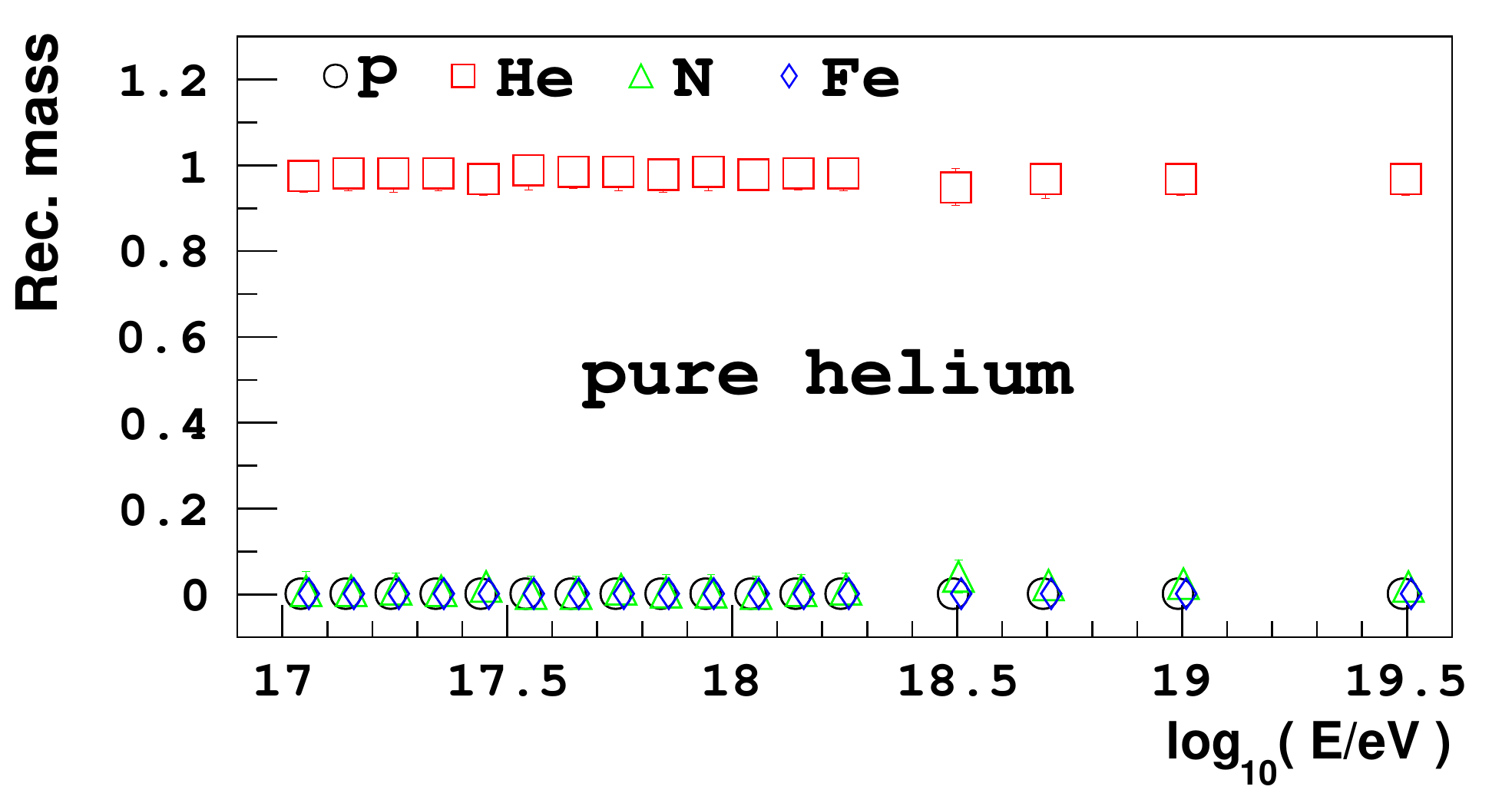}
  \caption{Same as Fig.\,~\ref{fig:EPOSLHC_pureP}, but assuming a {\bf{Helium}} primary composition over the whole energy range.}
  \label{fig:EPOSLHC_pureHe}
\end{figure}

\begin{figure}[htb!]
  \includegraphics[width = 0.48\textwidth]{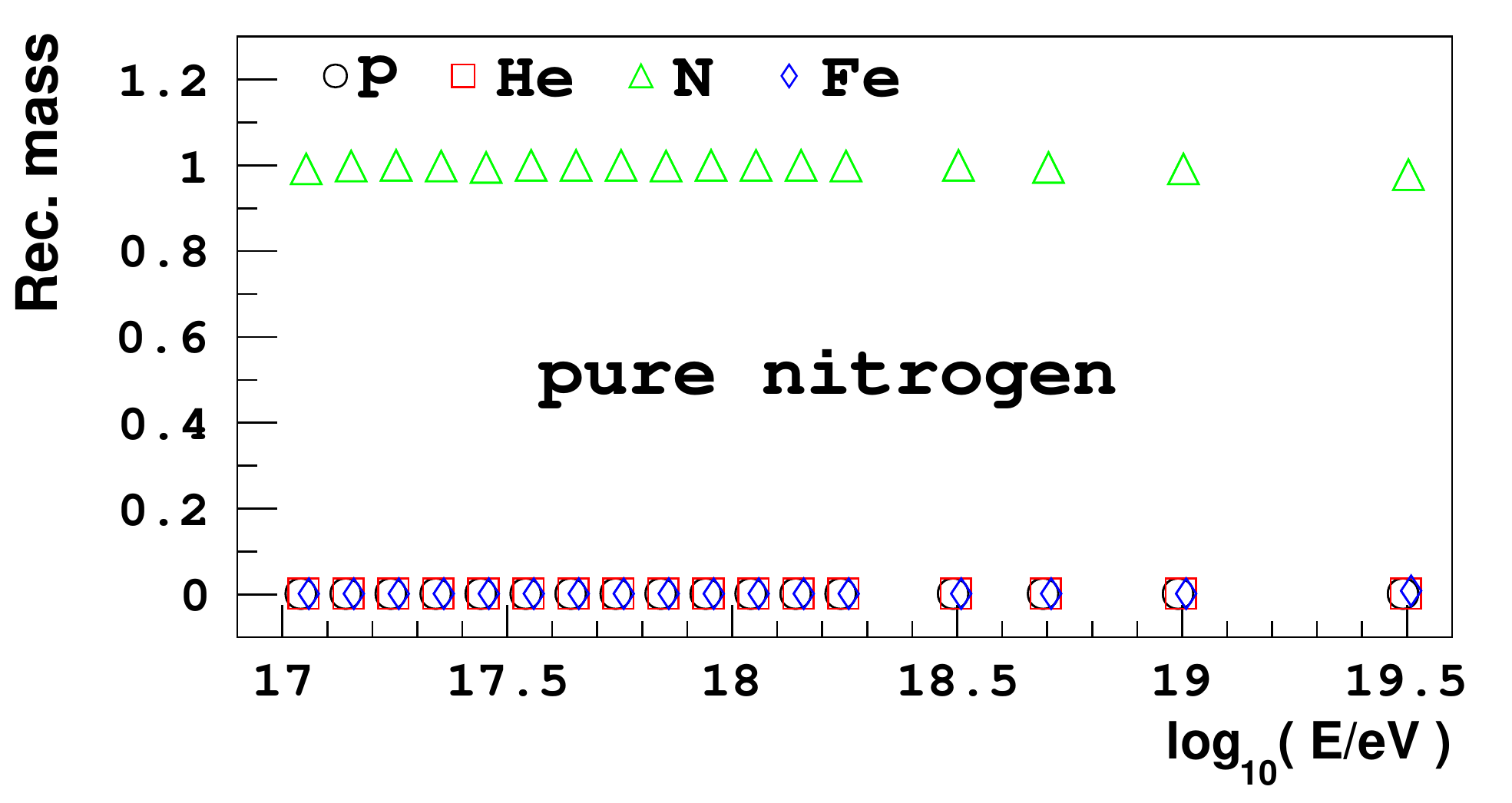}
  \caption{Same as Fig.\,~\ref{fig:EPOSLHC_pureP} but assuming a {\bf{Nitrogen}} primary composition over the whole energy range.}
  \label{fig:EPOSLHC_pureN}
\end{figure}

\begin{figure}[htb!]
  \includegraphics[width = 0.48\textwidth]{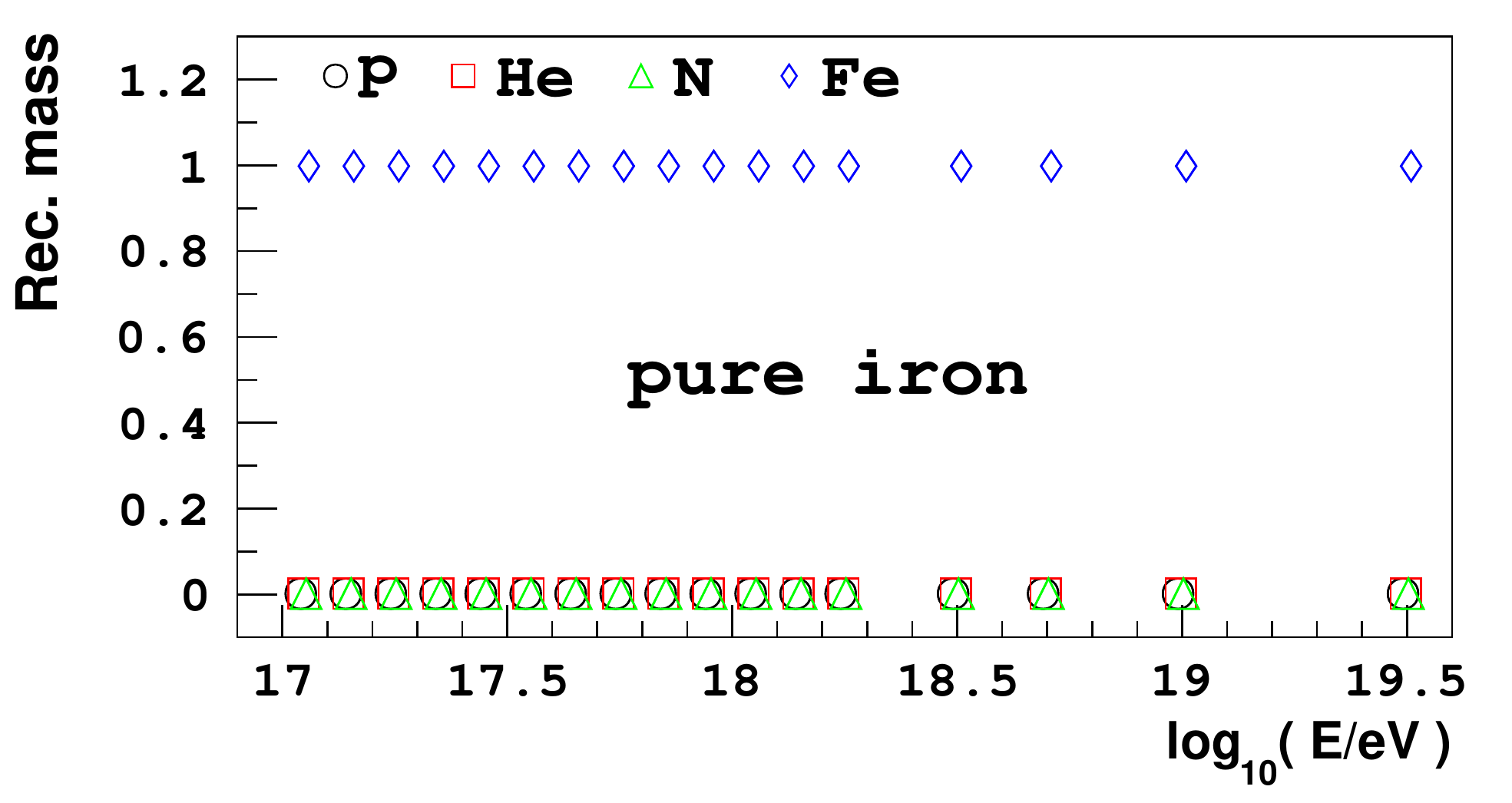}
  \caption{Same as Fig.\,~\ref{fig:EPOSLHC_pureP}, but assuming an {\bf{Iron}} primary composition over the whole energy range.}
  \label{fig:EPOSLHC_pureFe}
\end{figure}

\begin{figure}[htb!]
  \includegraphics[width = 0.48\textwidth]{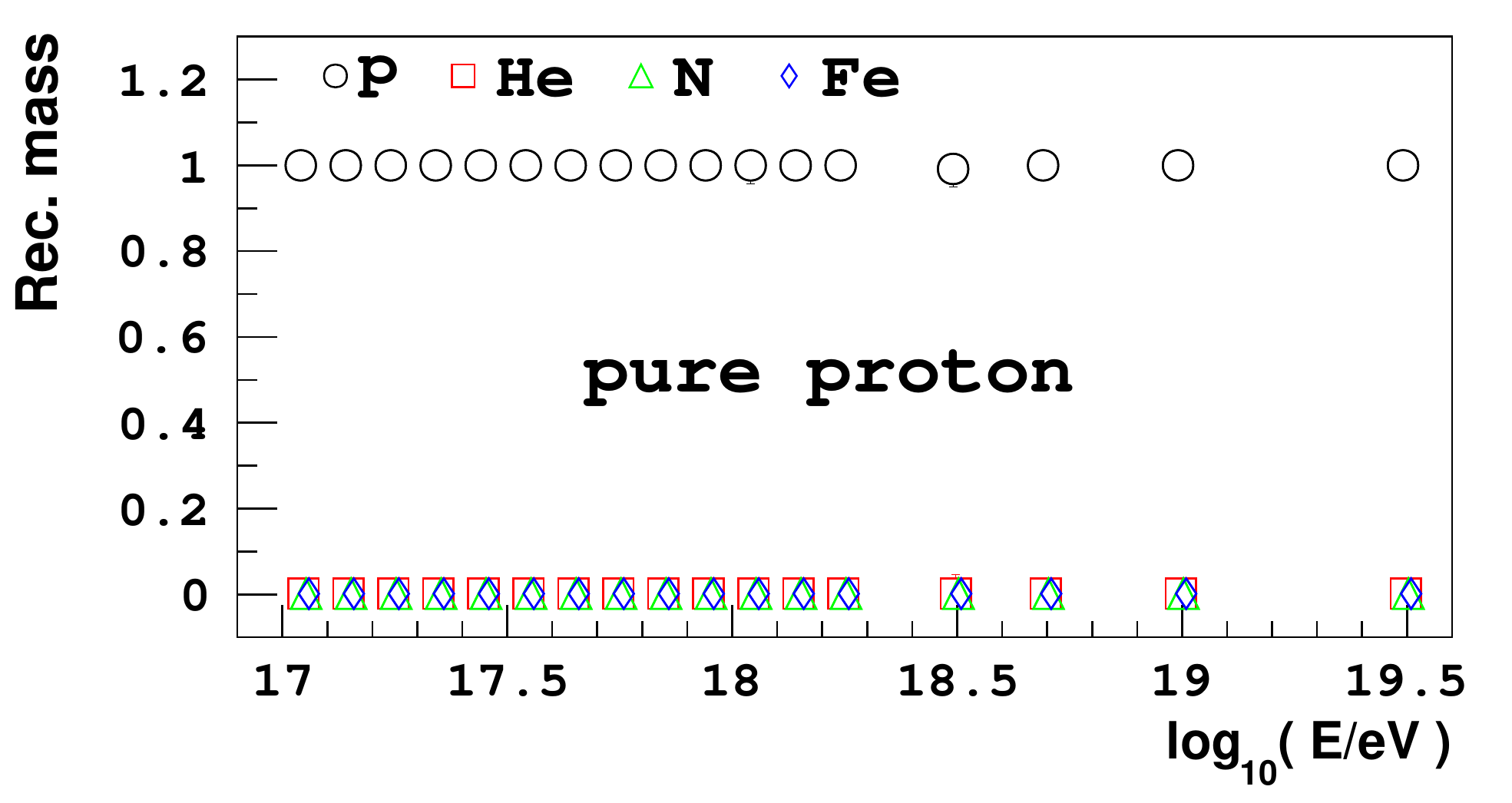}
  \caption{ Fitting only the mass fractions to mock data sets of \Xmax distributions.  The data sets have been generated using the \qgs model and assuming a {\bf{proton}} primary composition over the whole energy range. The composition fits were performed using our \Xmax parameterisations for the \qgs model predictions.}
  \label{fig:QGSJET_pureP}
\end{figure}

\begin{figure}[htb!]
  \includegraphics[width = 0.48\textwidth]{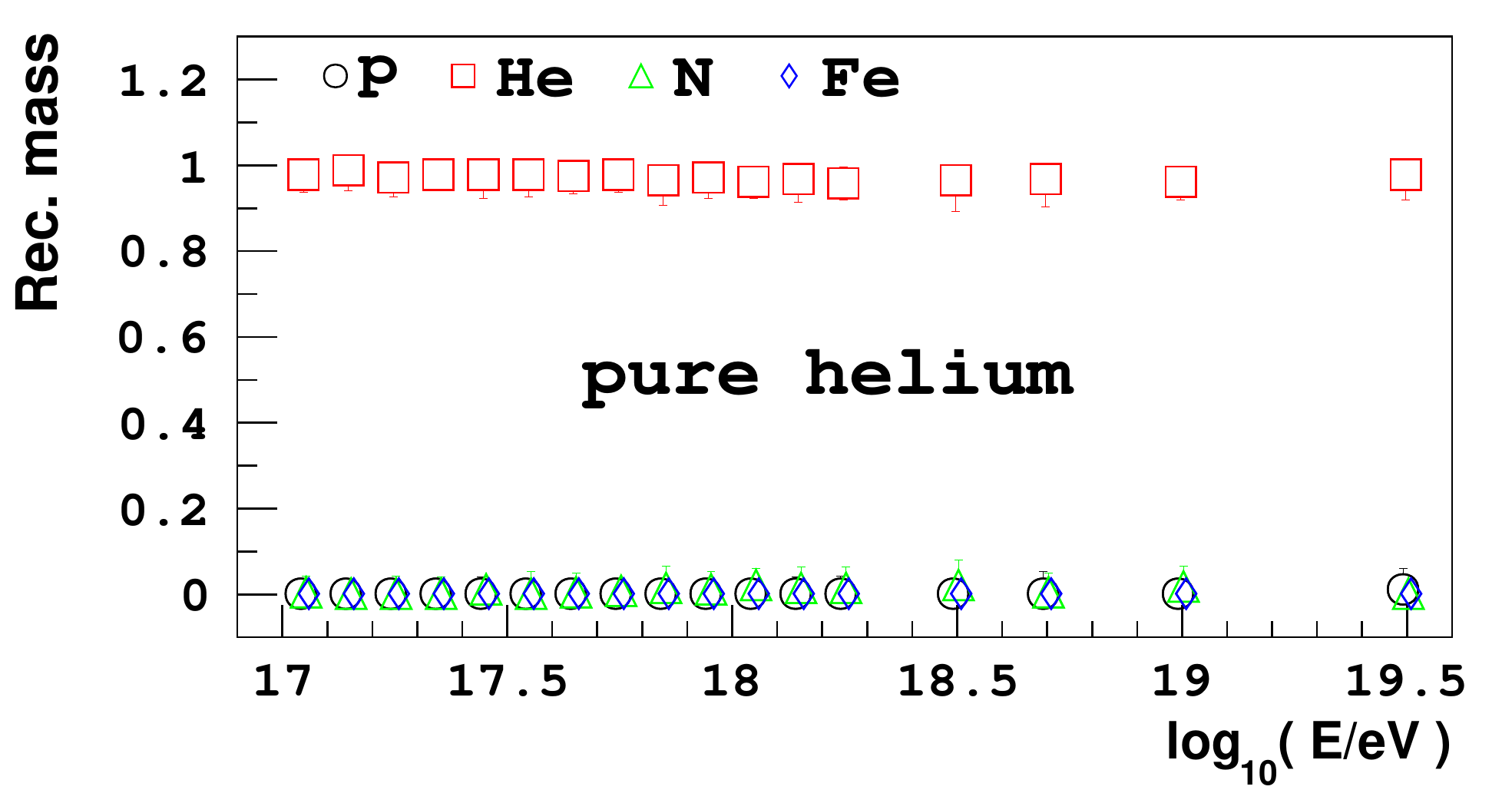}
  \caption{Same as Fig.\,~\ref{fig:QGSJET_pureP}, but assuming a {\bf{Helium}} primary composition over the whole energy range.}
  \label{fig:QGSJET_pureHe}
\end{figure}

\begin{figure}[htb!]
  \includegraphics[width = 0.48\textwidth]{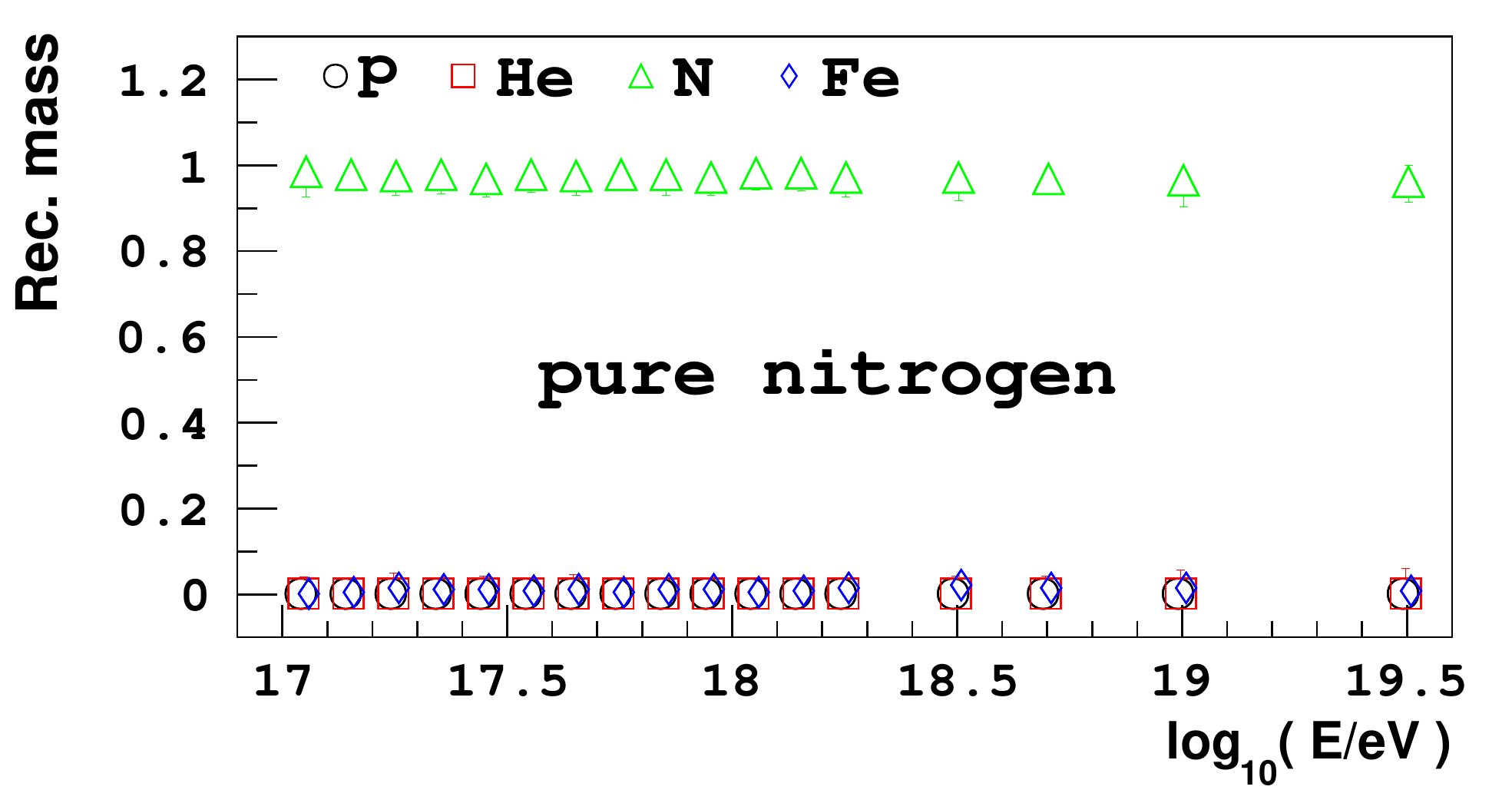}
  \caption{Same as Fig.\,~\ref{fig:QGSJET_pureP} but assuming a {\bf{Nitrogen}} primary composition over the whole energy range.}
  \label{fig:QGSJET_pureN}
\end{figure}

\begin{figure}[htb!]
  \includegraphics[width = 0.48\textwidth]{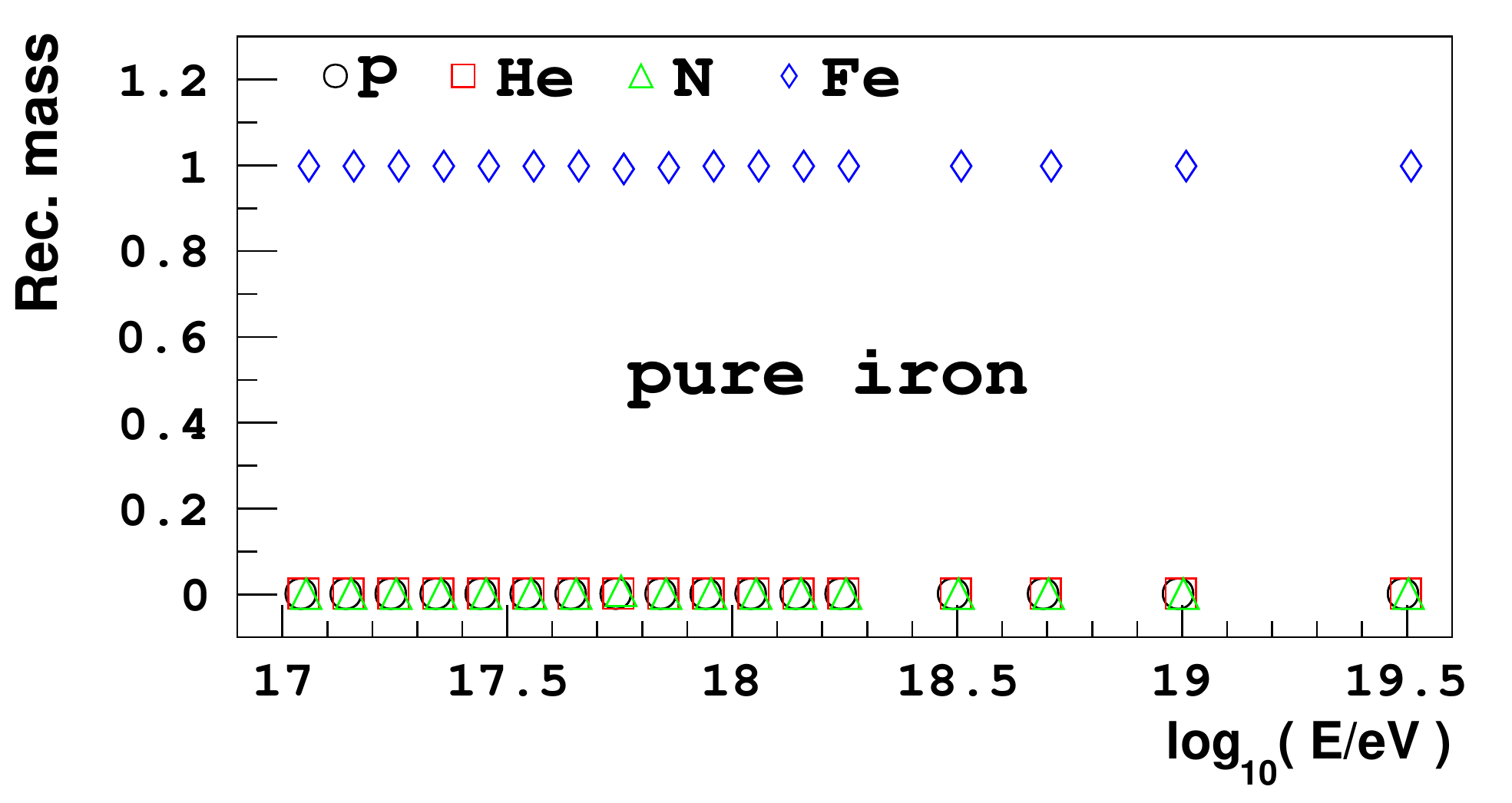}
  \caption{Same as Fig.\,~\ref{fig:QGSJET_pureP}, but assuming an {\bf{Iron}} primary composition over the whole energy range.}
  \label{fig:QGSJET_pureFe}
\end{figure}

We have fitted only the mass fractions (all coefficients from the \Xmax parameterisation were kept fixed) to data of a single primary generated with the same hadronic interaction model the parameterisation fitted is based on. Figs.~\ref{fig:EPOSLHC_pureP} to \ref{fig:EPOSLHC_pureFe} summarises the results (of these 100 fits) for the \epos hadronic model and  Figs.~\ref{fig:QGSJET_pureP} to \ref{fig:QGSJET_pureFe} for the \qgs model. The markers represent the medians of the fitted mass fractions, and the error bars represent the standard deviation. The results show that our \Xmax parameterisations are an accurate description of the expected \Xmax distribution of a primary according to the \epos or \qgs hadronic interaction models. Both our \epos and \qgs \Xmax parameterisation fits can accurately determine the mass composition of data from the same hadronic model.

\begin{figure}[htbp!]
  \includegraphics[width=0.48\textwidth]{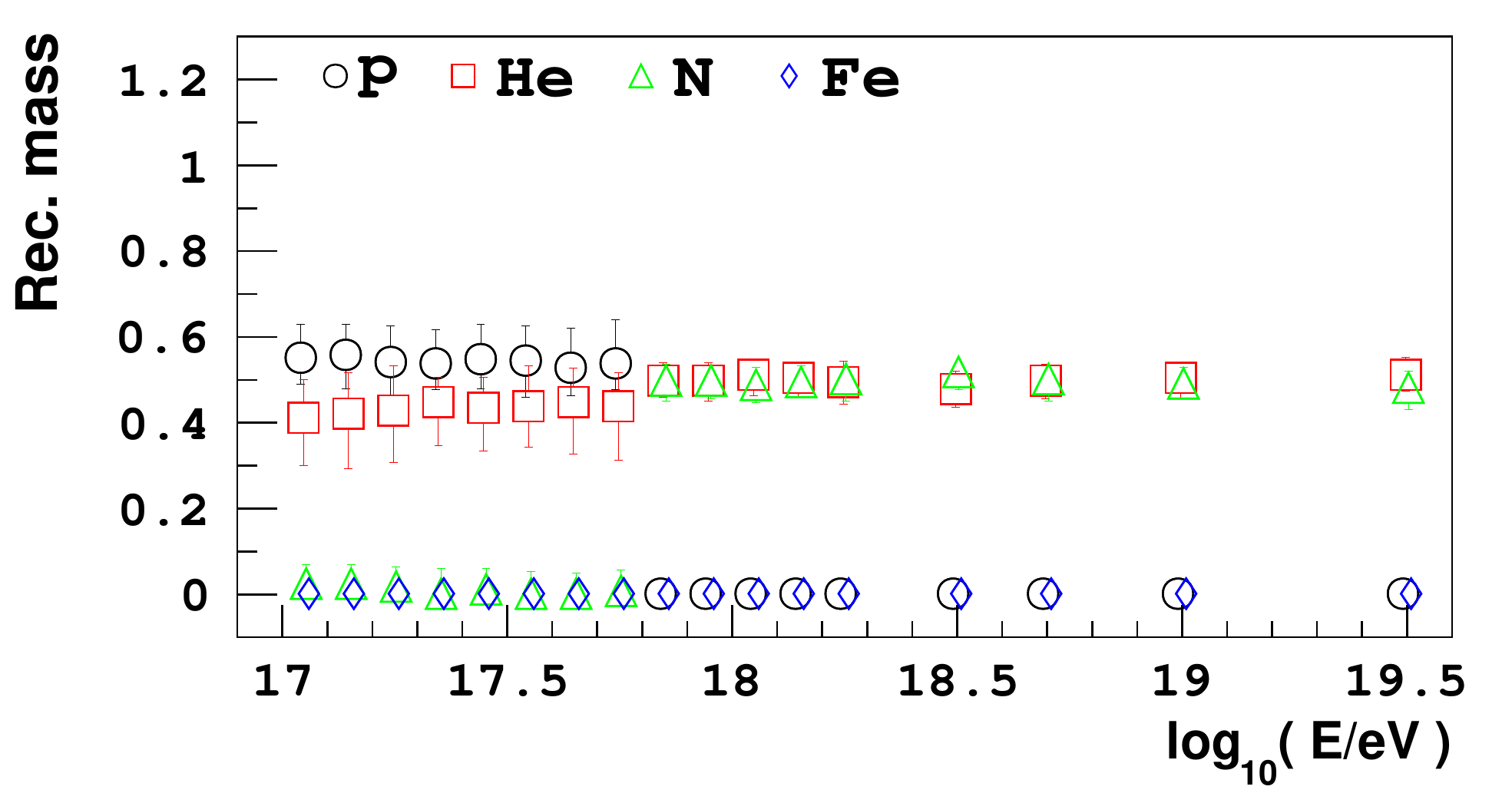}%
  \caption{Fitting only the mass fractions of our \epos parameterisation to \epos \Xmax data.}
  \label{fig:50proton50He_50He50N_EPOS_30_fitmass} 
\end{figure}         

\begin{figure}[htbp!]
  \includegraphics[width=0.48\textwidth]{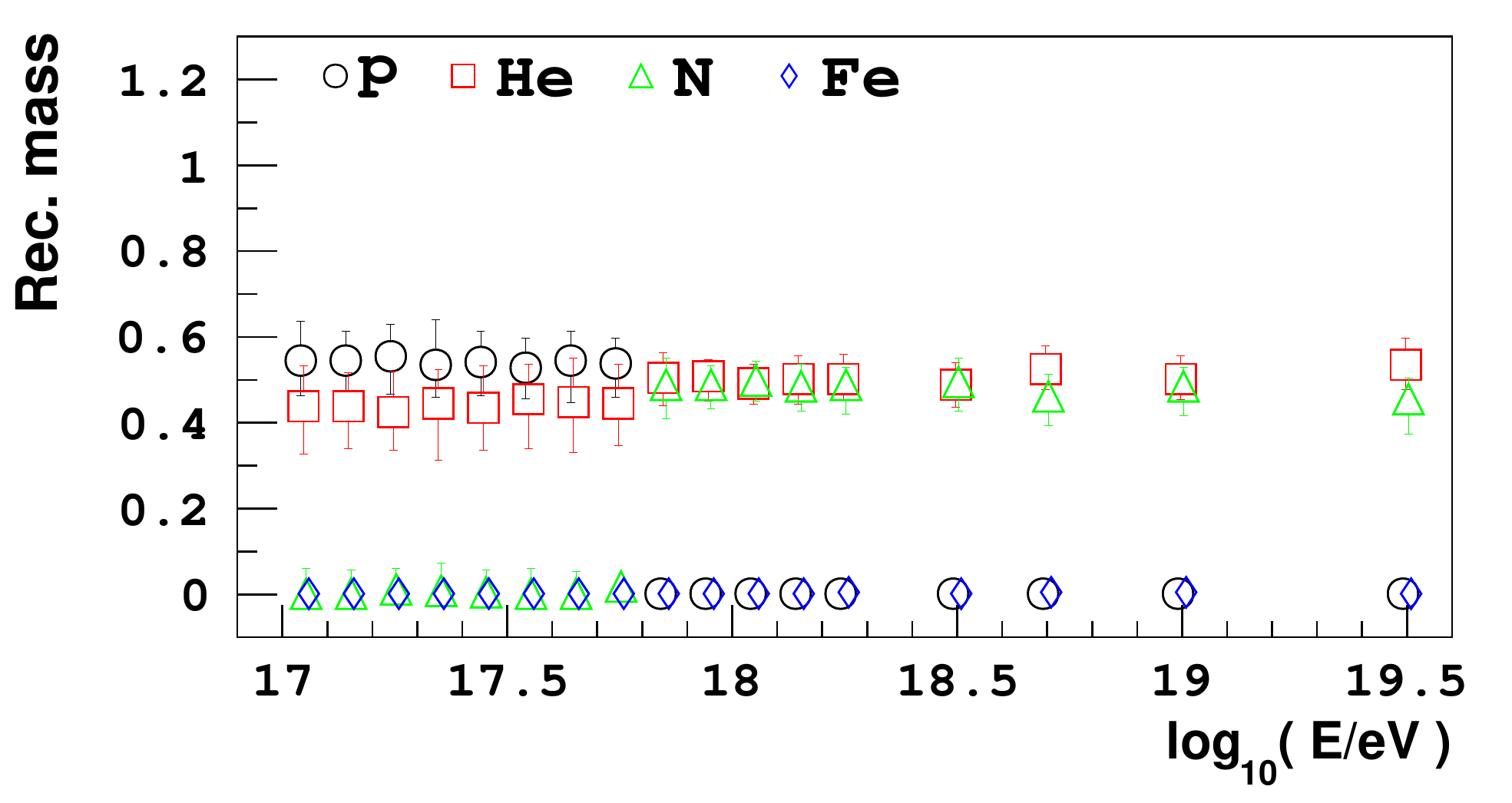}%
  \caption{Fitting only the mass fractions of our \qgs parameterisation to \qgs \Xmax data.}
  \label{fig:50proton50He_50He50N_QGSJET_31_fitmass}
\end{figure}

\begin{figure}[htbp!]
  \includegraphics[width=0.48\textwidth]{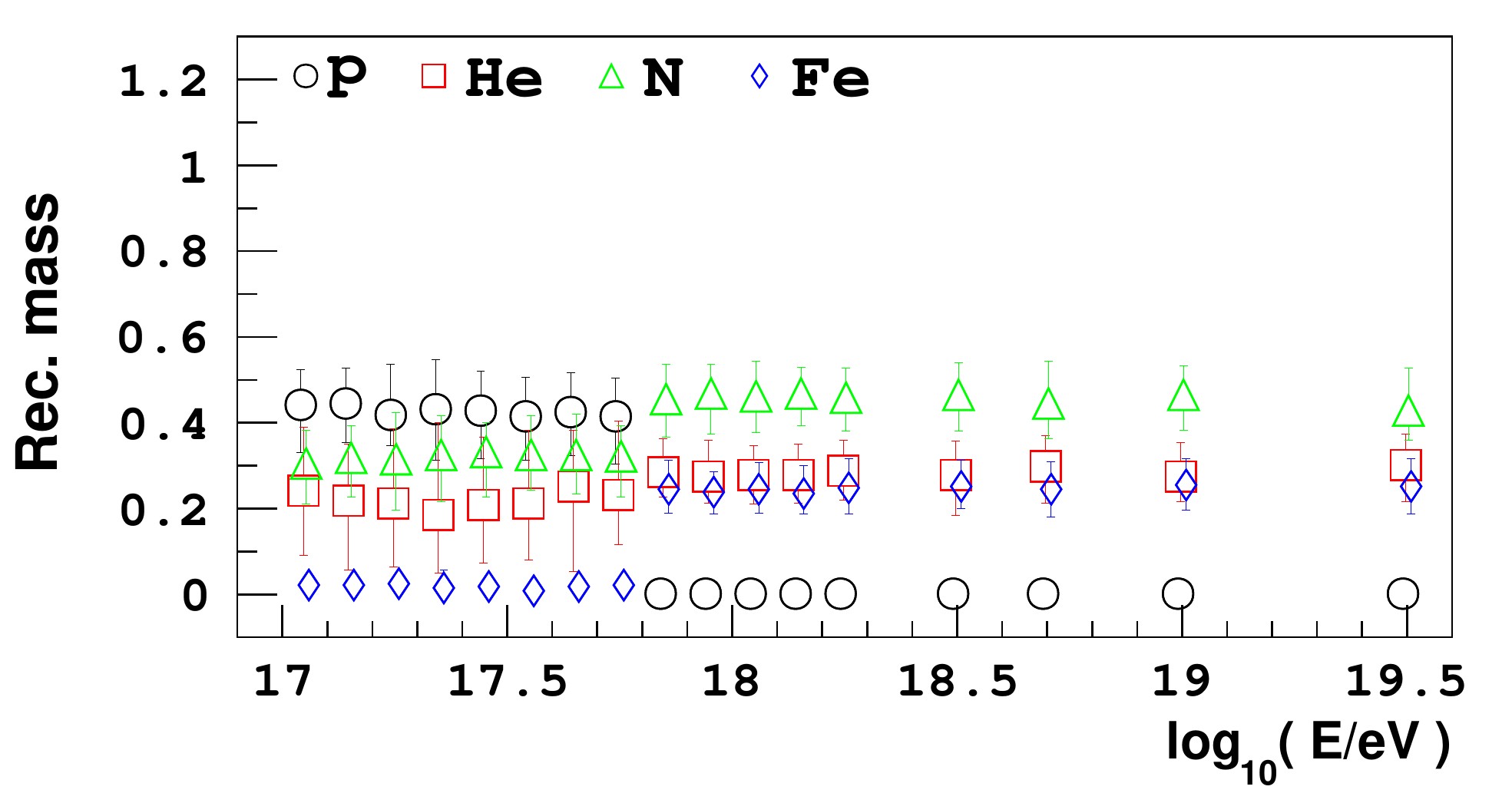}%
  \caption{Fitting \tnorm, \sigmanorm and the mass fractions of our \qgs parameterisation to \qgs \Xmax data.}
  \label{fig:50proton50He_50He50N_QGSJET_31_fitt0sigma} 
\end{figure}         

\begin{figure}[htbp!]
  \includegraphics[width=0.48\textwidth]{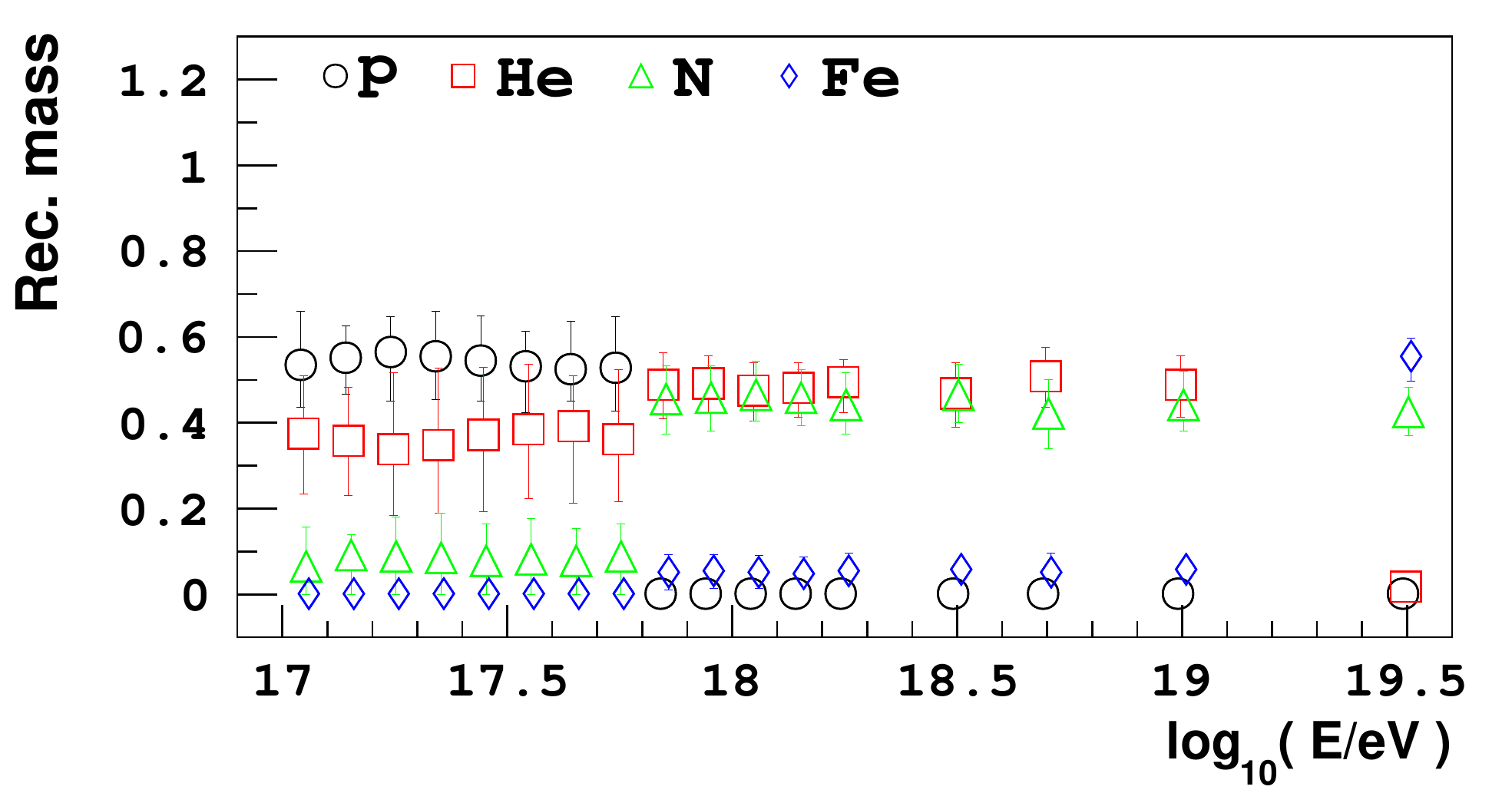}
  \caption{Fitting \tnorm, \sigmanorm and the mass fractions of our \qgs parameterisation to \qgs \Xmax data. Helium has been replaced by Iron in the last energy bin to increase the mass dispersion.}
  \label{fig:50proton50He_50He50N_50N50Fe_QGSJET_31_fitt0sigma} 
\end{figure}

\begin{figure}[htbp!]
  \includegraphics[width=0.48\textwidth]{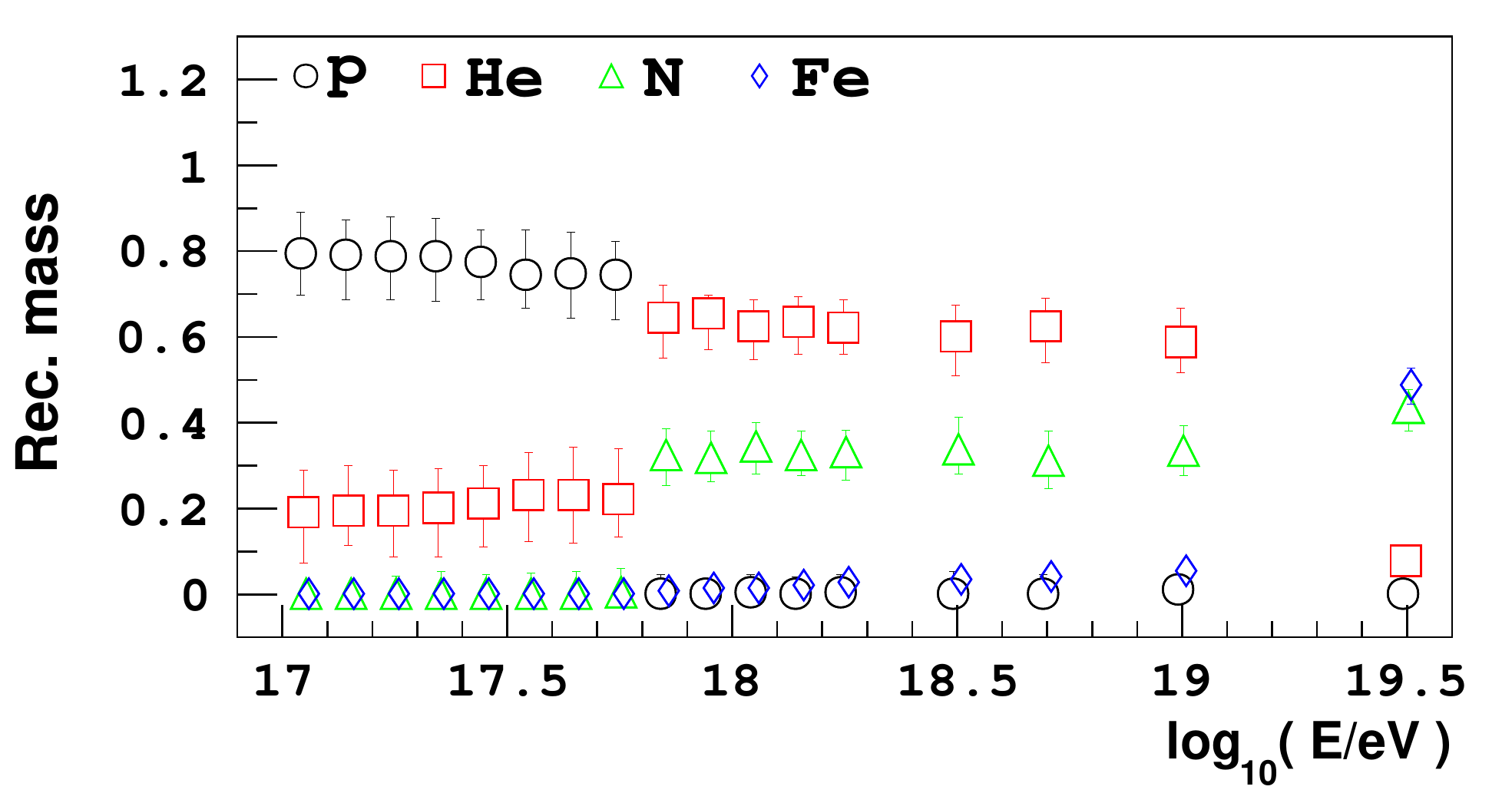}
  \caption{Fitting \tnorm, \sigmanorm and the mass fractions of our \epos parameterisation to \qgs \Xmax data. Helium has been replaced by Iron in the last energy bin to increase the mass dispersion.}
  \label{fig:50proton50He_50He50N_50N50Fe_QGSJET_30_fitt0sigma}
  
\end{figure}

\begin{figure}[htbp!]
  \includegraphics[width=0.48\textwidth]{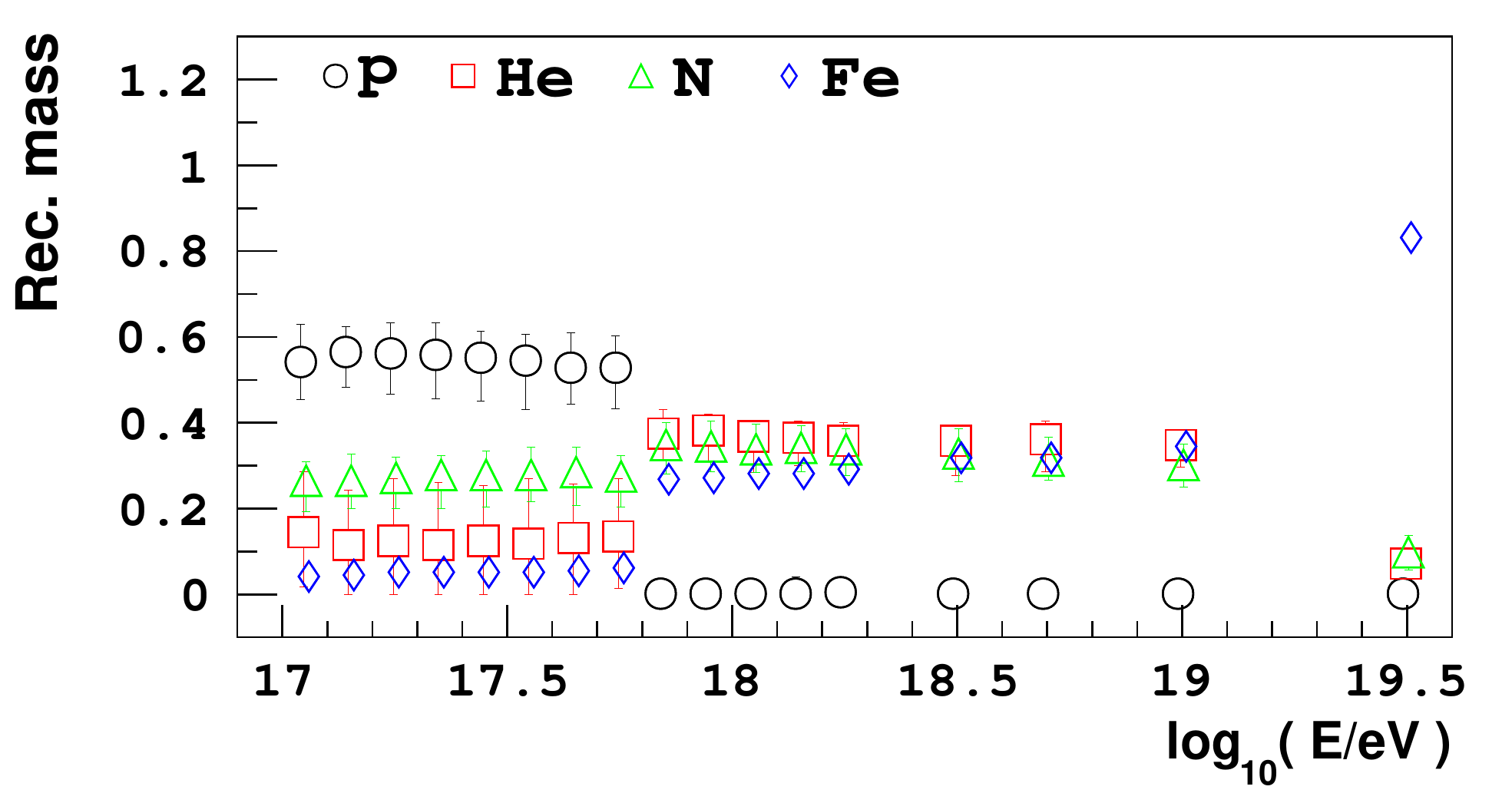}
  \caption{Fitting only the mass fractions (i.e. \tnorm and \sigmanorm are kept fixed) of our \epos parameterisation to \qgs \Xmax data. Compare this Fig. with \fig{fig:50proton50He_50He50N_50N50Fe_QGSJET_30_fitt0sigma} where \tnorm and \sigmanorm were included in the fit.}
  \label{fig:50proton50He_50He50N_50N50Fe_QGSJET_30_fitmass}
\end{figure}

\fig{fig:50proton50He_50He50N_EPOS_30_fitmass} to \fig{fig:50proton50He_50He50N_QGSJET_31_fitt0sigma} summarises the results of fits to 100 \Xmax data sets with a true mass composition consisting of $50\%$ proton and helium in the first 8 energy bins, and $50\%$ helium and nitrogen in the remaining 9 energy bins. When fitting only the mass fractions (i.e. keeping fixed the coefficients of the \Xmax distribution parameterisation) of our parameterisations to \conex \Xmax data based on the same model, the fits are able to reconstruct the mass composition to within an absolute offset in the median of $10\%$ from the true mass (as seen in \figs{fig:50proton50He_50He50N_EPOS_30_fitmass}{fig:50proton50He_50He50N_QGSJET_31_fitmass}). 

\fig{fig:50proton50He_50He50N_QGSJET_31_fitt0sigma} shows the results of fitting  \tnorm and \sigmanorm,  in addition to the mass fractions, of the \qgs parameterisation to \qgs data. These \qgs \Xmax distributions do not provide sufficient constraints on our fitted parameterisation, resulting in a mass composition reconstruction that does not resemble the true mass composition. In order to successfully fit \tnorm and \sigmanorm to data of a similar distribution, a wider range of primary masses over the energy range of the data is required (wider than the one in the given example). For example, in \fig{fig:50proton50He_50He50N_50N50Fe_QGSJET_31_fitt0sigma} we have increased the range of primary masses by replacing helium with iron in the last energy bin. The resulting fit of the mass fractions (with \tnorm and \sigmanorm also fitted) have an absolute offset in the median of less than $\sim 15\%$ from the true values, which is comparable to a fit of only the mass fractions to data of a similar composition.


\begin{figure}
  \includegraphics[width=0.48\textwidth]{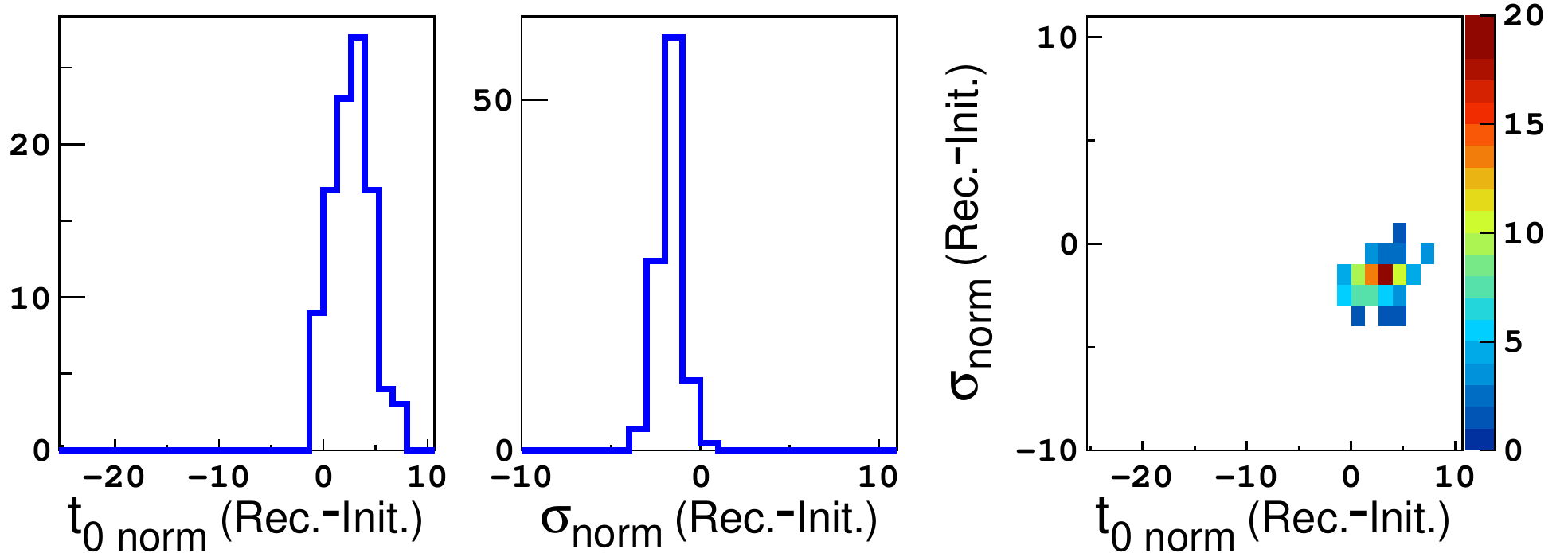}
  \caption{Change in \tnorm and \sigmanorm for protons from the fits in \fig{fig:50proton50He_50He50N_50N50Fe_QGSJET_31_fitt0sigma}.}
  \label{fig:hist_50proton50He_50He50N_50N50Fe_QGSJET_31_fitt0sigma} 
\end{figure}

\begin{figure}
  \includegraphics[width=0.48\textwidth]{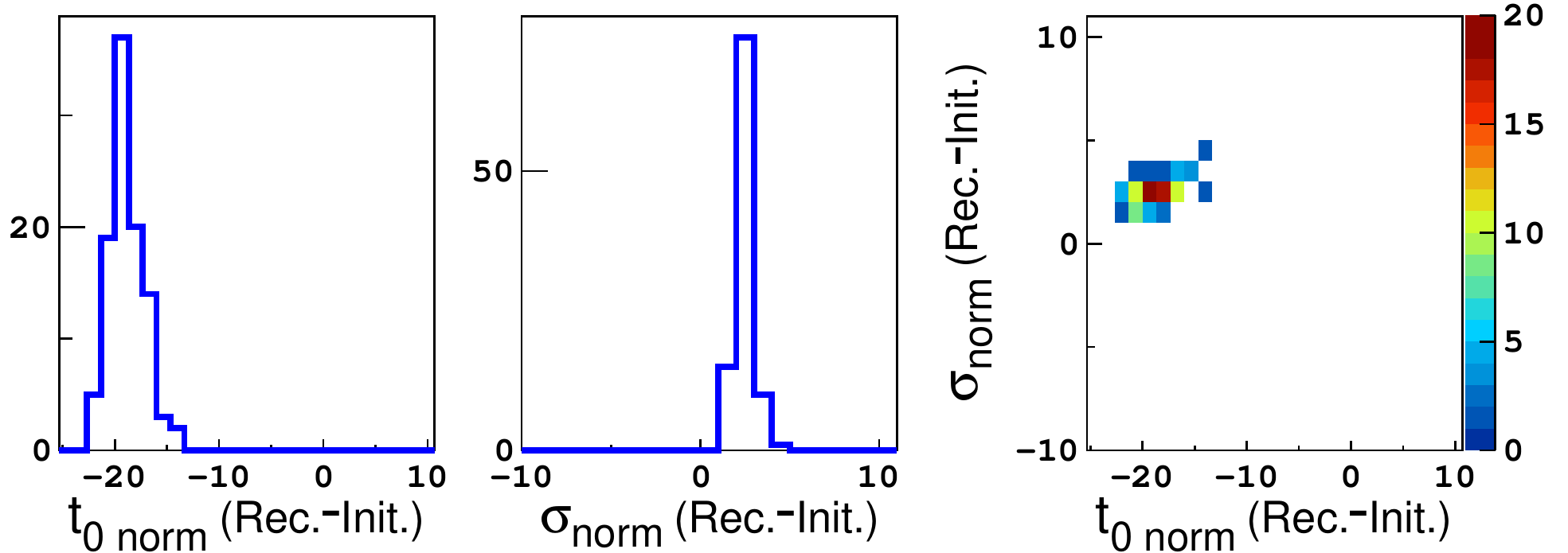}
  \caption{Change in \tnorm and \sigmanorm for protons from the fits in \fig{fig:50proton50He_50He50N_50N50Fe_QGSJET_30_fitt0sigma}.}
  \label{fig:hist_50proton50He_50He50N_50N50Fe_QGSJET_30_fitt0sigma} 

\end{figure}

\subsection{Fitting data originating from a different model.}

Compare \fig{fig:50proton50He_50He50N_50N50Fe_QGSJET_30_fitt0sigma} with \fig{fig:50proton50He_50He50N_50N50Fe_QGSJET_30_fitmass}, which shows the composition fits when using the \epos parameterisation to fit \qgs data,  with \tnorm and \sigmanorm fitted in the former, and \tnorm and \sigmanorm fixed in the latter. Fitting these two coefficients is enough to result in a reconstructed mass much closer to the true mass, despite the fitted data originating from a different model. By fitting \tnorm and \sigmanorm, there is no longer a significant iron component where there should only be $50\%$ helium and nitrogen, and in the $50\%$ proton and helium range there is no longer a fitted nitrogen component larger than the helium fraction.

\figs{fig:hist_50proton50He_50He50N_50N50Fe_QGSJET_31_fitt0sigma}{fig:hist_50proton50He_50He50N_50N50Fe_QGSJET_30_fitt0sigma} show the difference between the fitted values and initial values of \tnorm  and \sigmanorm (and their correlation) when fitted to the data with iron added in the last energy bin. \fig{fig:hist_50proton50He_50He50N_50N50Fe_QGSJET_31_fitt0sigma} displays the results of fitting \qgs data with our \qgs parameterisation, and as expected the difference between the reconstructed and initial values of our coefficients is minimal. \fig{fig:hist_50proton50He_50He50N_50N50Fe_QGSJET_30_fitt0sigma} displays the results of fitting the same \qgs data with our \epos parameterisation (the reconstructed mass is shown in \fig{fig:50proton50He_50He50N_50N50Fe_QGSJET_30_fitt0sigma}), and we see that \tnorm and \sigmanorm are shifted towards the \qgs values for these coefficients. The initial \epos proton \tnorm and \sigmanorm values are $\sim \depth{703}$ and $\sim \depth{22}$ respectively, while the initial \qgs proton \tnorm and \sigmanorm values (and therefore the approximate values of the \qgs MC data) are $\sim \depth{688}$ and $\sim \depth{25}$ respectively. 

Notice that in  \fig{fig:50proton50He_50He50N_EPOS_30_fitmass}  to \fig{fig:50proton50He_50He50N_50N50Fe_QGSJET_30_fitt0sigma} the bins containing a helium and nitrogen mix are reconstructed better than the bins containing a proton and helium mix. Proton and helium distributions are harder to reconstruct due to their wider spread and their larger overlap. A wider spread means that for a given number of events, less events will populate individual \Xmax bins. Therefore, proton and helium fits have larger statistical uncertainties. Additionally, the \Xmax parameterisations for lighter masses do not describe the \conex \thinspace \epos and \qgs simulated data as accurately. \fig{fig:EPOS_QGS_diff_mean_sigma_all} in Appendix \ref{AppA} illustrates that as the primary mass of the distribution increases, the \Xmax parameterisations reproduce the true \meanXmax and \sigmaXmax of the distributions with better accuracy.  Appendix \ref{AppA} shows that for proton and helium data especially, the fits of Equation~\eqref{eq:Xmaxbasic} to MC data of either hadronic model tend to overestimate the number of events at the mode of the distribution. When fitting mixes of protons and helium, our fits tend to have a reconstruction bias towards protons. 

As the absolute separation between $\sigma$ for different primaries is similar in the \epos and \qgs parameterisations (like $t_0$), marginally better results would be obtained in \fig{fig:50proton50He_50He50N_50N50Fe_QGSJET_30_fitt0sigma} if instead of fitting \sigmanorm such that the initial ratios of $\sigma$ among primaries are conserved, \sigmanorm was fitted such that the initial separation between \sigmanorm among primaries was conserved (like \tnorm). However, conserving the initial ratios of $\sigma$ is the more physical approach, because if \sigmanorm for protons changes by \depth{10}, we would not expect that \sigmanorm for iron would also change by \depth{10}. Additionally, nature does not necessarily conform to the \epos or \qgs predictions of the absolute separation of \sigmanorm among primaries.

\section{\tnorm and \sigmanorm parameter space scan of the Auger FD \Xmax data}
\label{tnorm and sigmanorm parameter space scan of the Auger FD Xmax data}

\begin{figure}[!htb]
  \centering 
  \includegraphics[width=0.48\textwidth]{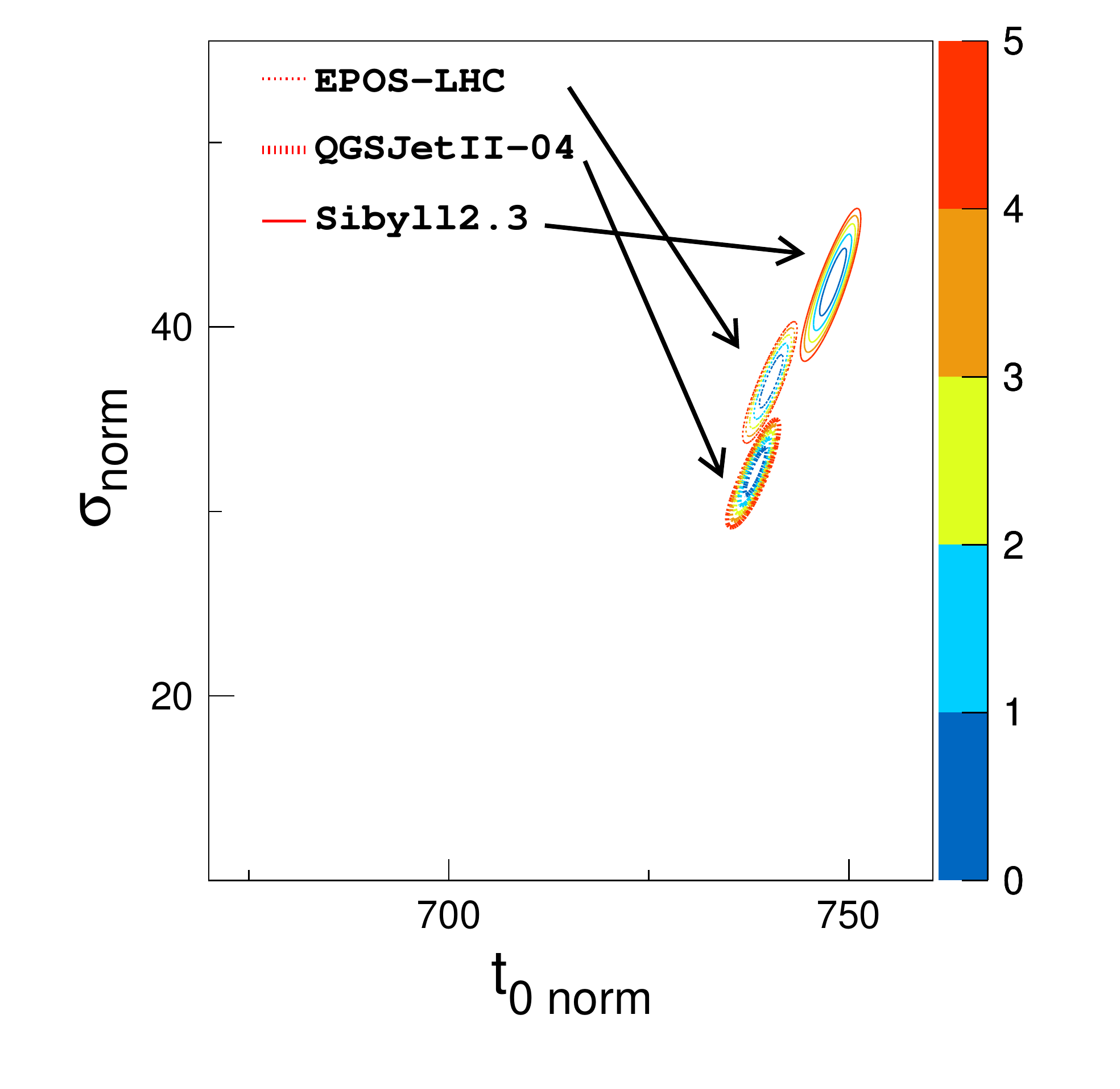}
      \caption{The \tnorm and \sigmanorm parameter space scan over the Auger FD \Xmax data. For each model parameterisation, at specific values of \tnorm and \sigmanorm, the mass fractions are fitted to the data, and the first $5\sigma$ contours of the minimised Poisson log likelihood are shown. The scanned shape coefficient values for proton are shown. The coefficient values of the heavier nuclei change (relative to protons) in the way the shape coefficient would be fitted, outlined in Section~\ref{sec.method}.}
       \label{fig:FDdata_t0sigma_scan}
    \end{figure}

\fig{fig:FDdata_t0sigma_scan} shows the minimised Poisson log likelihood space of the mass fraction fit of a parameterisation to Auger FD \Xmax data, where \tnorm and \sigmanorm have been fixed to some particular value (indicated by the x and y axes). The z-axis shows the difference between the minimised probability for some value of \tnorm and \sigmanorm, and the absolute minimised probability obtained from the \tnorm and \sigmanorm values which best fitted the data for a particular parameterisation. A difference of 1 in the minimised Poisson log likelihood corresponds to $1\sigma$. The absolute minima of the \epos and \qgs fits to the Auger FD data correspond to a similar value of \tnorm for protons, whereas the absolute minimum of the \sib fit is located at a significantly larger value of \tnorm for protons. Between the three fitted parameterisations, when estimating the heavier nuclei \tnorm values there is more similarity. This is because the separation between the proton $t_0$ prediction and heavier nuclei is larger in the \sib parameterisation than \epos or \qgs (see \figsThree{fig:model_diff_comparison_eq}{fig:model_diff_comparison_es}{fig:model_diff_comparison_sq}). This is also true for $\sigma$.

These scans show that the fits of the Auger FD \Xmax data performed in Section~\ref{sec.res} did not become stuck in a local minimum. The scans can also reveal secondary solutions which are not as deep as the deepest minimum.
\section{Evaluating the fit performance for a mass composition consistent with the Auger results}
\label{sec.augerperformance}

The performance of fitting \tnorm, \sigmanorm and the mass fractions of our parameterisations to the Auger FD \Xmax data is evaluated by fitting mock \Xmax data sets that resemble the Auger FD \Xmax distributions. This was achieved by fitting \tnorm, \sigmanorm and the mass fractions of a particular parameterisation to the Auger FD \Xmax data, and then using this fitted parameterisation to generate the mock data sets. Appendix~\ref{AppB} displays the \tnorm and \sigmanorm values fitted to the Auger data, values which correspond to the absolute minima found from the scans in Section~\ref{tnorm and sigmanorm parameter space scan of the Auger FD Xmax data}. These mock data sets have a true mass composition which is defined by the parameterisation used to generate them, therefore we can evaluate the ability of our \tnorm, \sigmanorm and mass fraction fit to accurately reconstruct the true mass fractions. The binning of the mock Auger \Xmax distributions is \depth{20}.

The measured FD \Xmax distributions are broadened by the \Xmax resolution of the detector, and are affected by the detector acceptance, therefore the mock \Xmax data generated from the fitted parameterisation are convolved with the same detector effects. The \Xmax resolution and acceptance of the Auger data is taken into account when fitting this mock Auger \Xmax data. Our mock \Xmax distributions and the \Xmax distributions measured by Auger are treated with exactly the same approach.

\subsection{Fitting \tnorm, \sigmanorm and the mass fractions}

\begin{figure}[htb!]
  \includegraphics[width=0.48\textwidth]{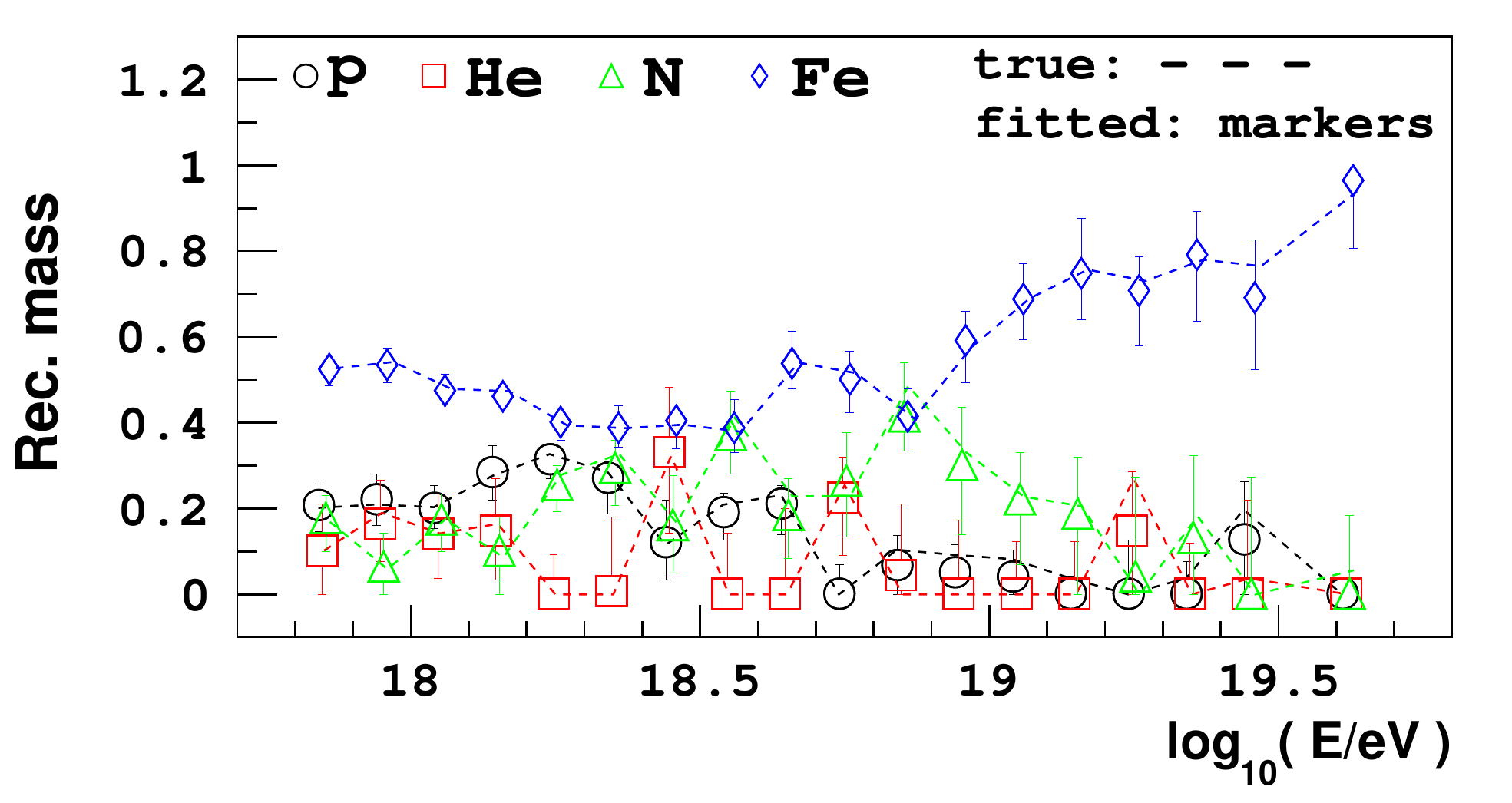}%
  \caption{\epos fit of \Xmax data generated from the \epos parameterisation fit of Auger data. }
  \label{fig:EPOS_30_fitt0sigma_mod}
\end{figure}

\begin{figure}[htb!]
        \includegraphics[width=0.48\textwidth]{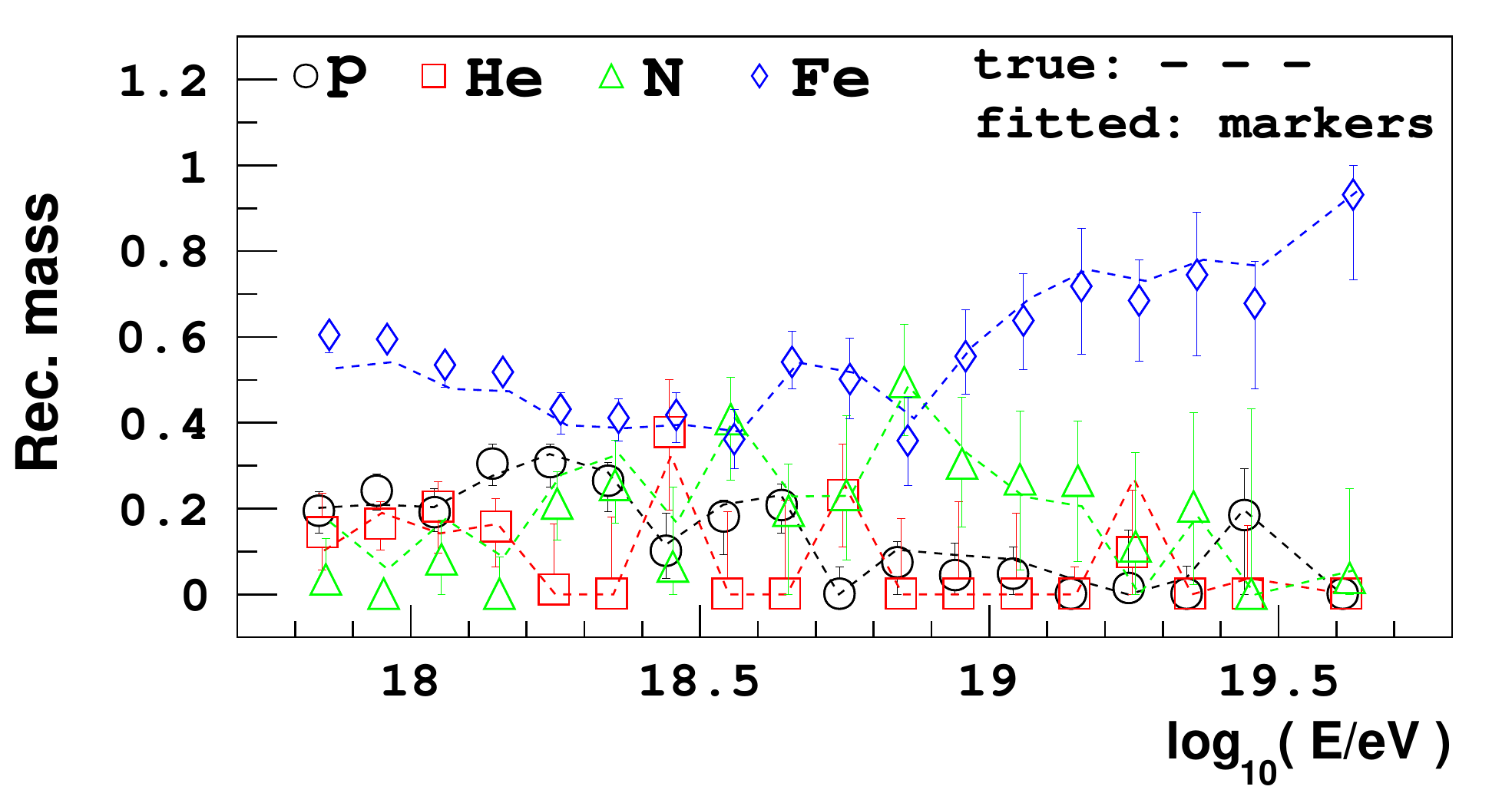}%
        \caption{\qgs fit of \Xmax data generated from the \epos parameterisation fit of Auger data.}
        \label{fig:EPOS_31_fitt0sigma_mod}
\end{figure}

\begin{figure}[htb!]
  \includegraphics[width=0.48\textwidth]{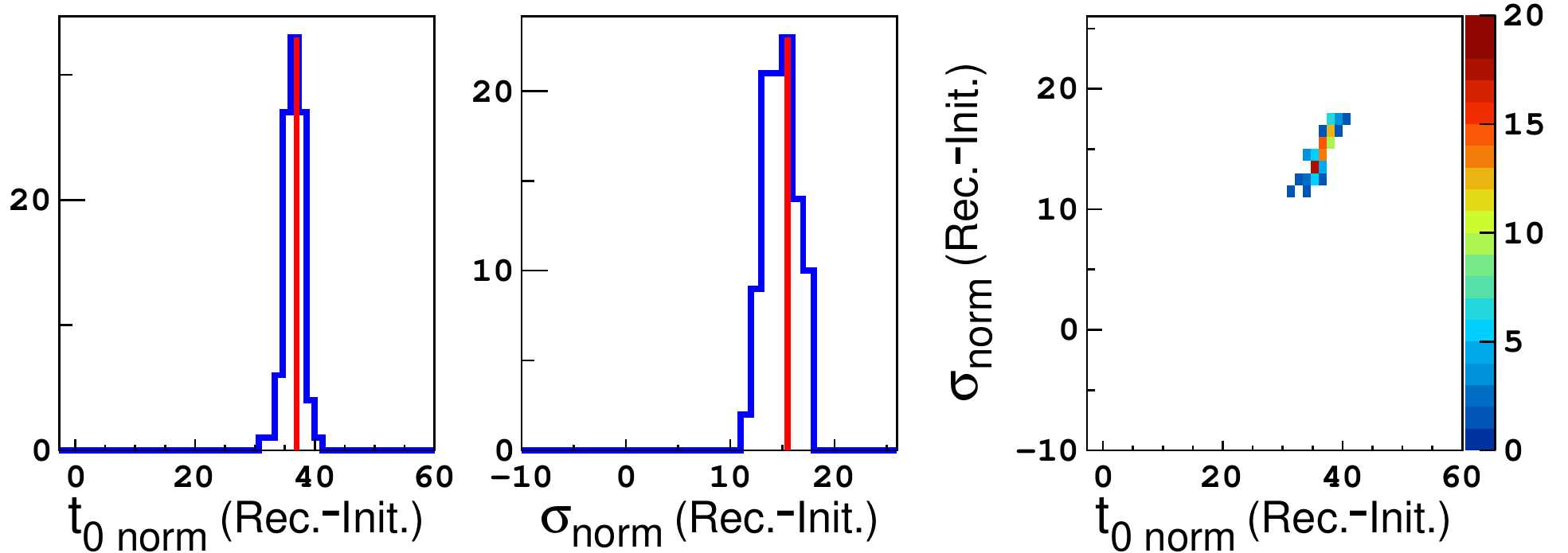}%
  \caption{Change in \tnorm and \sigmanorm for protons from the fits in \fig{fig:EPOS_30_fitt0sigma_mod}.}  
  \label{fig:hist_EPOS_30_fitt0sigma}
\end{figure}

\begin{figure}[htb!]
  \includegraphics[width=0.48\textwidth]{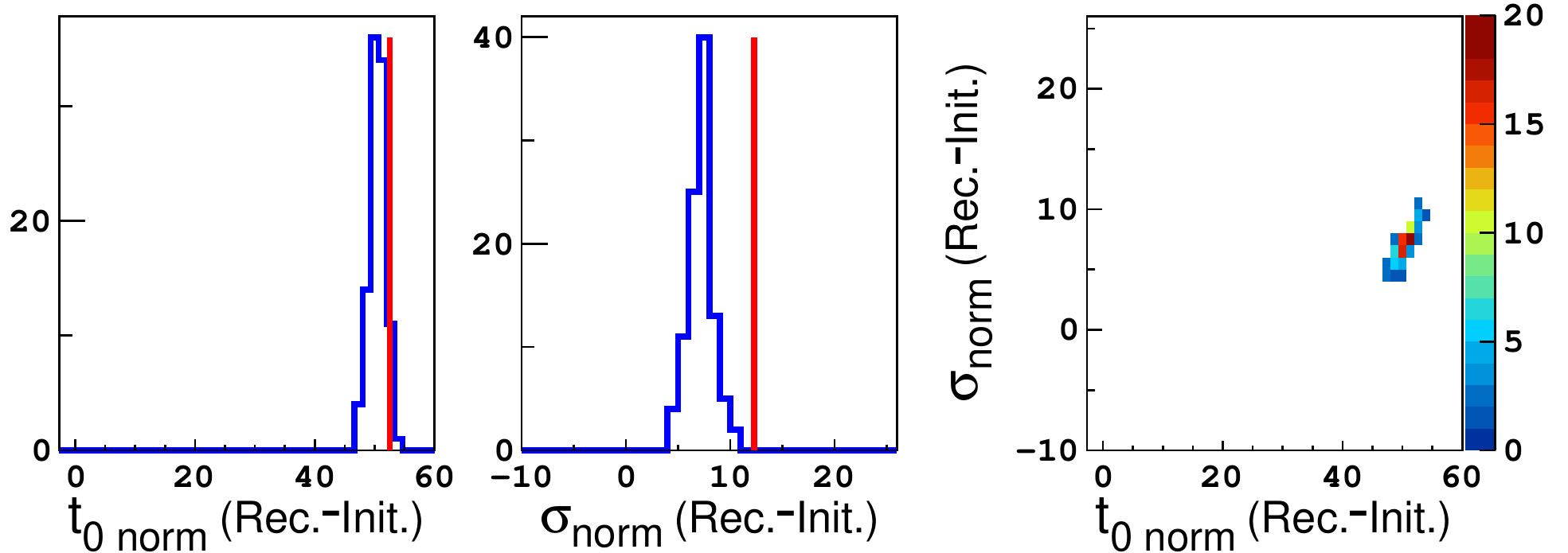}%
  \caption{Change in \tnorm and \sigmanorm for protons from the fits in \fig{fig:EPOS_31_fitt0sigma_mod}.}  
  \label{fig:hist_EPOS_31_fitt0sigma}
\end{figure}

\figsThree{fig:EPOS_30_fitt0sigma_mod}{fig:EPOS_31_fitt0sigma_mod}{fig:EPOS_32_fitt0sigma_mod} display the mass composition results from fitting the mass fractions, \tnorm and \sigmanorm of either the \epos, \qgs or \sib parameterisations respectively, to 100 data sets generated from the parameterisation which resulted when the mass fractions, \tnorm and \sigmanorm of the {\bf\epos} parameterisation were fitted to Auger FD \Xmax data (as will be shown in Section~\ref{sec.res}). The true mass composition of the mock data is therefore the mass composition which resulted from the \epos fit to the Auger FD \Xmax data. \figsThree{fig:hist_EPOS_30_fitt0sigma}{fig:hist_EPOS_31_fitt0sigma}{fig:hist_EPOS_32_fitt0sigma} display the fitted proton values of \tnorm and \sigmanorm relative to the original values of the model applied, compared to the change required to match the true proton values of the mock data. The red lines indicate the mock data input values and the blue histograms are the reconstructed values. The correlations between the reconstructed \tnorm and \sigmanorm are also shown in \figsThree{fig:hist_EPOS_30_fitt0sigma}{fig:hist_EPOS_31_fitt0sigma}{fig:hist_EPOS_32_fitt0sigma}. There are no reconstruction systematics when using the \epos parameterisation to fit \epos generated data (\fig{fig:hist_EPOS_30_fitt0sigma}), but there are some systematics when using the \qgs or \sib parameterisations to fit \epos generated data (\figs{fig:hist_EPOS_31_fitt0sigma}{fig:hist_EPOS_32_fitt0sigma}). These systematics in \tnorm and \sigmanorm translate into relative small systematics of the reconstructed mass fractions (as seen in \figs{fig:EPOS_31_fitt0sigma_mod}{fig:EPOS_32_fitt0sigma_mod}).

\figs{fig:EPOS_31_fitt0sigma_mod}{fig:hist_EPOS_31_fitt0sigma} show that despite the differences between the \epos and \qgs parameterisations (which are not limited to different \tnorm and \sigmanorm predictions), by allowing \tnorm and \sigmanorm of the \qgs \Xmax parameterisation to be fitted to mock data based on the \epos parameterisation, the true mass fractions are reconstructed with an overall accuracy comparable to the \epos fits of \epos data. The absolute offsets in the median mass fractions from the true mass are less than $10\%$ in most energy bins. This demonstrates that fitting \tnorm and \sigmanorm significantly reduces the differences between the \epos and \qgs \Xmax parameterisations. As we are fitting the \qgs parameterisation to mock data based on the \epos parameterisation, we do not expect the average fitted values of \tnorm and \sigmanorm to be centred on the red lines even if no systematic offset was present in the mass fractions reconstruction. This is because the separation of these coefficients between masses differs between the \epos and \qgs parameterisations, thus if the fitted \qgs value of \tnorm for protons was equal to the \epos value of \tnorm for protons, the accordingly adjusted \tnorm values of other masses would differ between these parameterisations. 

\begin{figure}[htb!]
        \includegraphics[width=0.48\textwidth]{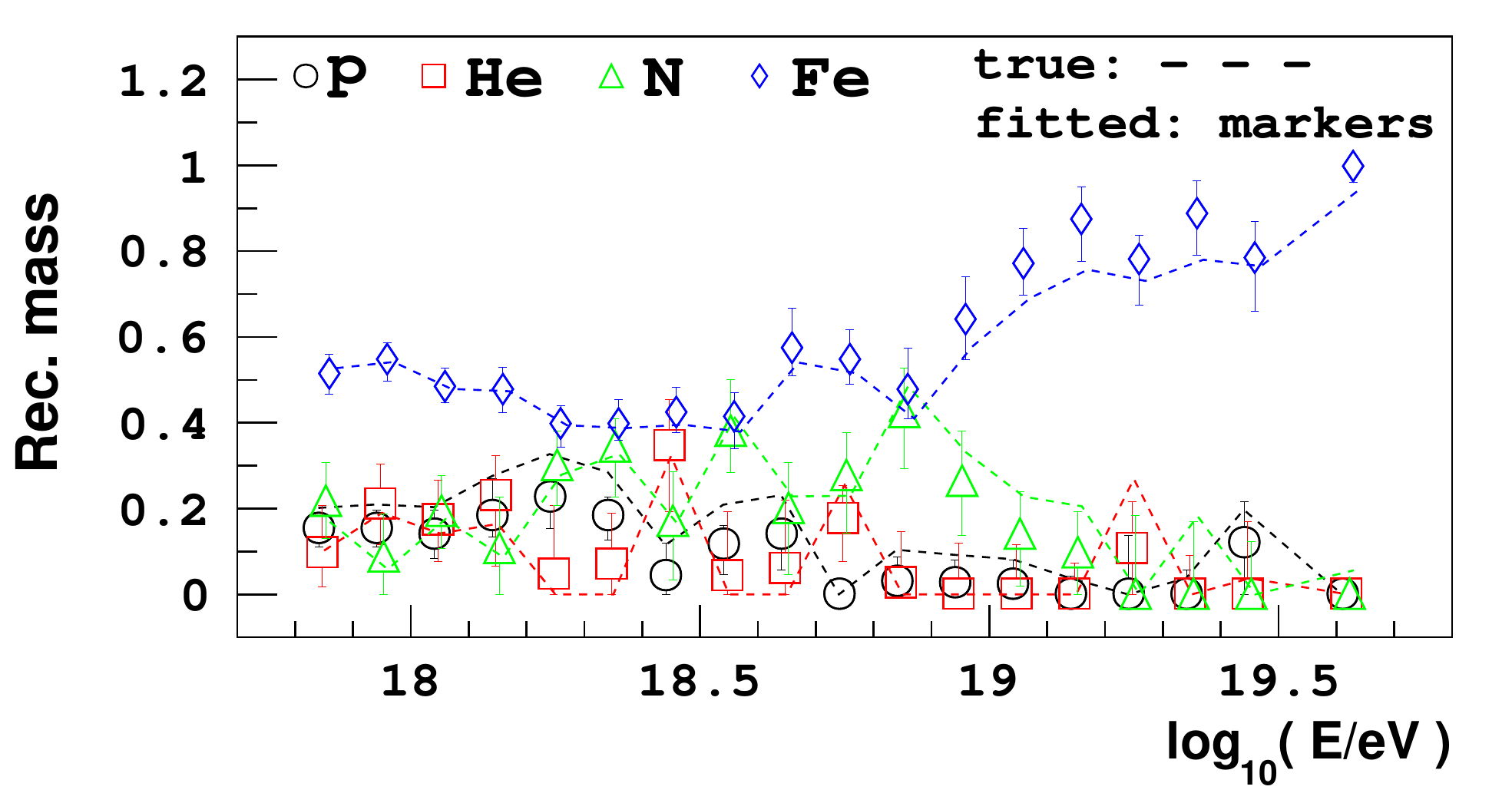}%
        \caption{\sib fit of \Xmax data generated from the \epos parameterisation fit of Auger data.}
        \label{fig:EPOS_32_fitt0sigma_mod}
\end{figure}

\begin{figure}[htb!]
  \includegraphics[width=0.48\textwidth]{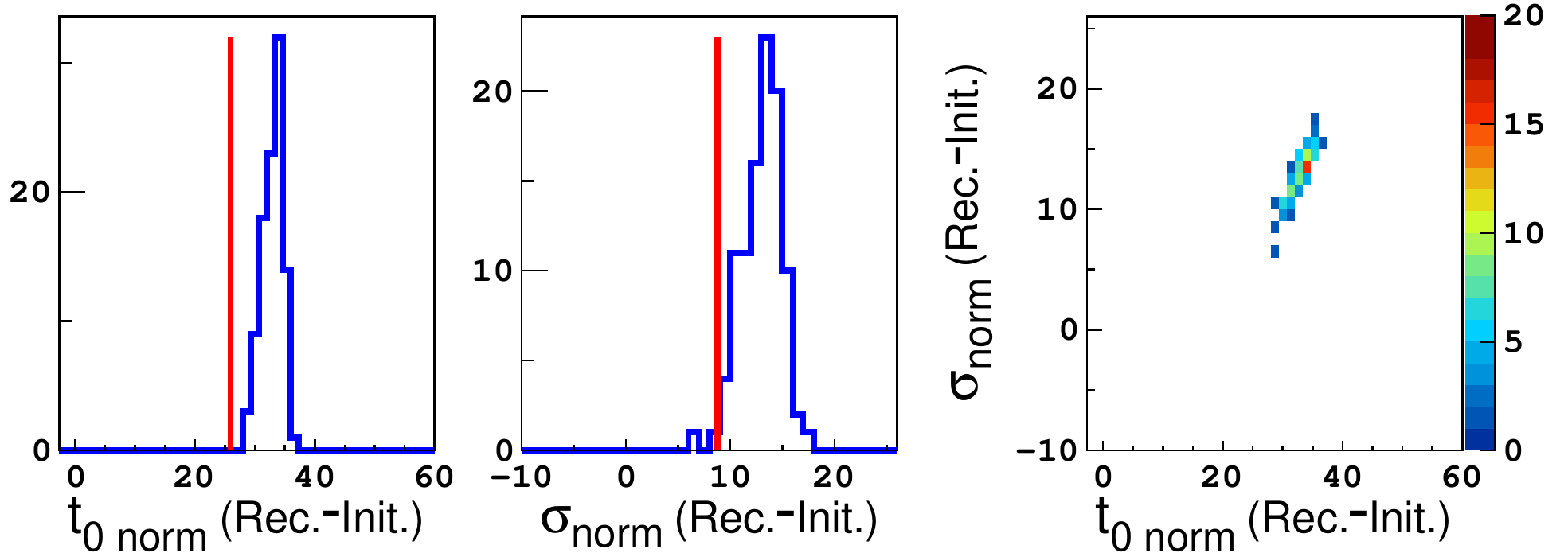}%
  \caption{Change in \tnorm and \sigmanorm for protons from the fits in \fig{fig:EPOS_32_fitt0sigma_mod}.}  
  \label{fig:hist_EPOS_32_fitt0sigma}
\end{figure}

The mass composition reconstruction accuracy of the \epos fit to \epos based data changes less with energy than the accuracy of the \qgs fit to the \epos data. This is because the \epos $t_0$ parameterisation fit to the \epos based data is offset by a constant value at all energies from the true $t_0$ of the mock data, whereas the difference between the fitted \qgs $t_0$ parameterisation and the true $t_0$ of the mock data (based on \epos) changes with energy.

\fig{fig:EPOS_32_fitt0sigma_mod} shows the \sib fit to the \epos data results in a reconstructed mass that is very representative of the true mass, but this mass reconstruction is not as accurate as the \epos and \qgs fits to this data. This is because a \tnorm and \sigmanorm shift of the \sib parameterisation does not align the \sib $t_0$ and $\sigma$ parameterisations with the \epos (or \qgs) descriptions as adequately as the \epos or \qgs descriptions can be aligned with each other (compare \figsThree{fig:model_diff_comparison_eq}{fig:model_diff_comparison_es}{fig:model_diff_comparison_sq}). Larger differences in the $\lambda$ \sib parameterisation relative to the other parameterisations further hinders an accurate mass reconstruction of data based on these other parameterisations. 

\begin{figure}[htb!]
  \includegraphics[width=0.48\textwidth]{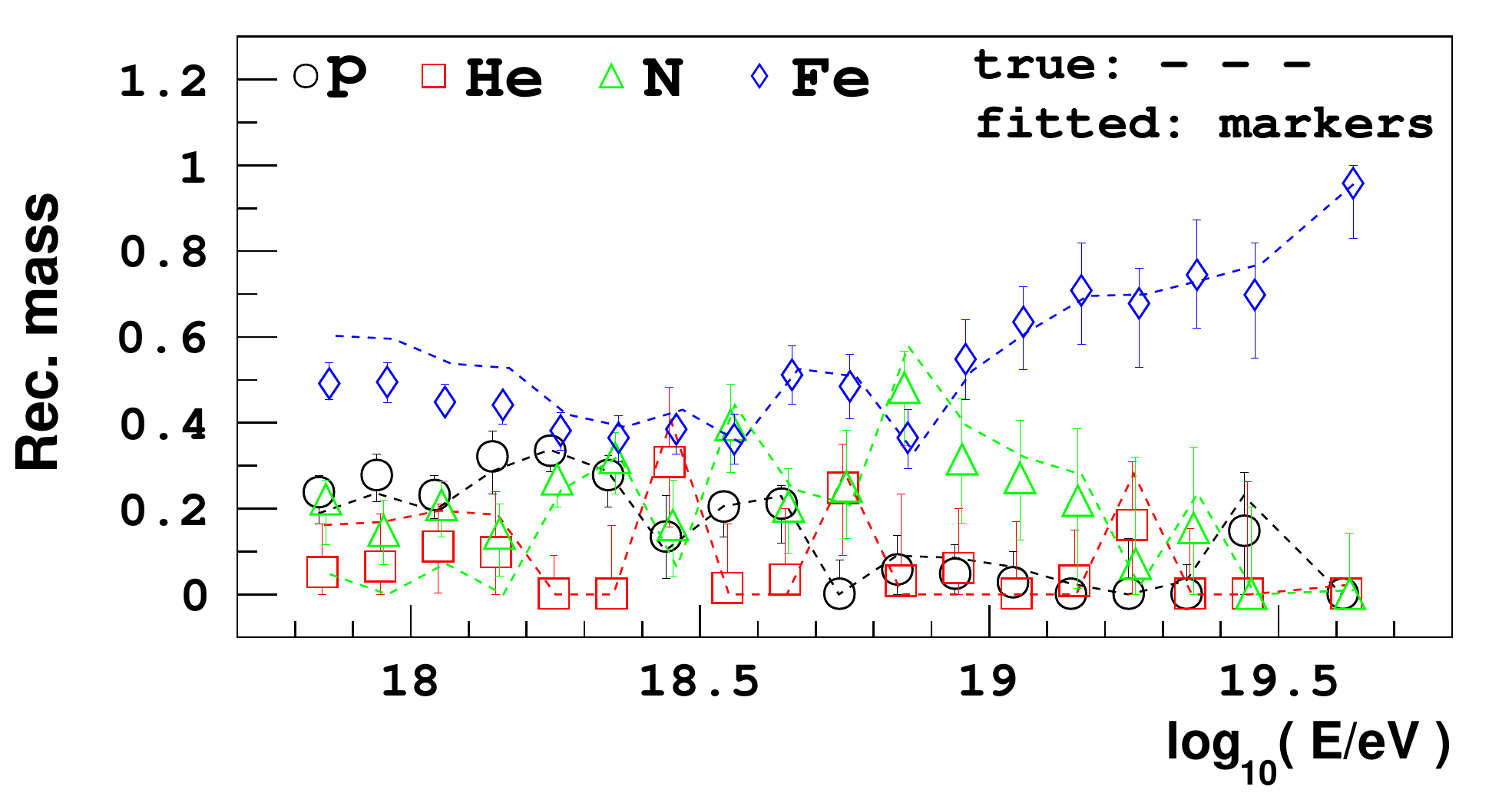}%
  \caption{\epos fit of \Xmax data generated from the \qgs parameterisation fit of Auger data. }
  \label{fig:QGS_30_fitt0sigma_mod}
\end{figure}

\begin{figure}[htb!]
        \includegraphics[width=0.48\textwidth]{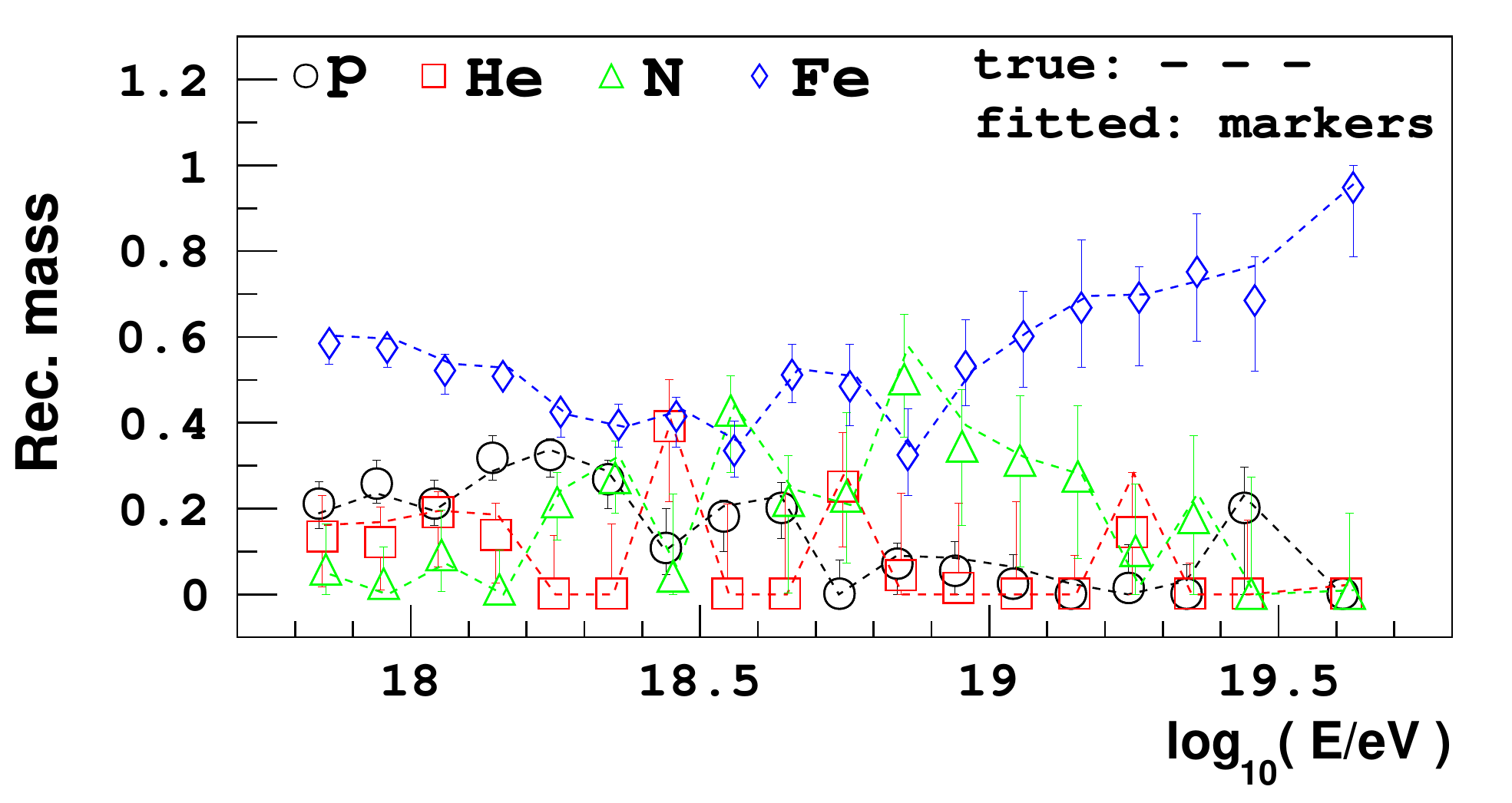}%
        \caption{\qgs fit of \Xmax data generated from the \qgs parameterisation fit of Auger data.}
        \label{fig:QGS_31_fitt0sigma_mod}
\end{figure}

\begin{figure}[htb!]
        \includegraphics[width=0.48\textwidth]{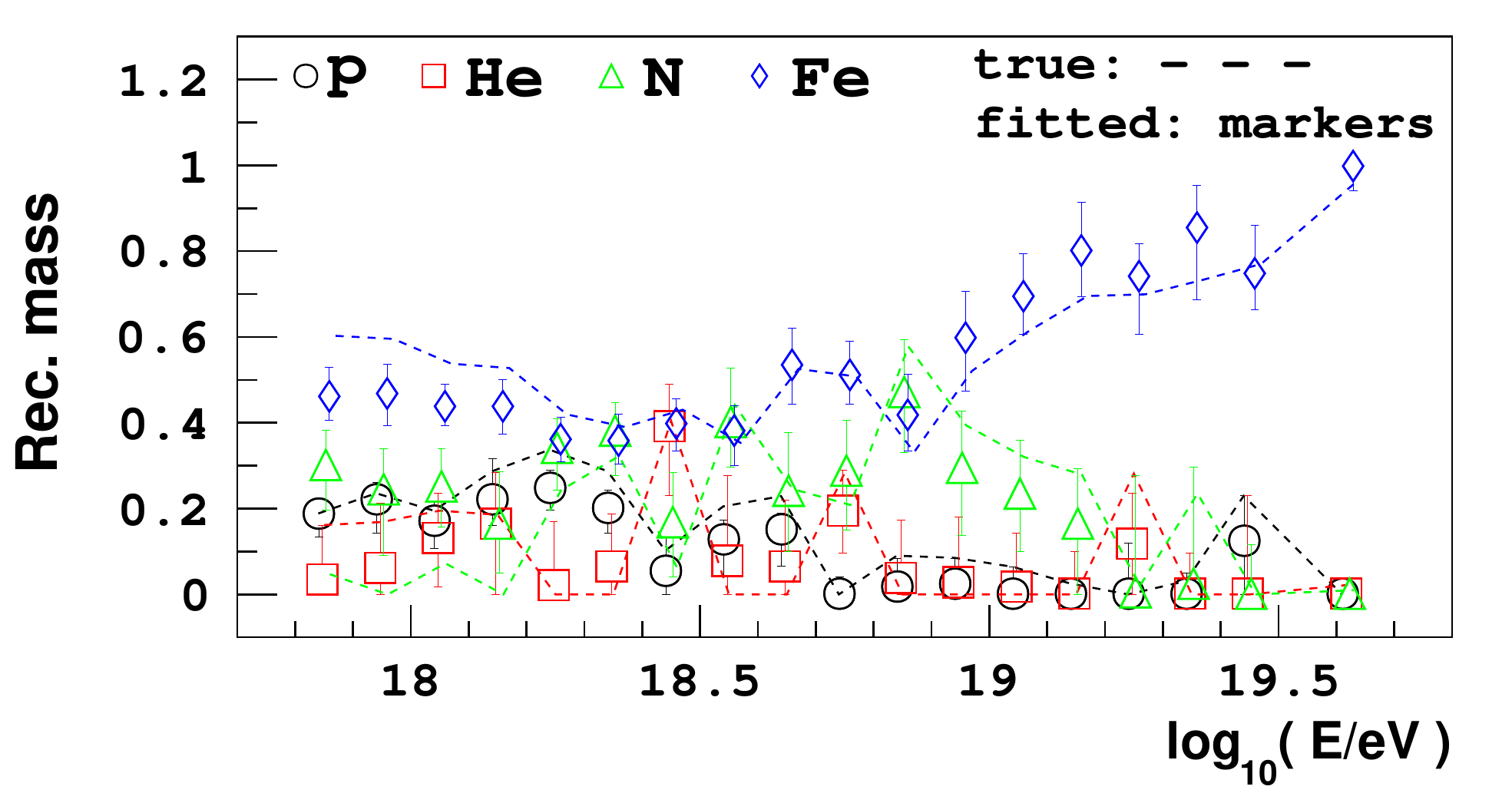}%
        \caption{\sib fit of \Xmax data generated from the \qgs parameterisation fit of Auger data.}
        \label{fig:QGS_32_fitt0sigma_mod}
\end{figure}

\begin{figure}[htb!]
  \includegraphics[width=0.48\textwidth]{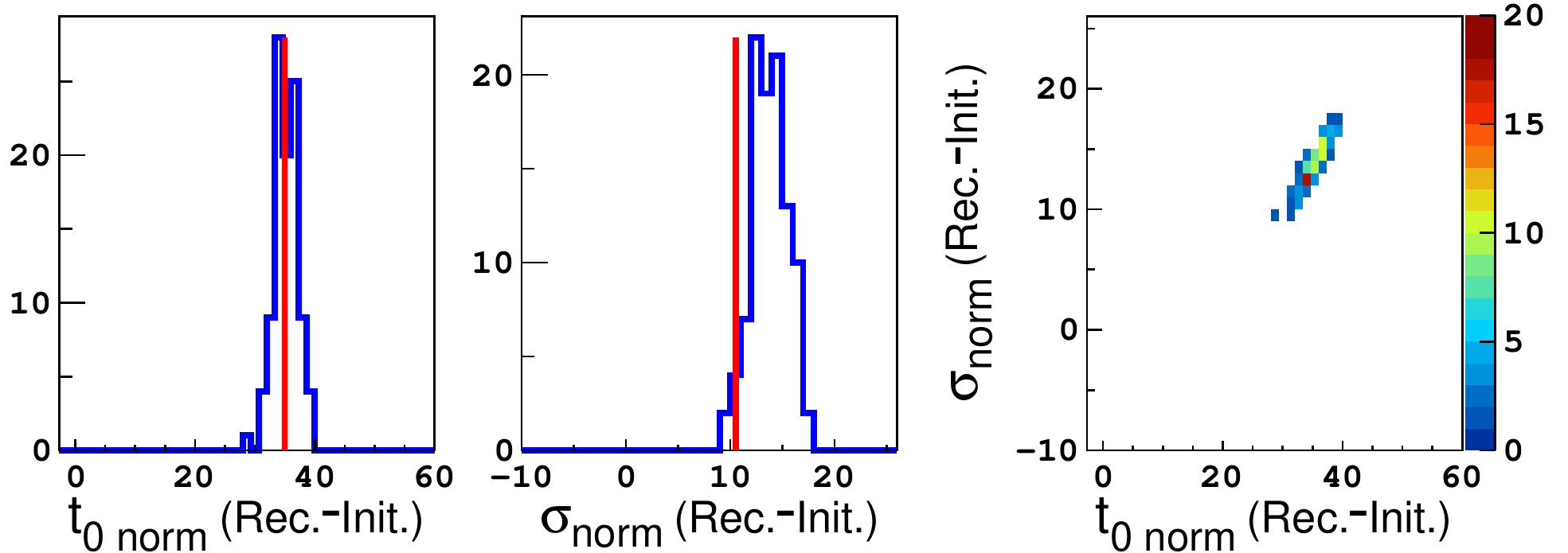}%
  \caption{Change in \tnorm and \sigmanorm for protons from the fits in \fig{fig:QGS_30_fitt0sigma_mod}.}  
  \label{fig:hist_QGS_30_fitt0sigma}
\end{figure}

\begin{figure}[htb!]
  \includegraphics[width=0.48\textwidth]{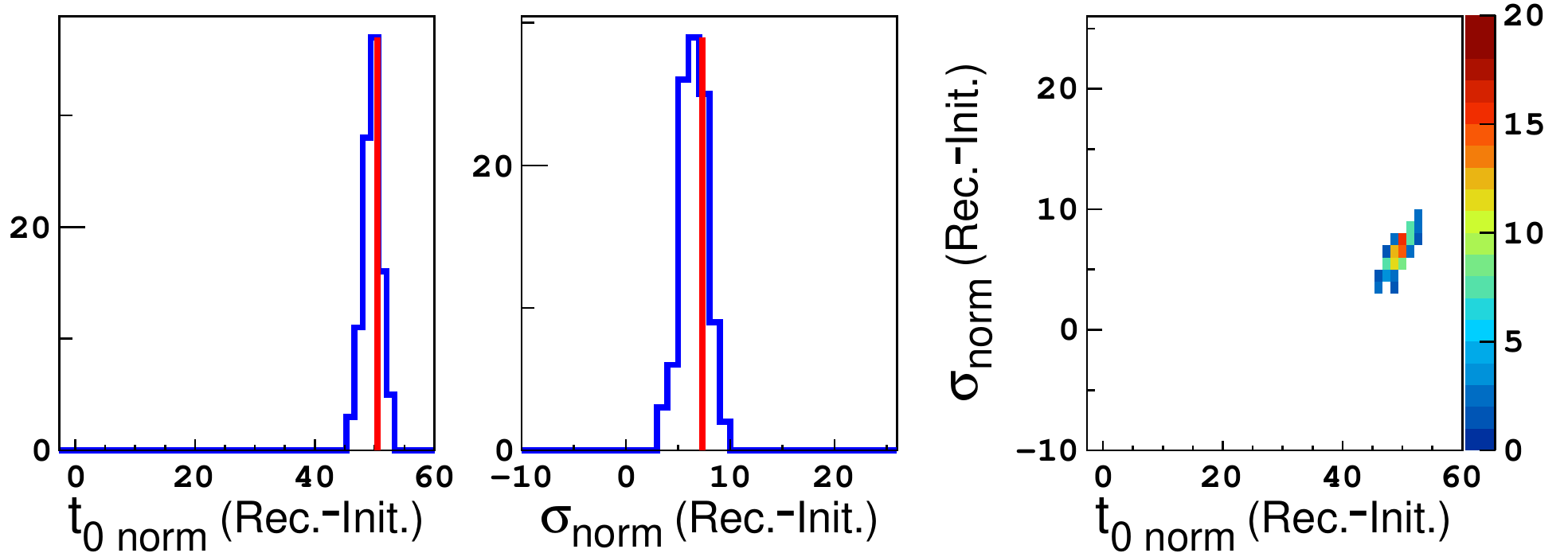}%
  \caption{Change in \tnorm and \sigmanorm for protons from the fits in \fig{fig:QGS_31_fitt0sigma_mod}.}  
  \label{fig:hist_QGS_31_fitt0sigma}
\end{figure}

\begin{figure}[htb!]
  \includegraphics[width=0.48\textwidth]{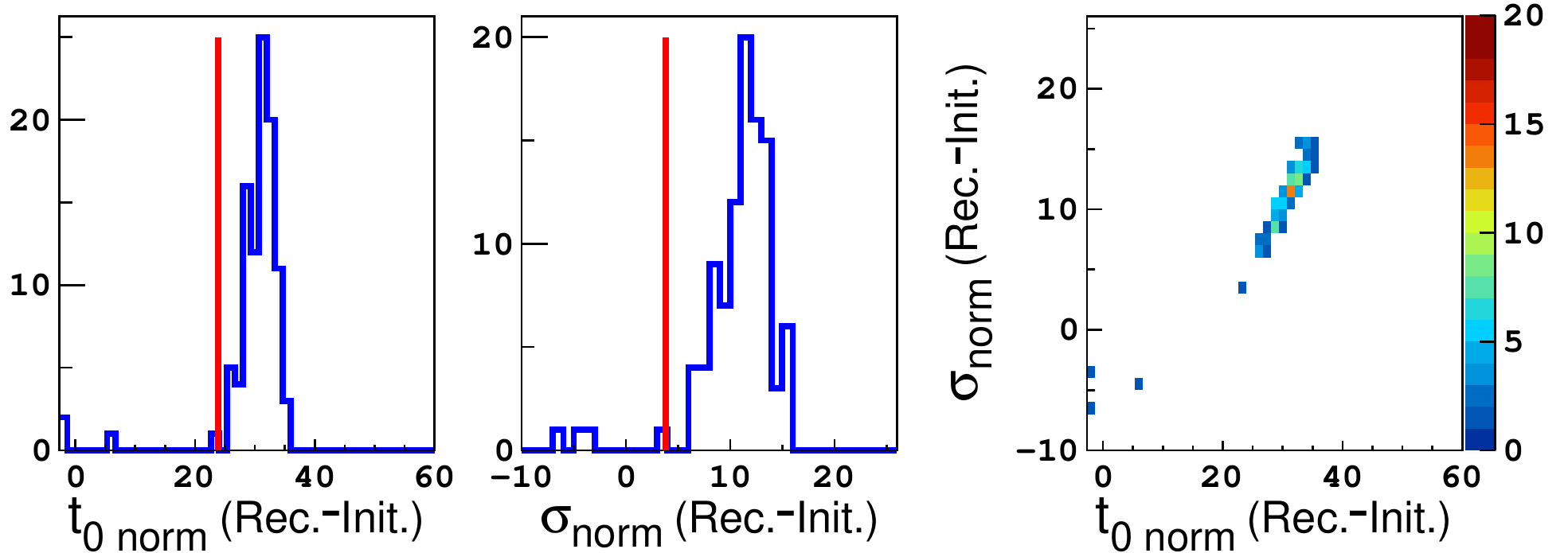}%
  \caption{Change in \tnorm and \sigmanorm for protons from the fits in \fig{fig:QGS_32_fitt0sigma_mod}.}  
  \label{fig:hist_QGS_32_fitt0sigma}
\end{figure}

Similar to the earlier figures presented, \figsThree{fig:QGS_30_fitt0sigma_mod}{fig:QGS_31_fitt0sigma_mod}{fig:QGS_32_fitt0sigma_mod} display the mass composition results from fitting the mass fractions, \tnorm and \sigmanorm of either the \epos, \qgs or \sib parameterisations respectively, to 100 data sets generated from the parameterisation which resulted when the mass fractions, \tnorm and \sigmanorm of the \textbf{\qgs} parameterisation were fitted to Auger FD \Xmax data. The true mass composition of the mock data is the mass composition from this \qgs fit to the Auger FD \Xmax data. The \qgs based mock \Xmax distributions will be slightly different to the \epos based mock distributions,  because the \Xmax parameterisations do not perfectly fit the Auger data, and the respective parameterisations consist of differences which can not be compensated for by an appropriate \tnorm and \sigmanorm shift. \figsThree{fig:hist_QGS_30_fitt0sigma}{fig:hist_QGS_31_fitt0sigma}{fig:hist_QGS_32_fitt0sigma} display the fitted values of \tnorm and \sigmanorm for the \epos, \qgs or \sib fits respectively to the \qgs based data.

The fits to \qgs based mock data produce similar results to the fits of \epos based mock data. The mass fraction, \tnorm and \sigmanorm fit of the \epos parameterisation to \qgs based mock data reconstructs the mass composition above \energy{18.2} with an accuracy almost as good as the \qgs parameterisation fit to the same data. For both the \epos and \qgs fits, the absolute offsets in the median mass fractions from the true mass are less than $10\%$ in most energy bins. As noted before, due to the differences between the \epos and \qgs $t_0$ descriptions as a function of energy, the mass reconstruction accuracy of the \epos fit varies more with energy than the \qgs fit. Again the \sib fit, in this case to \qgs based data, does not reconstruct the mass composition as accurately as the \epos or \qgs fits.

\begin{figure}[htb!]
  \includegraphics[width=0.48\textwidth]{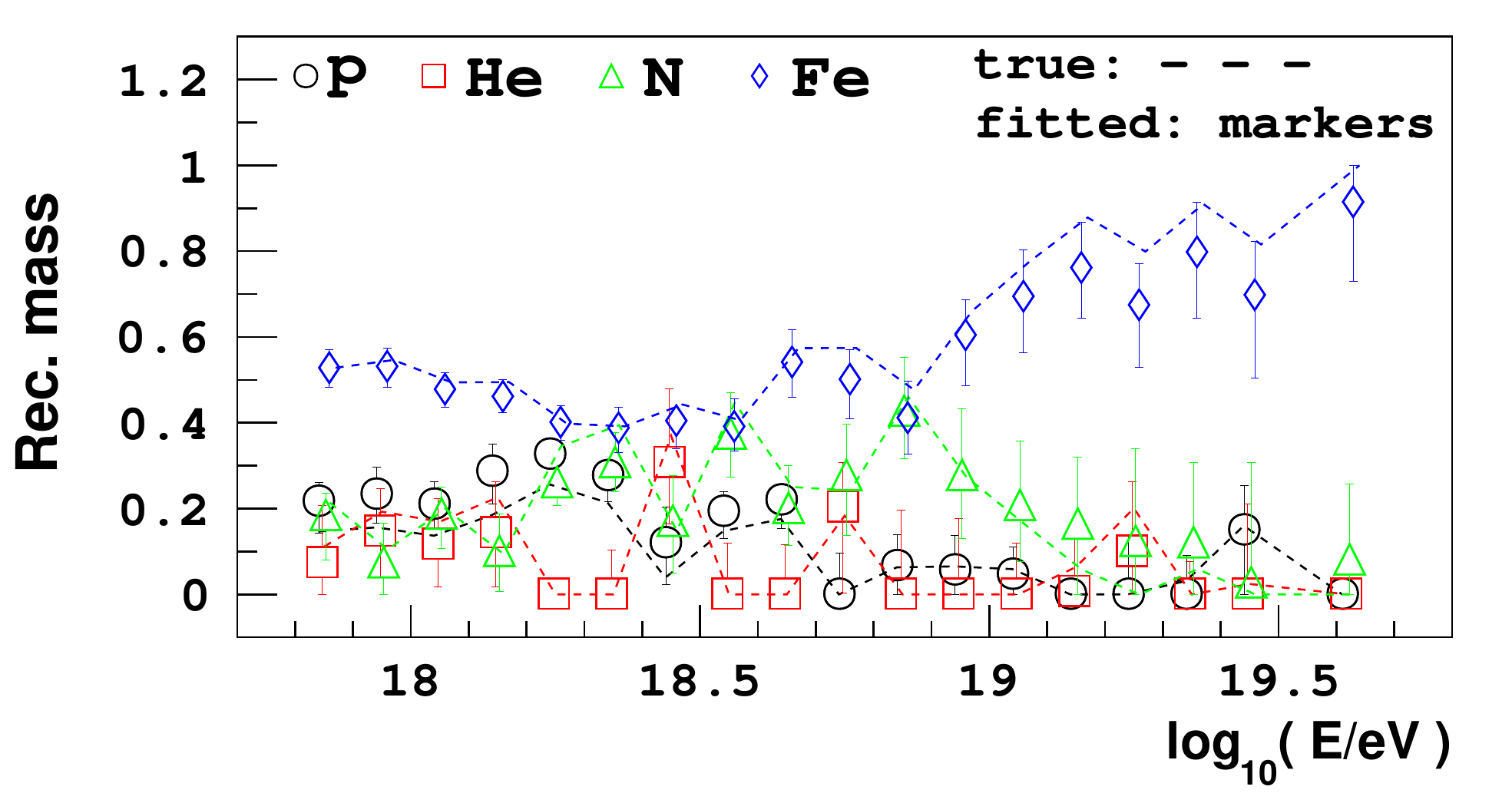}%
  \caption{\epos fit of \Xmax data generated from the \sib parameterisation fit of Auger data.}
  \label{fig:SIB_30_fitt0sigma_mod}
\end{figure}

\begin{figure}[htb!]
        \includegraphics[width=0.48\textwidth]{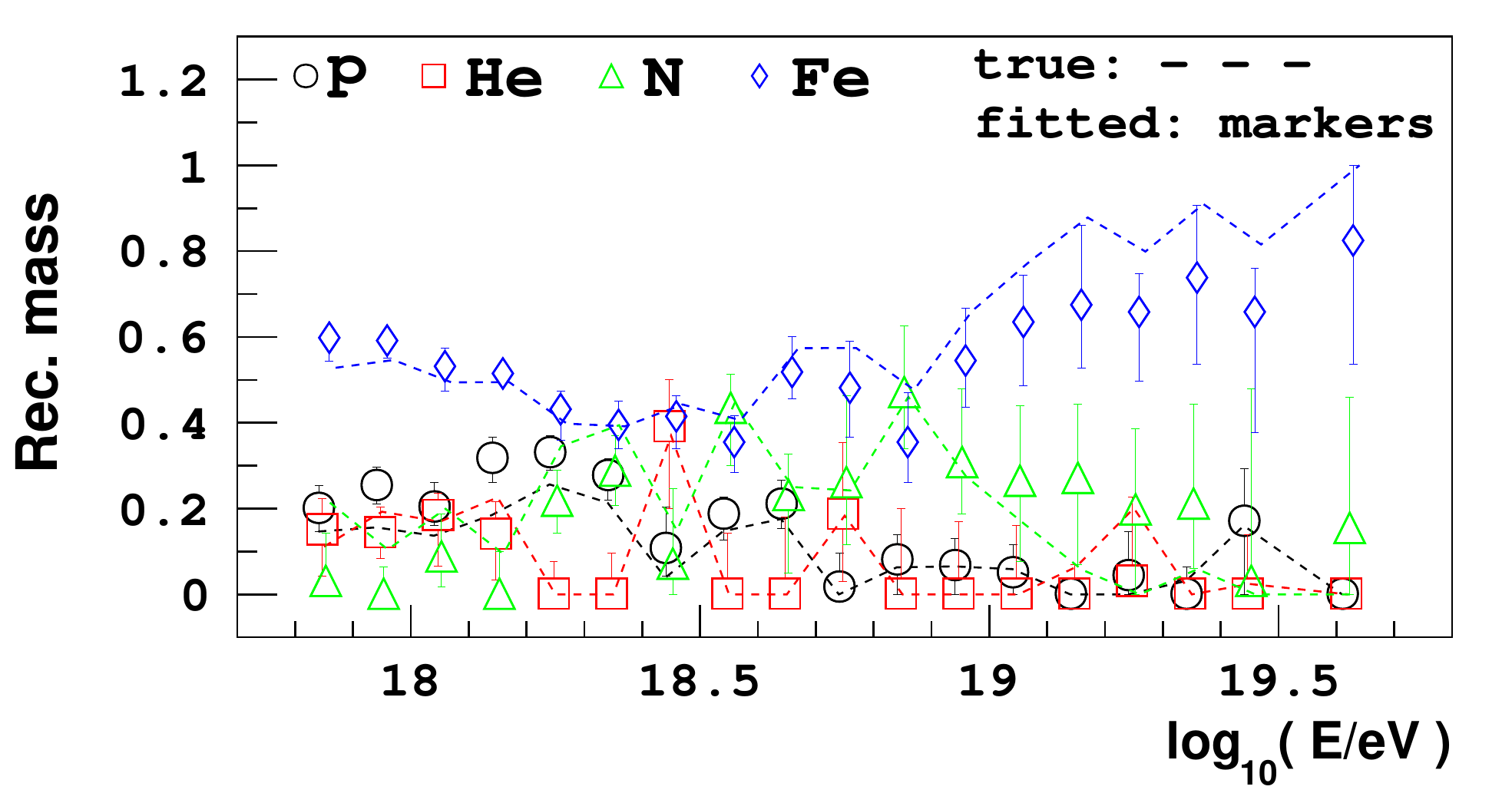}%
        \caption{\qgs fit of \Xmax data generated from the \sib parameterisation fit of Auger data.}
        \label{fig:SIB_31_fitt0sigma_mod}
\end{figure}

\begin{figure}[htb!]
        \includegraphics[width=0.48\textwidth]{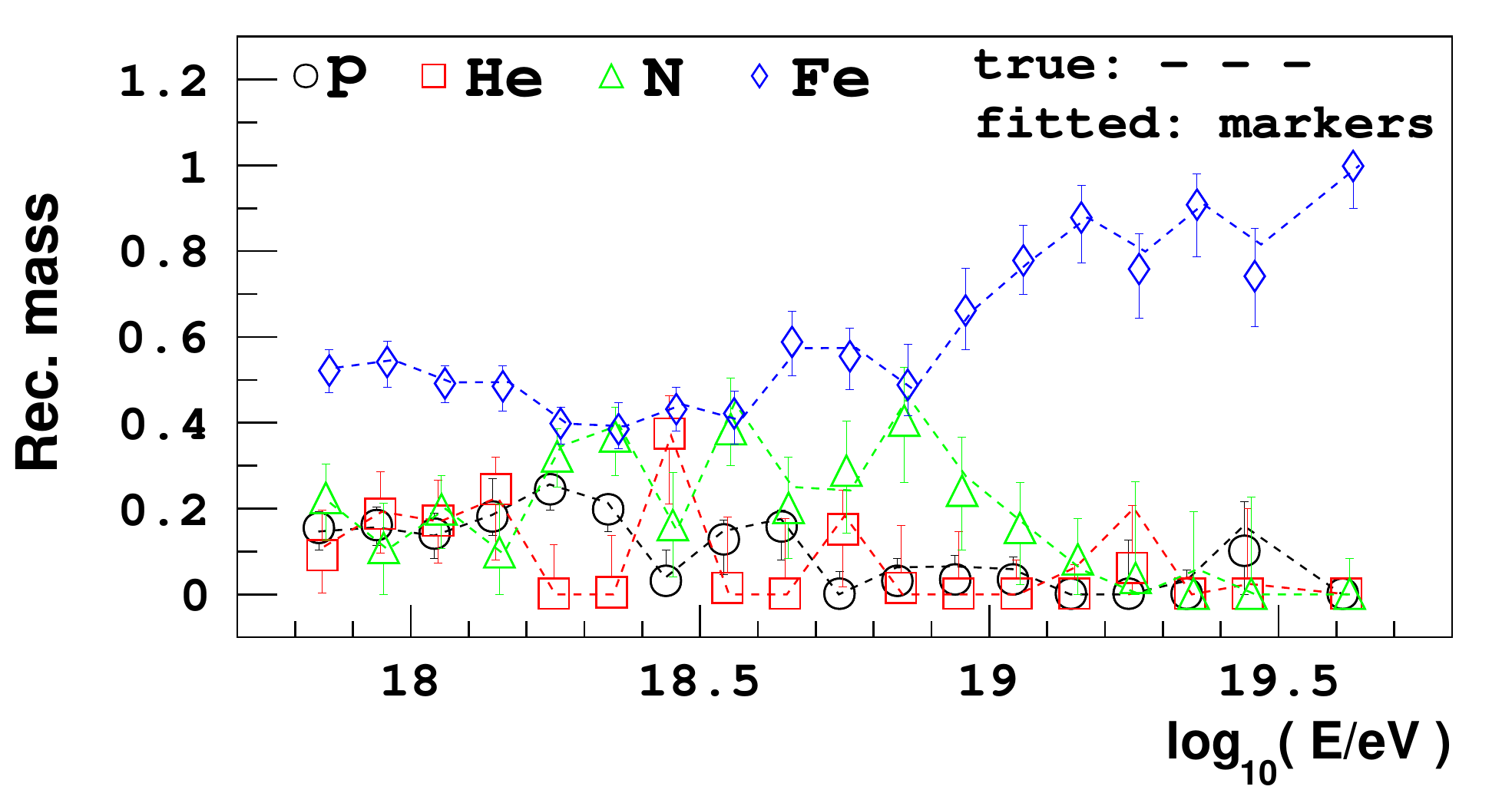}%
        \caption{\sib fit of \Xmax data generated from the \sib parameterisation fit of Auger data.}
        \label{fig:SIB_32_fitt0sigma_mod}
\end{figure}

\begin{figure}[htb!]
  \includegraphics[width=0.48\textwidth]{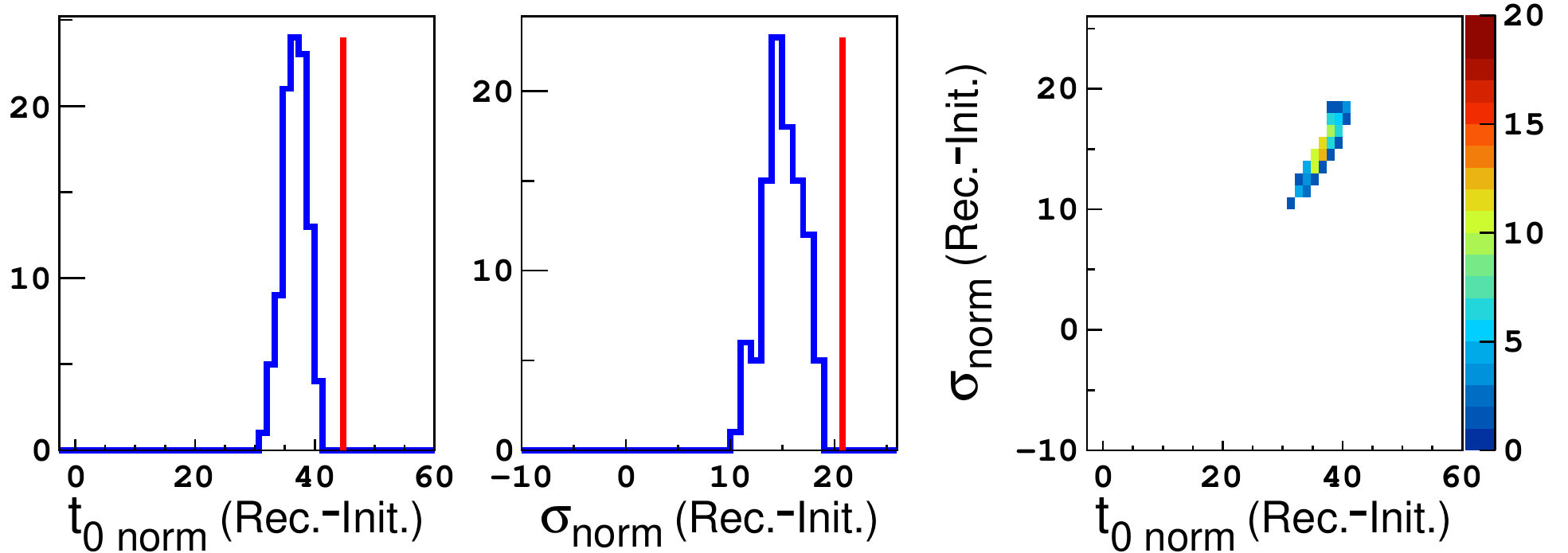}%
  \caption{Change in \tnorm and \sigmanorm for protons from the fits in \fig{fig:SIB_30_fitt0sigma_mod}.}  
  \label{fig:hist_SIB_30_fitt0sigma}
\end{figure}

\begin{figure}[htb!]
  \includegraphics[width=0.48\textwidth]{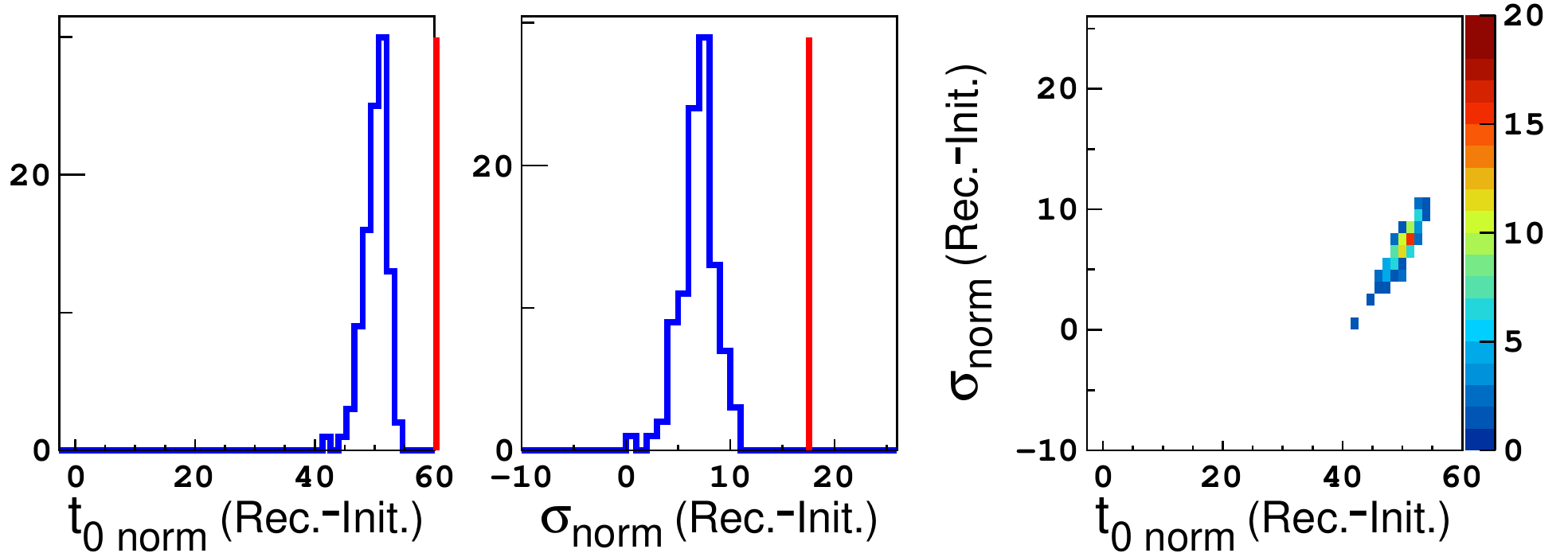}%
  \caption{Change in \tnorm and \sigmanorm for protons from the fits in \fig{fig:SIB_31_fitt0sigma_mod}.}  
  \label{fig:hist_SIB_31_fitt0sigma}
\end{figure}

\begin{figure}[htb!]
  \includegraphics[width=0.48\textwidth]{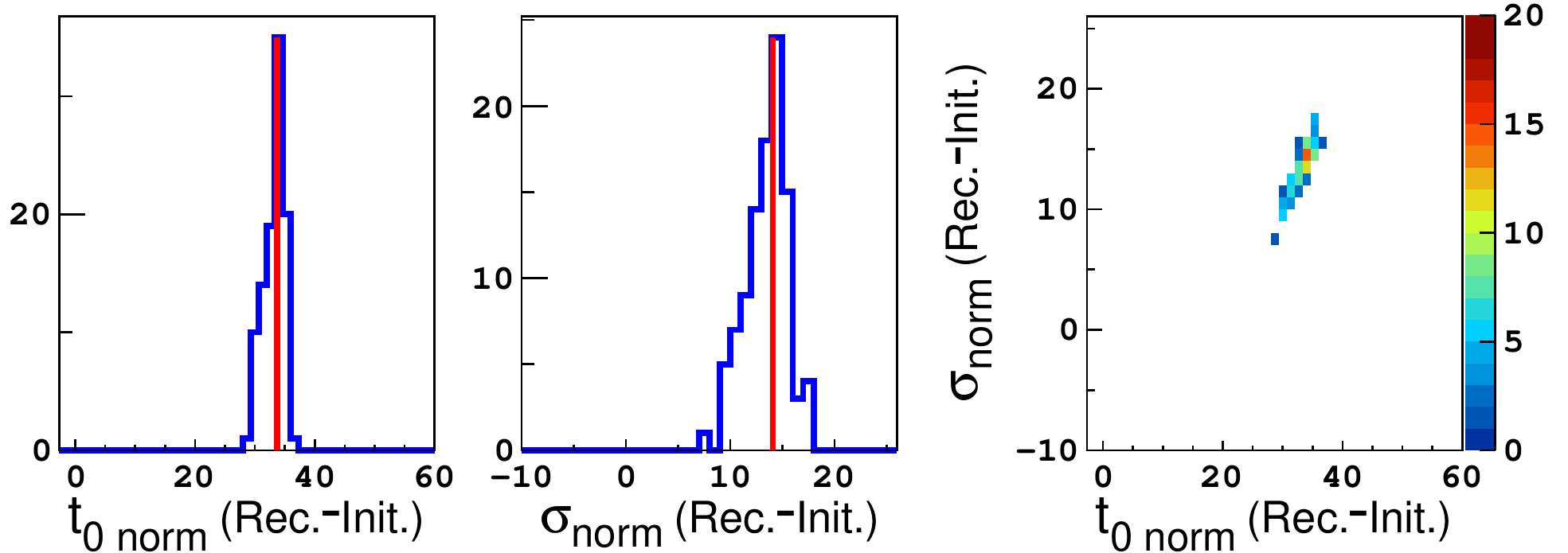}%
  \caption{Change in \tnorm and \sigmanorm for protons from the fits in \fig{fig:SIB_32_fitt0sigma_mod}.}  
  \label{fig:hist_SIB_32_fitt0sigma}
\end{figure}

\figsThree{fig:SIB_30_fitt0sigma_mod}{fig:SIB_31_fitt0sigma_mod}{fig:SIB_32_fitt0sigma_mod} display the mass composition results from fitting the mass fractions, \tnorm and \sigmanorm of either the \epos, \qgs or \sib parameterisations respectively, to 100 data sets generated from the parameterisation which resulted when the mass fractions, \tnorm and \sigmanorm of the \textbf{\sib} parameterisation were fitted to Auger FD \Xmax data. \figsThree{fig:hist_SIB_30_fitt0sigma}{fig:hist_SIB_31_fitt0sigma}{fig:hist_SIB_32_fitt0sigma} display the respective \tnorm and \sigmanorm from these fits. The \epos and \qgs fits to the \sib based data do not reconstruct the true mass composition as accurately as the \sib fit, but they do accurately represent the general transition of the mass composition. The \sib fit to \sib based data (see \figs{fig:SIB_32_fitt0sigma_mod}{fig:hist_SIB_32_fitt0sigma}) results in absolute offsets in the median mass fractions from the true mass of less than $10\%$.

The data fitted in this section sufficiently constrains the fitted values of \tnorm and \sigmanorm, regardless of the parameterisation fitted. If different populations of \tnorm and \sigmanorm were present in a histogram plot, it would indicate the data is unable to adequately constrain the fit, due to the degeneracy between the fitted shape coefficients and the mass fractions. 

Data consisting of predominantly iron, such as the data sets fitted in this section, are easier to fit than data consisting of predominately protons and helium. 

The ability of a \tnorm and \sigmanorm fit of these parameterisations to reconstruct the general mass composition trend of data based on any of these three parameterisations, indicates that the normalisations of $t_0$ and $\sigma$ are the most relevant differences between these parameterisations in regards to reconstructing the mass composition. The results of the \tnorm, \sigmanorm and mass fraction fits of the Auger FD \Xmax data ~\cite{Aab:2014kda} are presented in Section~\ref{sec.res}.

\subsection{Fitting \tnorm, $B$, \sigmanorm and the mass fractions}

The coefficient $B$ (which defines the energy dependence of $t_0$) can also be fitted with \tnorm and \sigmanorm provided the data consists of an adequate dispersion of masses and statistics. This three-coefficient fit will generally be less precise than the two-coefficient fit of only \tnorm and \sigmanorm. Fitting additional coefficients increases the degeneracy between the fitted variables, unless there is significant mass diversity and statistics. Our \epos, \qgs and \sib predictions of $B$ are fairly similar among primaries, therefore we do not expect to see a significant improvement in the systematics of the reconstructed mass composition when adding $B$ to our parameterisation fits of data based on any of these three models. However, it is possible that nature has a different energy dependence for $t_0$ (different from the three models), so by including $B$ in the fit we reduce considerably the model dependence of the mass composition interpretation of the \Xmax distributions. 

\begin{figure}[htb!]
        \includegraphics[width=0.48\textwidth]{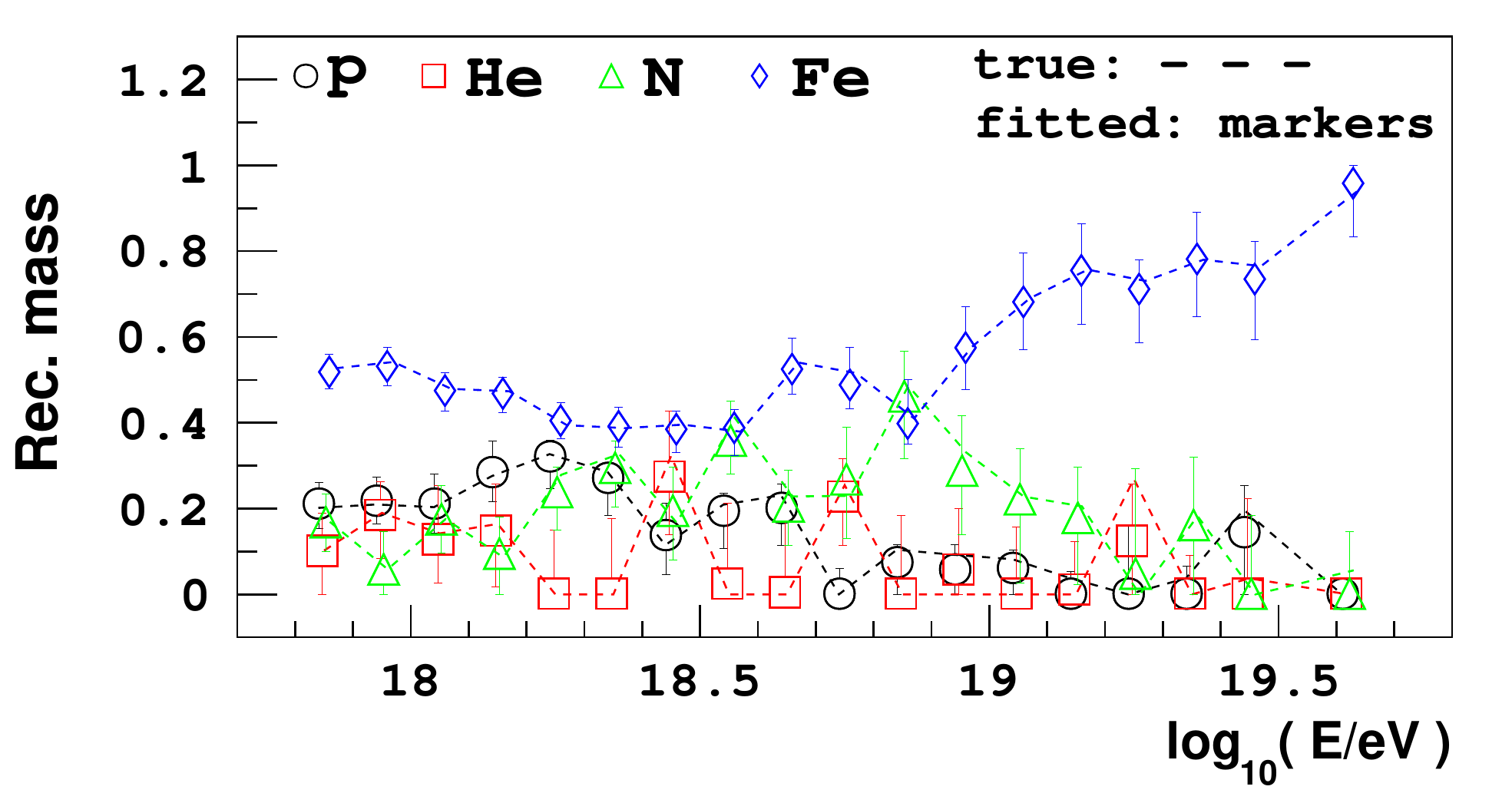}%
        \caption{\epos fit of \Xmax data generated from the \epos parameterisation fit of Auger data.}
        \label{fig:EPOS_30_fitt0Asigma_mod}
\end{figure}

\begin{figure}[htb!]
  \includegraphics[width=0.48\textwidth]{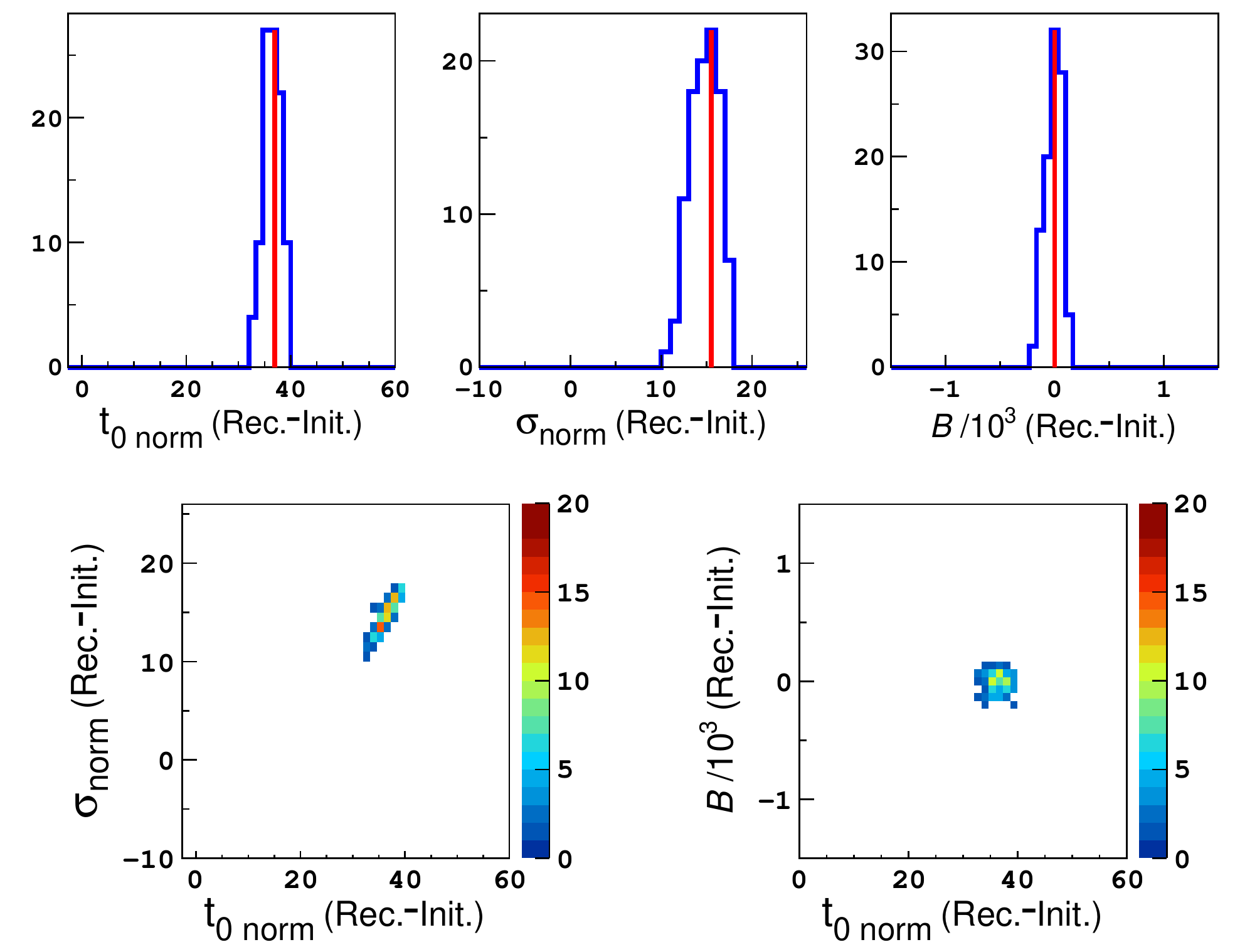}%
  \caption{Change in \tnorm, $B$ and \sigmanorm for protons from the fits in \fig{fig:EPOS_30_fitt0Asigma_mod}.}  
  \label{fig:hist_EPOS_30_fitt0Asigma}
\end{figure}

 \figs{fig:EPOS_30_fitt0Asigma_mod}{fig:hist_EPOS_30_fitt0Asigma} display the reconstructed mass composition and fitted coefficient values from fitting \tnorm, $B$ and \sigmanorm of our \epos parameterisations to data generated from the \epos \tnorm and \sigmanorm fit of the FD \Xmax data set. Comparing this result to \fig{fig:EPOS_30_fitt0sigma_mod}, the systematic offsets in the median reconstructed mass composition from the true mass for the three-coefficient fit are similar to the two-coefficient fit.  \fig{fig:hist_EPOS_30_fitt0Asigma} shows that the three fitted shape coefficients are accurately fitted and are well constrained.
 
However, as mentioned previously, data consisting of predominantly iron are easier to fit than data consisting of predominately proton and helium. The \tnorm, $B$, \sigmanorm  and mass fraction fit of the latter data can result in a reconstructed mass composition which is considerably less accurate than a fit where $B$ is fixed to the true value of the data. This is because the degeneracy between the fitted parameters can result in the fitted shape coefficients shifting away from the true values.

\subsection{Effect of \Xmax systematic uncertainties when fitting \tnorm and \sigmanorm}
\begin{figure}[htb!]
  \includegraphics[width=0.48\textwidth]{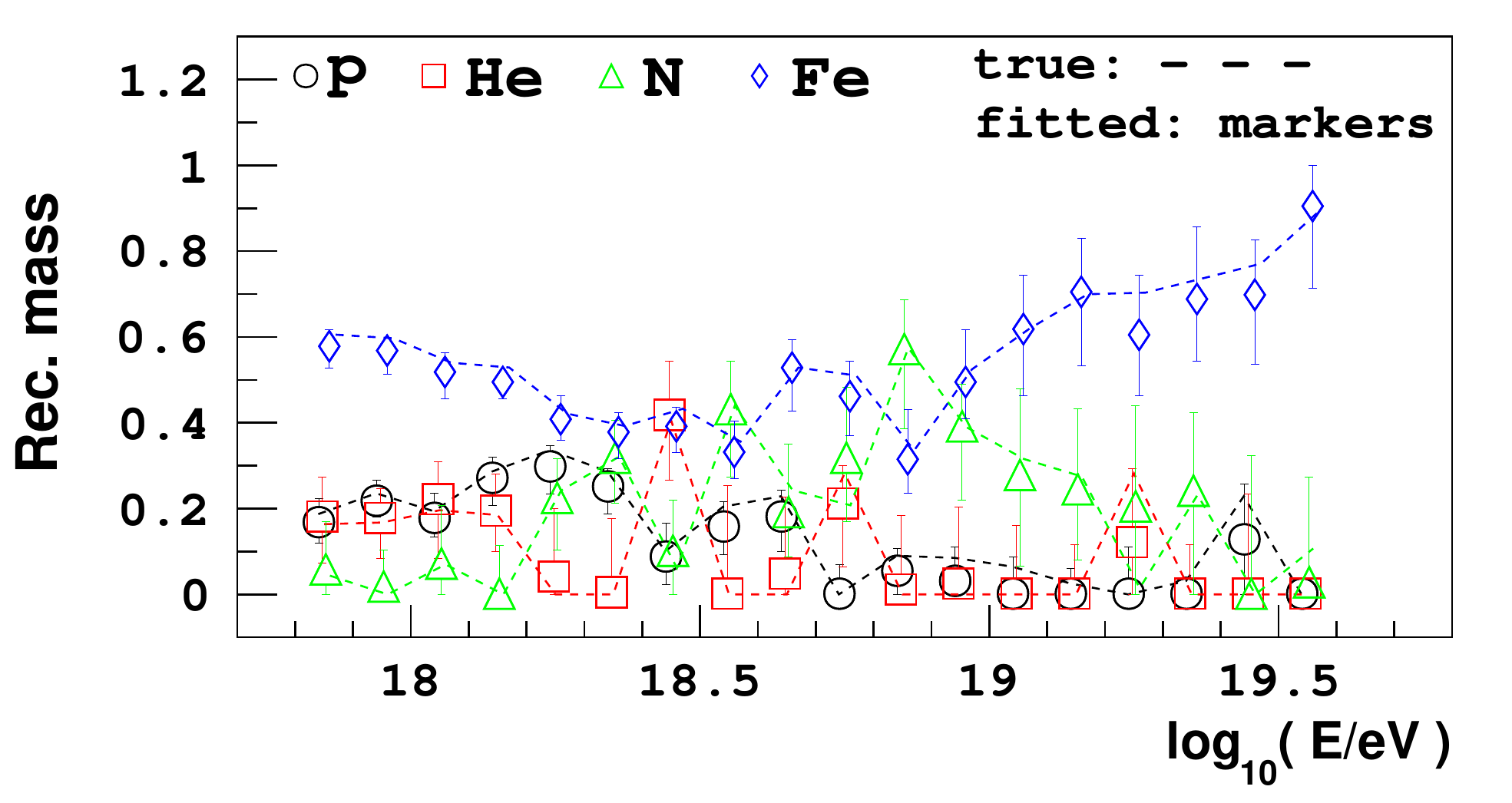}%
  \caption{Fits of \tnorm and \sigmanorm to \Xmax data consisting of a $\SI{-10}{\gcm}$ systematic offset in \Xmax. The \Xmax data was also smeared by a Gaussian distributed random variable of $\sigma = \SI{10}{\gcm}$, which was unaccounted for in the initial \Xmax parameterisation fitted.}
  \label{fig:performance31fit1100_twentyrealdata_extendlimitmodel31old_fitt0normsigmanorm_Xmaxoffset10_Xmaxres10}
\end{figure}

\begin{figure}[htb!]
  \includegraphics[width=0.48\textwidth]{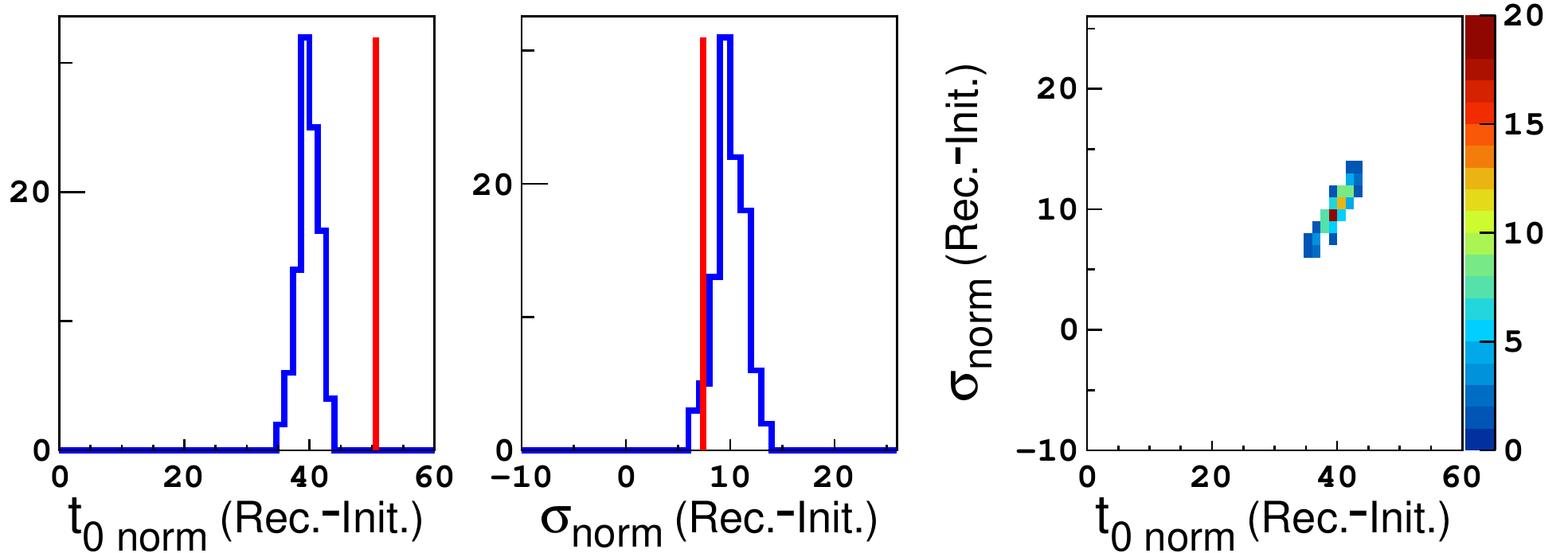}%
  \caption{Change in \tnorm and \sigmanorm for protons from the fits in \fig{fig:performance31fit1100_twentyrealdata_extendlimitmodel31old_fitt0normsigmanorm_Xmaxoffset10_Xmaxres10}.}
  \label{fig:histcorr_performance31fit1100_twentyrealdata_extendlimitmodel31old_fitt0normsigmanorm_Xmaxoffset10_Xmaxres10}
\end{figure}

Fitting \tnorm can compensate for systematic offsets in \Xmax, while fitting \sigmanorm can compensate for systematic errors in the estimation of the detector resolution of \Xmax. \figs{fig:performance31fit1100_twentyrealdata_extendlimitmodel31old_fitt0normsigmanorm_Xmaxoffset10_Xmaxres10}{fig:histcorr_performance31fit1100_twentyrealdata_extendlimitmodel31old_fitt0normsigmanorm_Xmaxoffset10_Xmaxres10} shows the results of fitting the mass fractions, \tnorm and \sigmanorm of our \qgs parameterisation to 100 data sets generated from the parameterisation which resulted when the mass fractions, \tnorm and \sigmanorm of the \qgs parameterisation were fitted to Auger FD \Xmax data. Across the whole energy range, the mock data was shifted by a systematic offset of \depth{-10}, and also smeared by a Gaussian distributed random variable of $\sigma = \depth{10}$ (this additional smearing is not accounted for in the resolution of the applied \Xmax parameterisation), to test if the fit of \tnorm and \sigmanorm can compensate for these systematics. The red lines in \fig{fig:histcorr_performance31fit1100_twentyrealdata_extendlimitmodel31old_fitt0normsigmanorm_Xmaxoffset10_Xmaxres10} indicate the true \tnorm and \sigmanorm values of the data (relative to the initial \qgs parameterisation being fitted) before the \Xmax systematics were applied. 

The mean shift in the fitted \tnorm values from the original \tnorm values of the data is $\sim$\;\depth{-10} (\fig{fig:histcorr_performance31fit1100_twentyrealdata_extendlimitmodel31old_fitt0normsigmanorm_Xmaxoffset10_Xmaxres10}), to compensate mainly for the \depth{-10} \Xmax systematic offset applied to the data. As $t_0$ changes by the same amount for each primary when \tnorm is fitted, and the \Xmax systematic was applied consistently to all data, the \tnorm fit is capable of completely accounting for the \Xmax systematic offset. However, \sigmanorm for each primary is changed by different absolute amounts when fitting this coefficient, but all of the data is smeared (all masses are consistently smeared), consequently the correct \sigmanorm cannot be fitted for each primary, which may also effect the fit of \tnorm. The shift in \sigmanorm for protons from the original \sigmanorm is only $\sim$\;\depth{+2}. Despite the fit of \sigmanorm being unable to thoroughly account for the \depth{10} systematic in the resolution, the absolute offsets in the median reconstructed mass fractions from the true mass are less than $10\%$ in most energy bins, due to a combined shift of \tnorm and \sigmanorm in the appropriate directions. 


The accuracy of the reconstructed mass fractions from the fit of this shifted and smeared data is similar to the same fit of the un-shifted and un-smeared data in \fig{fig:QGS_31_fitt0sigma_mod}. Reasonable detector resolution systematics and systematic offsets in \Xmax will not significantly effect the accuracy of the reconstructed mass composition.

If the data was not smeared by a Guassian random variable, and only shifted by a constant \Xmax offset, the \tnorm and \sigmanorm fit of this shifted data would result in a change in the fitted \tnorm (compared to the \tnorm fitted to the un-shifted data) which is very close to the value of the \Xmax offset. Shifting the \Xmax data by a constant value has essentially the same effect on the fit as shifting the parameterisation by a constant value, with a very minuscule difference arising if the detector acceptance of \Xmax is not shifted by the same offset to account for the applied \Xmax offset (this is not an issue when fitting the measured Auger data).

 \section{\label{sec.res}Results}
We have applied our \epos, \qgs and \sib \Xmax parameterisations separately to \Xmax data measured by the Pierre Auger Observatory fluorescence detector (FD) \cite{Aab:2014kda}. 

\begin{figure*}
  \centering
  \includegraphics[width=0.9\textwidth]{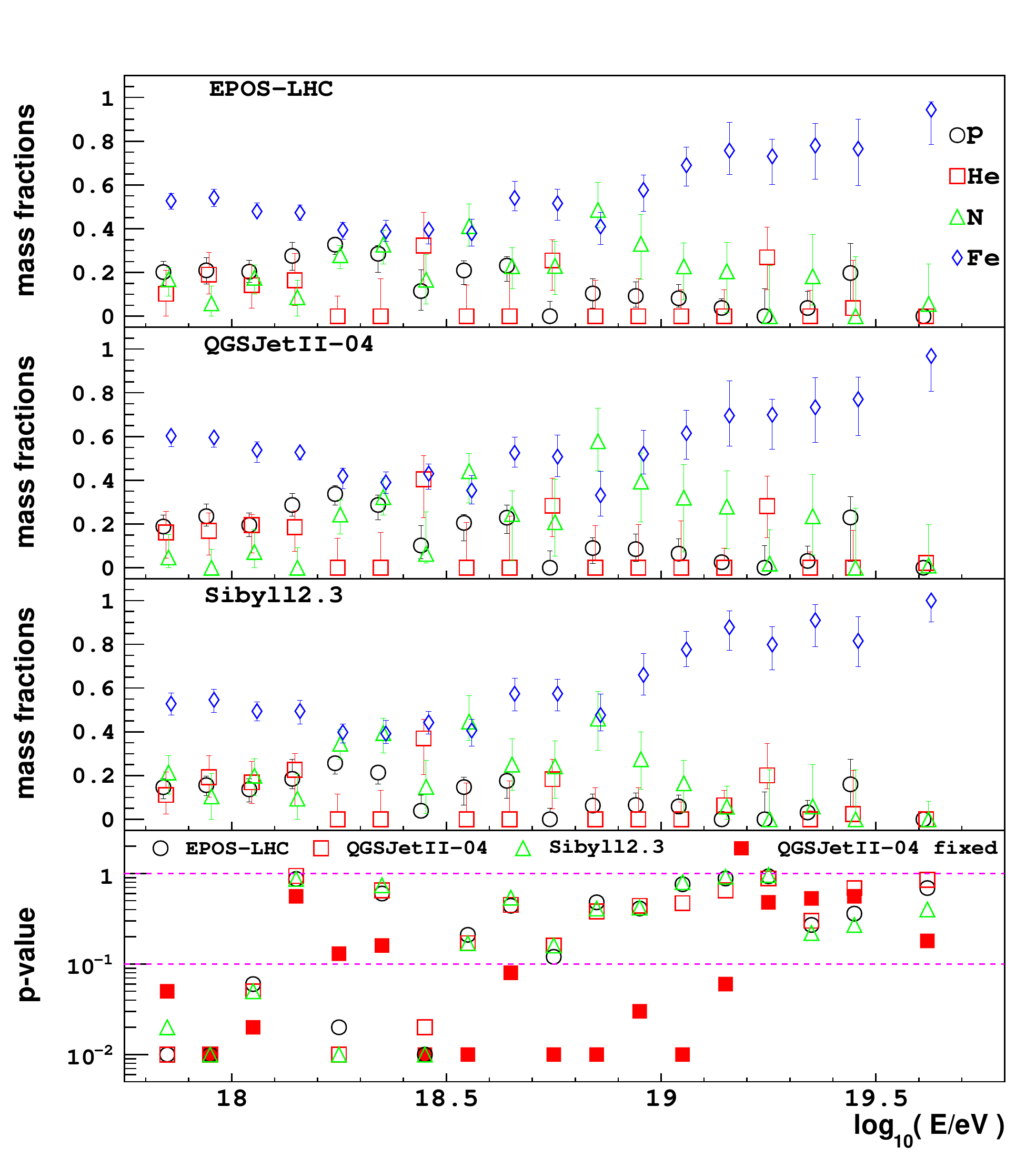}%
  \caption{ Fitting \tnorm, \sigmanorm and the mass fractions of our parameterisations to FD \Xmax data measured by the Pierre Auger Observatory. The fitted mass fractions and p-values for each fitted model are shown. The red solid squares show the p-values for \qgs when fitting only the mass fractions (\tnorm and \sigmanorm fixed).}
 
  \label{fig:realdata_comp_2_all}
\end{figure*}

  \begin{figure}[!htb]
  \centering 
      \includegraphics[width=0.48\textwidth]{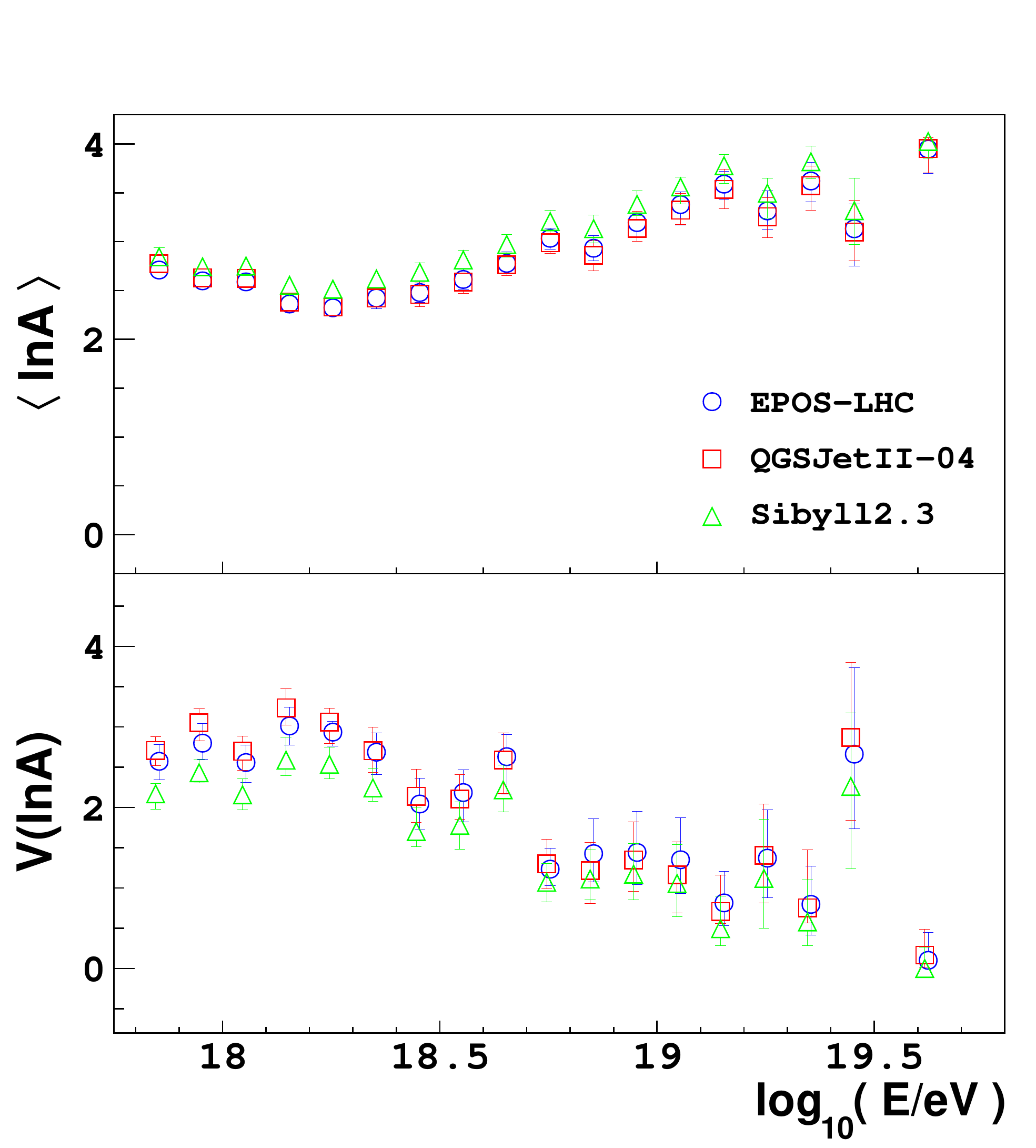}%
      \caption{First two moments of the \lnA distribution estimated from the fitted fractions of the \tnorm, \sigmanorm and mass fraction fit of the FD \Xmax distributions measured by the Pierre Auger Observatory.}
      \label{fig:lnAmoments}
\end{figure}

\begin{figure}[!ht]
  \centering
   \hbox{\hspace{-0.5cm}
 \includegraphics[width=0.48\textwidth]{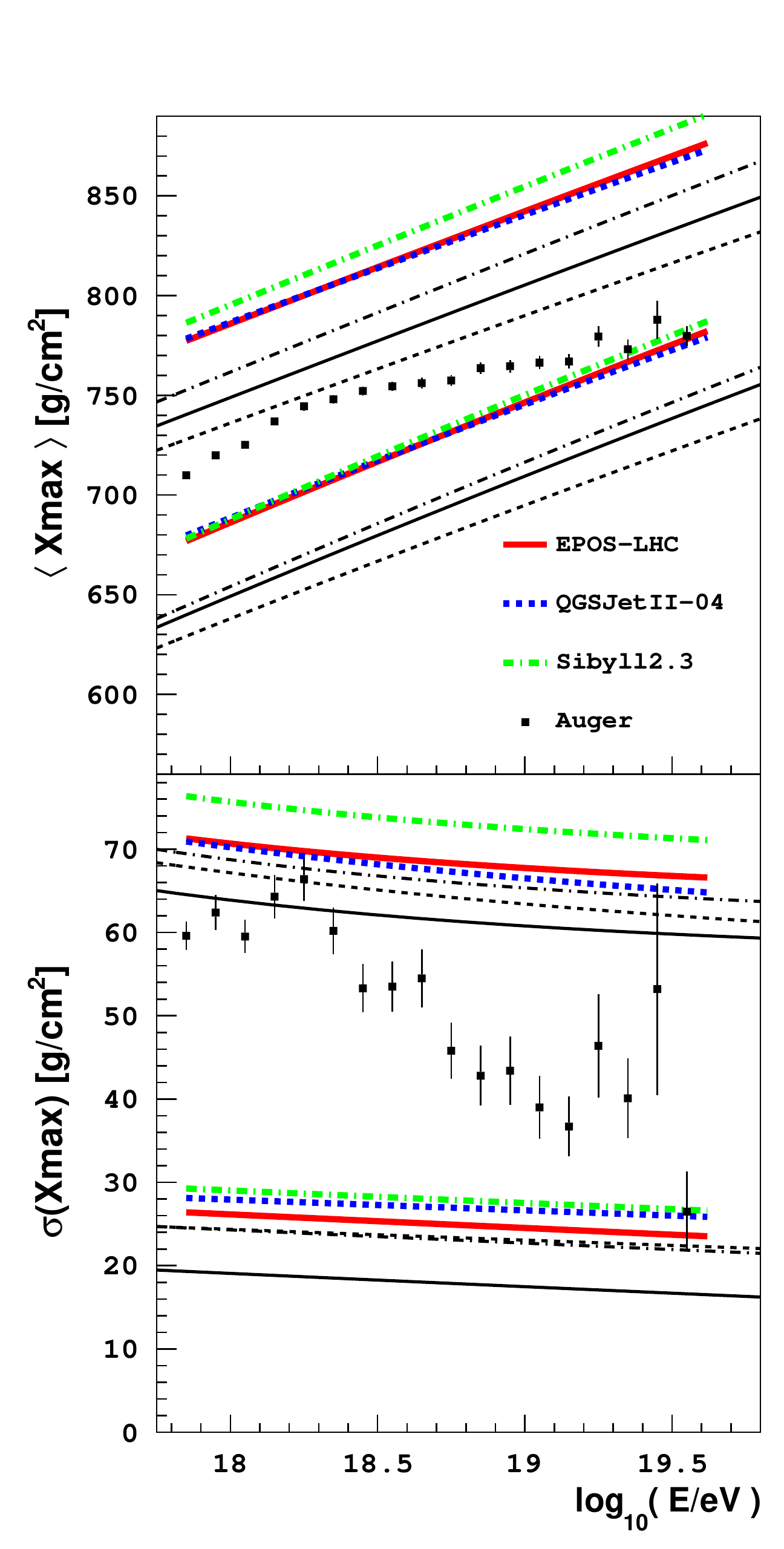}}
 \caption{ The black lines show the \meanXmax and \sigmaXmax initially predicted by the \Xmax parameterisations for proton and iron. The red, blue and green lines show the new predictions for the \meanXmax and \sigmaXmax after fits of \tnorm, \sigmanorm  and the mass fractions to FD \Xmax distributions measured by the Pierre Auger Observatory.}
   \label{fig:realdata_moments_2_all}
  \end{figure}

\fig{fig:realdata_comp_2_all} displays the results from fitting the mass fractions and the coefficients \tnorm and \sigmanorm of our \epos, \qgs and \sib \Xmax distribution parameterisations. The top three panels display the fitted mass fractions for each model, and the bottom panel shows the p-values for these fits. The fits of these parameterisations to the \Xmax distributions are shown in Appendix~\ref{AppFDfits_t0sigma}. 

 The p-value is defined as the probability of obtaining a worse fit (larger likelihood ratio $\mathcal{L}$) than that obtained with the data. The resulting parameterisation and fractions from the fit of the \Xmax distributions were used to generate sets of mock \Xmax distributions to determine the p-values, and to calculate the mass composition statistical errors. Fitting \tnorm and \sigmanorm improves the goodness of the fit of the \Xmax distributions (bottom panel \fig{fig:realdata_comp_2_all}). This is evident by comparing the \qgs p-values for the \tnorm and \sigmanorm fit to the \qgs p-values for the fit of only the mass fractions.

We find that the \epos, \qgs and \sib parameterisation fits of the \Xmax distributions give a consistent mass composition result. \fig{fig:lnAmoments} shows the corresponding moments of the \lnA distribution. The results suggest a composition consisting of predominantly iron. Below \energy{18.8}, the small proportions of proton, helium and nitrogen vary. Above \energy{18.8}, there is little proton or helium, and with increasing energy the nitrogen component gradually gives way to the growing iron component, which dominates at the highest energies. There does not appear to be a distinct feature near the ankle ($\sim \energy{18.2}$), where it is assumed cosmic rays transition from Galactic to extragalactic \cite{Linsley}. Considering the upper limits on the large scale anisotropy \cite{2012ApJS..203...34P} indicate protons below \energy{18.5} are most likely of extragalactic origin, the fitted proton fractions below the ankle are suitably small if cosmic rays below the ankle are Galactic. A significant modification of the hadronic models is required to accommodate a proton dominant composition above \energy{18} \cite{Berezinsky:1988wi}. 

The first two moments of the Auger \Xmax distributions from \cite{Aab:2014kda} and their predictions (for proton and Fe) as a function of energy are shown in \fig{fig:realdata_moments_2_all}. It shows that the \tnorm and \sigmanorm fits reduce the difference between the predictions from  the \epos and \qgs hadronic models. For $t_0$ and $\sigma$, the separation between the proton prediction and heavier nuclei is larger in the \sib parameterisation than the \epos or \qgs parameterisations, consequently the \sib proton predictions from the fit are in disagreement with the two other parameterisations. The values of the coefficients in Equation~\eqref{eq:Xmaxbasicshape} for proton, helium, nitrogen and iron primaries for the \epos, \qgs and \sib models (assuming a normalisation energy of $E_0 = \energy{18.24}$) can be found in \tab{tab:comp} of Appendix~\ref{AppB}. The values fitted to the data for \tnorm and \sigmanorm are also shown in \tab{tab:comp}. The statistical errors in the estimated value of \meanXmax for protons or iron over the energy range are the same as the statistical error in the fitted value of \tnorm, while for  \sigmaXmax the statistical error is less than \depth{1} for protons and iron.

The fitted values of \tnorm are much larger than the initial parameterisation predictions, consequently the predicted \meanXmax from the fits are much larger than the initial predictions. The fitted \sigmanorm values are also larger than the initial predictions, consequently the predicted \sigmaXmax from the fit is larger. After the fit of \tnorm and \sigmanorm, our \epos, \qgs and \sib parameterisations still have different predictions for the \Xmax distribution shape properties as a function of mass and energy, but despite this there is reasonable agreement on the reconstructed mass composition from these fits. An observed shift in the fitted values of \tnorm and \sigmanorm from the initial parameterisation prediction could be due to the initial parameterisation inadequately describing nature, systematics in the measured \Xmax values, or a combination of both factors. Degeneracy between the fitted parameters could also contribute to a shift in the fitted coefficients, however the performance analysis in Section~\ref{sec.augerperformance} indicates that the results presented here are unlikely to be affected by degeneracy.

\begin{figure*}[!htb]
  \centering
  \includegraphics[width=0.9\textwidth]{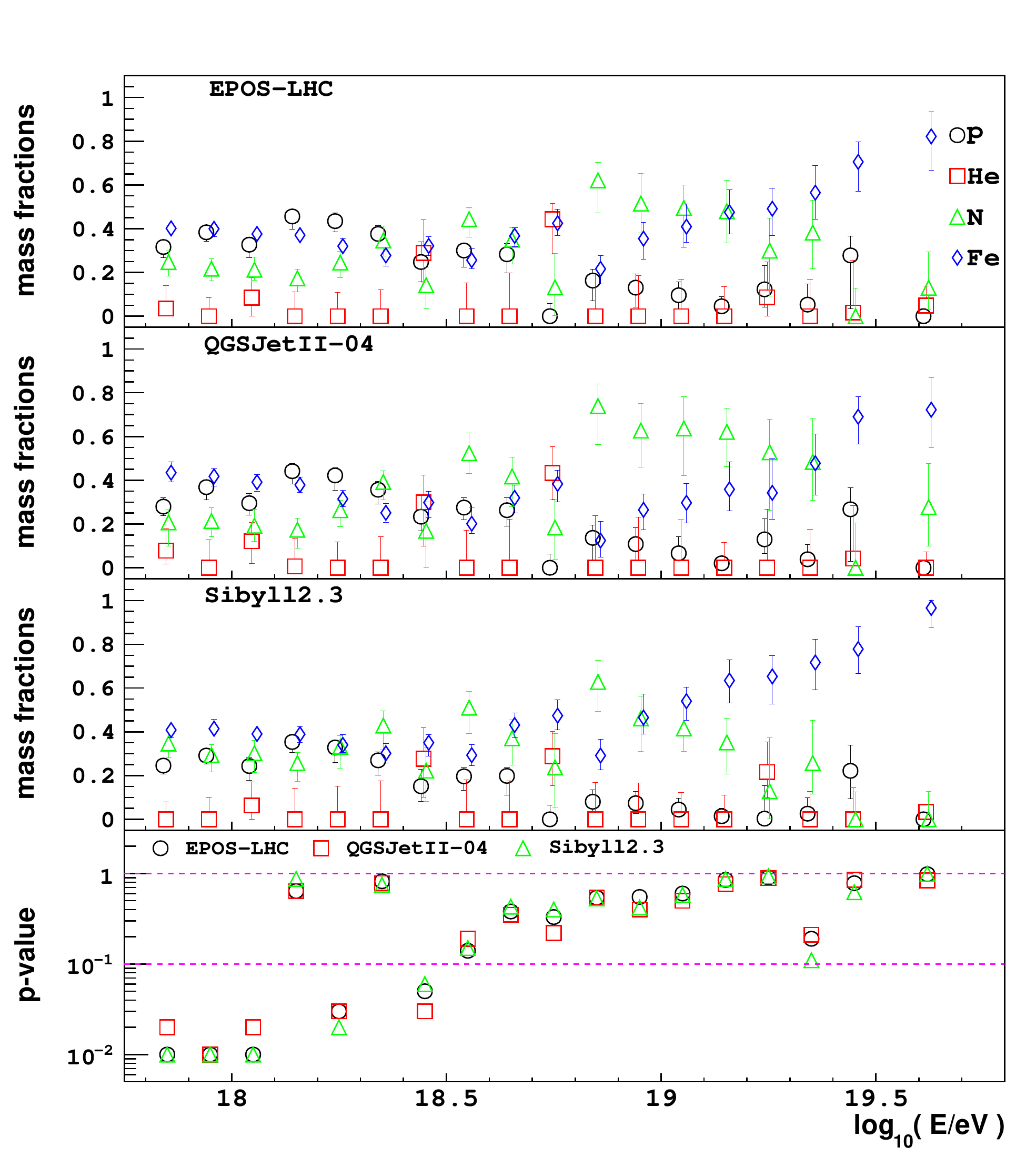}%
  \caption{ Fitting \tnorm and the mass fractions of our parameterisations to FD \Xmax data measured by the Pierre Auger Observatory. The fitted mass fractions and p-values for each fitted model are shown.}
 
  \label{fig:realdata_comp_2_all_t0}
\end{figure*}
 
The mass composition results are sensitive to the assumed values of the \Xmax distribution properties which are not affected by the fit of \tnorm and \sigmanorm (such as the elongation rate and the \meanXmax separation between p and Fe). The results are also sensitive to the fitting range limits. As our knowledge of the hadronic physics occurring at the highest energies progresses, the coefficients which are fitted and the fitting range limits applied may change. For example, a reduced upper limit of \tnorm would result in the \tnorm, \sigmanorm and mass fraction fit of the Auger data reconstructing a mass composition consisting of predominantly proton and helium. An increase in the statistics of the Auger \Xmax data, and/or an increased energy range, can reveal additional information regarding the shape coefficients. 

Using the fitted values of \tnorm and \sigmanorm, the parameters of the equations in \cite{Abreu:2013env}, to convert the \Xmax moments into \lnA moments, have been determined and are shown in Tables~\ref{tab:lnA_coeff_3} and \ref{tab:lnA_coeff_4} of Appendix~\ref{AppC}. 

  \begin{figure}[!htb]
  \centering 
      \includegraphics[width=0.48\textwidth]{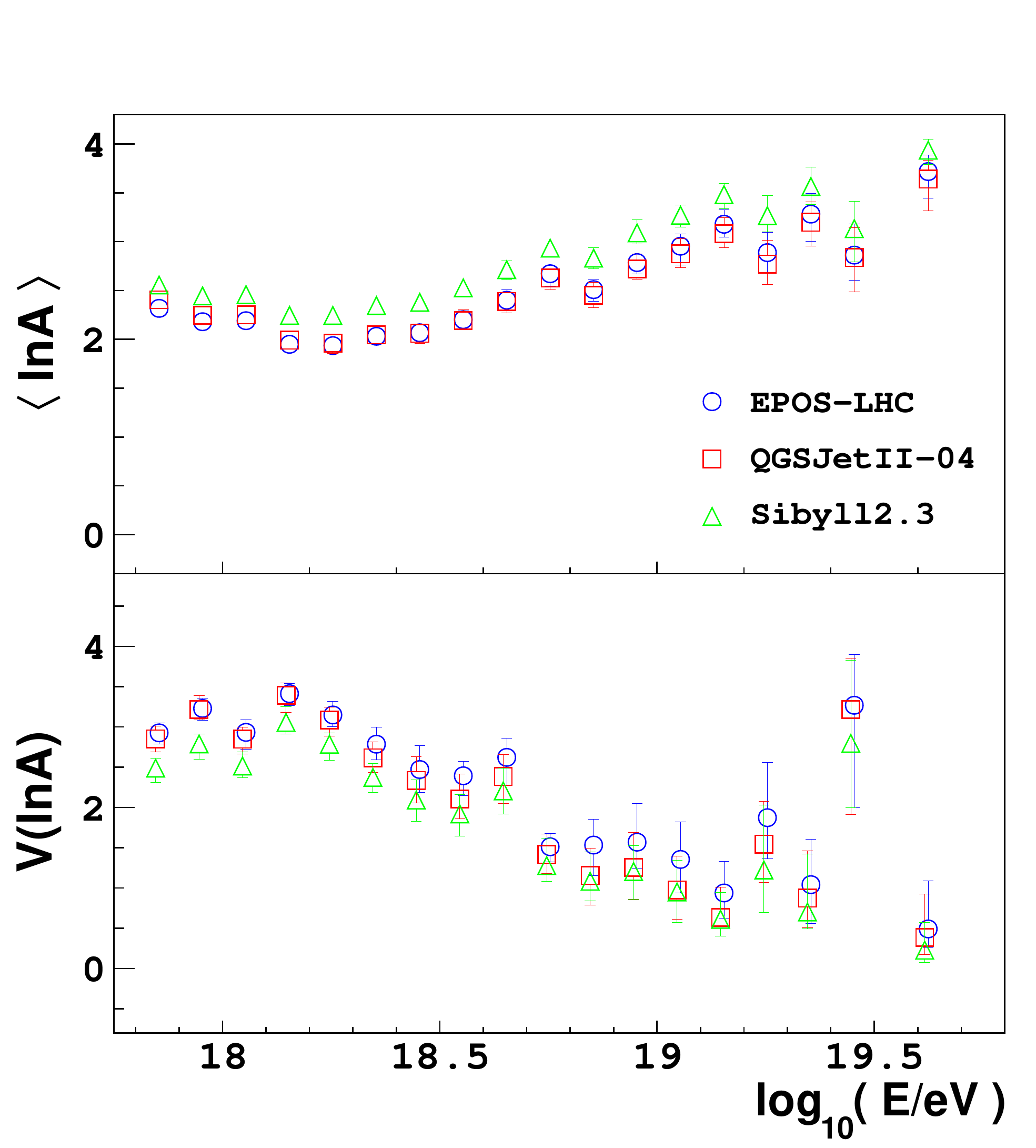}%
      \caption{First two moments of the \lnA distribution estimated from the fitted fractions of the \tnorm and mass fraction fit of the FD \Xmax distributions measured by the Pierre Auger Observatory.}
      \label{fig:lnAmoments_t0}
\end{figure}

\begin{figure}[!htb]
  \centering
   \hbox{\hspace{-0.5cm}
 \includegraphics[trim={0 0 0cm 0}, width=0.48\textwidth]{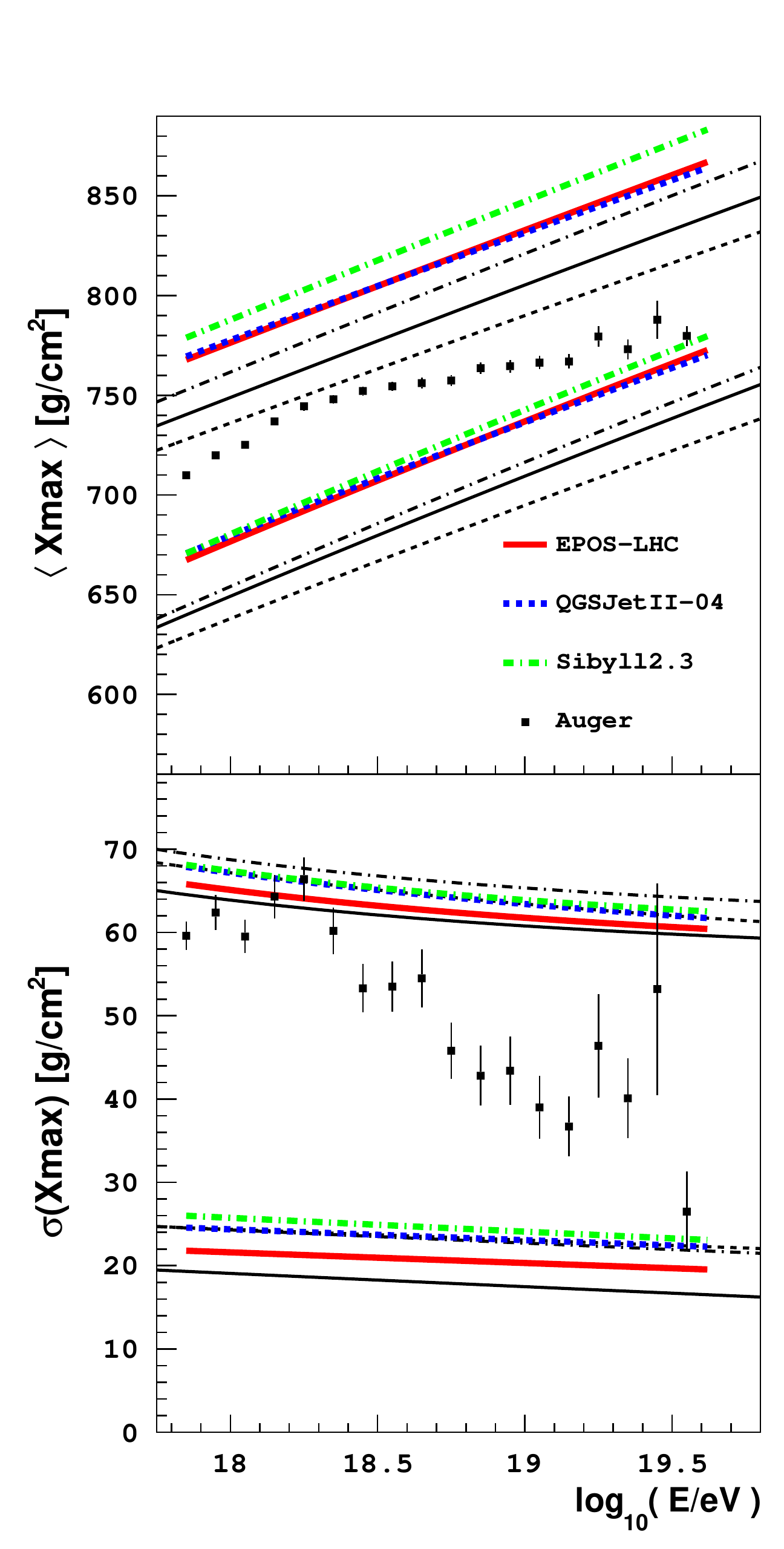}}
 \caption{The black lines show the \meanXmax and \sigmaXmax initially predicted by the \Xmax parameterisations for proton and iron. The red, blue and green lines show the new predictions for the \meanXmax and \sigmaXmax after fits of the mass fractions and \tnorm (applying the standard \qgs $\sigma$ prediction) to FD \Xmax distributions measured by the Pierre Auger Observatory.}
   \label{fig:realdata_moments_2_all_t0}
  \end{figure}
  
Given the large \tnorm and \sigmanorm values fitted to the Auger data when the mass fractions, \tnorm and \sigmanorm are fitted, a second set of fits were performed where only \tnorm and the mass fractions were fitted to the Auger data, using the same \tnorm fitting range. These fits of the three parameterisations each used the standard \qgs $\sigma$ prediction. The resulting mass composition, \lnA and \Xmax moments are shown in \figsThree{fig:realdata_comp_2_all_t0}{fig:lnAmoments_t0}{fig:realdata_moments_2_all_t0} respectively. The fitted values of \tnorm are shown in \tab{tab:comp} of Appendix~\ref{AppB}, and using these values the parameters of the equations in \cite{Abreu:2013env} have been determined and are shown in Tables~\ref{tab:lnA_coeff_5} and \ref{tab:lnA_coeff_6} of Appendix~\ref{AppC}. 

As the fitted values of \tnorm are not as large compared to the two-coefficient fit, the predicted \meanXmax of the fits are not as large, but still quite large compared to the initial parameterisation predictions. The reconstructed mass composition from the fits of only \tnorm (\fig{fig:realdata_comp_2_all_t0}) consists of a larger abundance of nitrogen and protons, at the expense of iron and helium, compared to that of the \tnorm and \sigmanorm fit (\fig{fig:realdata_comp_2_all}). The general transition of the mass composition for the three parameterisations is consistent between the one-coefficient and two-coefficient fits.

\clearpage
\section{Conclusions}
We have presented a novel method to estimate the mass composition (from \Xmax distributions) which is less dependent on hadronic models. The method uses parameterisations of \Xmax distributions according to different hadronic interaction models. Provided that the measured \Xmax distributions consist of different primary masses and sufficient statistics over a large energy range (which seems to be the case for the Auger \Xmax data), two shape coefficients, of the \Xmax distribution parameterisation, can be fitted together with the mass fractions, reducing the model dependency in the mass composition interpretation (we have tested the \epos, \qgs and \sib models). The main differences between the predicted \Xmax distributions from different models are the normalisation values of the mode and spread for each primary. So, by fitting two coefficients (\tnorm and \sigmanorm) which adjust the normalisation of the mode and spread for each primary in an appropriate manner, the resulting mass composition is consistent for the three hadronic models tested here.  A third coefficient, ``$B$'', which adjust the energy dependence of the \meanXmax  can be fitted, further reducing the systematic model uncertainty in the fitted mass composition. However, given the current statistics and limited energy range of the published Auger \Xmax distributions and the possible distribution of masses, fitting this third parameter may introduce large systematic uncertainties in the composition.

The mass fraction, \tnorm and \sigmanorm fits reconstruct a mass composition trend with energy that is consistent between the three models. There is a dominant abundance of iron over the energy range, particularly at the highest energies where there is almost pure iron. By fitting only \tnorm and adopting the \qgs $\sigma$ prediction for the three models, the relative abundance of protons increases. 

The results are sensitive to the other model parameters that we keep fixed, such as the elongation rate and the \meanXmax separation between p and Fe. It is important to note that systematics in the measured \Xmax values are absorbed by the fits of \tnorm and \sigmanorm. Thus, the composition fractions are not significantly affected by systematics in \Xmax.


%

\onecolumngrid
\newpage
\appendix
\section{ Fits to \Xmax distributions}\label{AppA}
\begin{figure}[!htb]
\centering \includegraphics[width=0.98\linewidth]{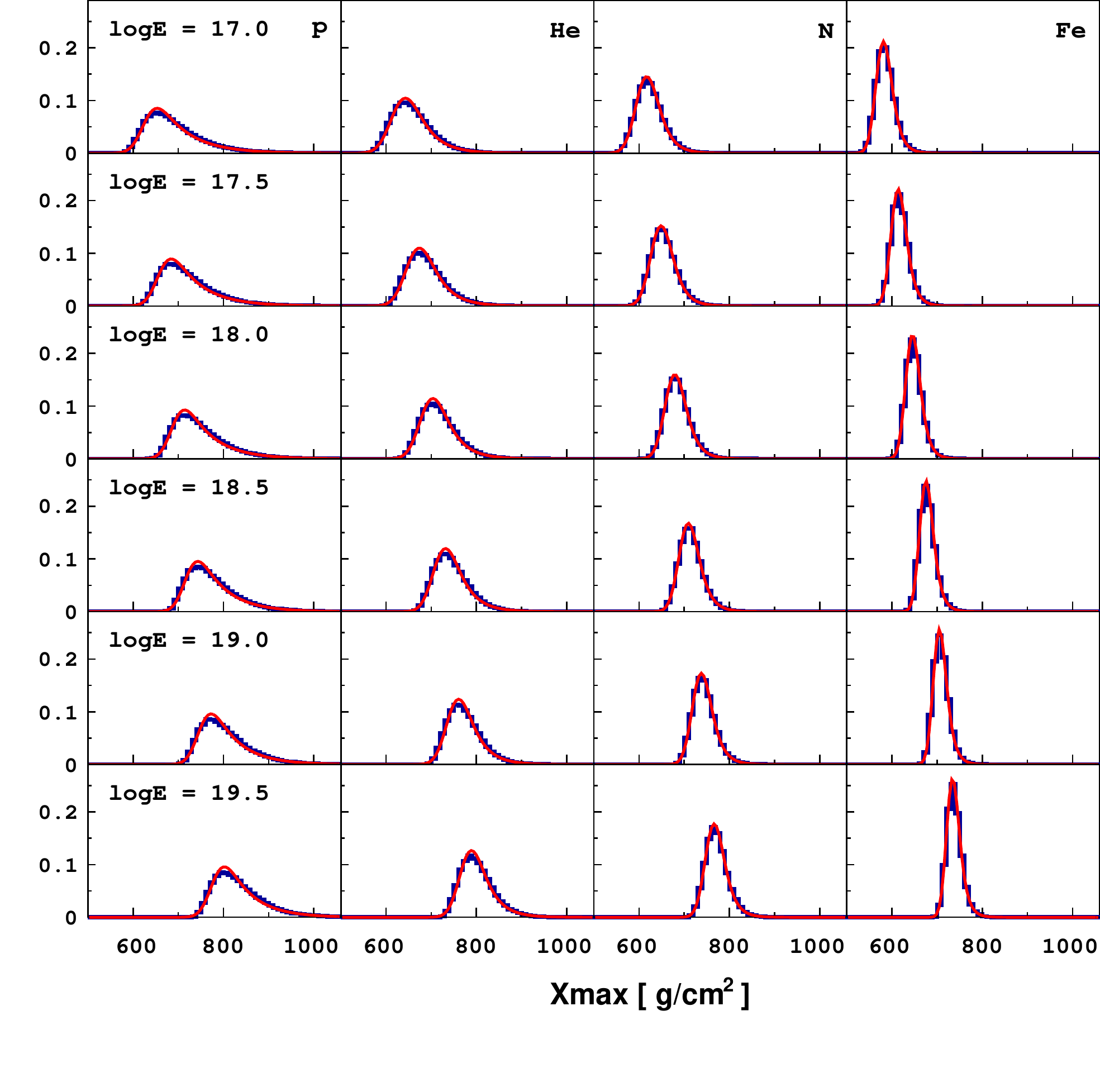}
\caption{Energy binned \epos \Xmax distributions (blue line) fitted with Equation~\eqref{eq:Xmaxbasic} (red line).}
\label{fig:EPOS_all}
\end{figure}

\begin{figure*}[!htb]
\centering \includegraphics[width=0.5\linewidth]{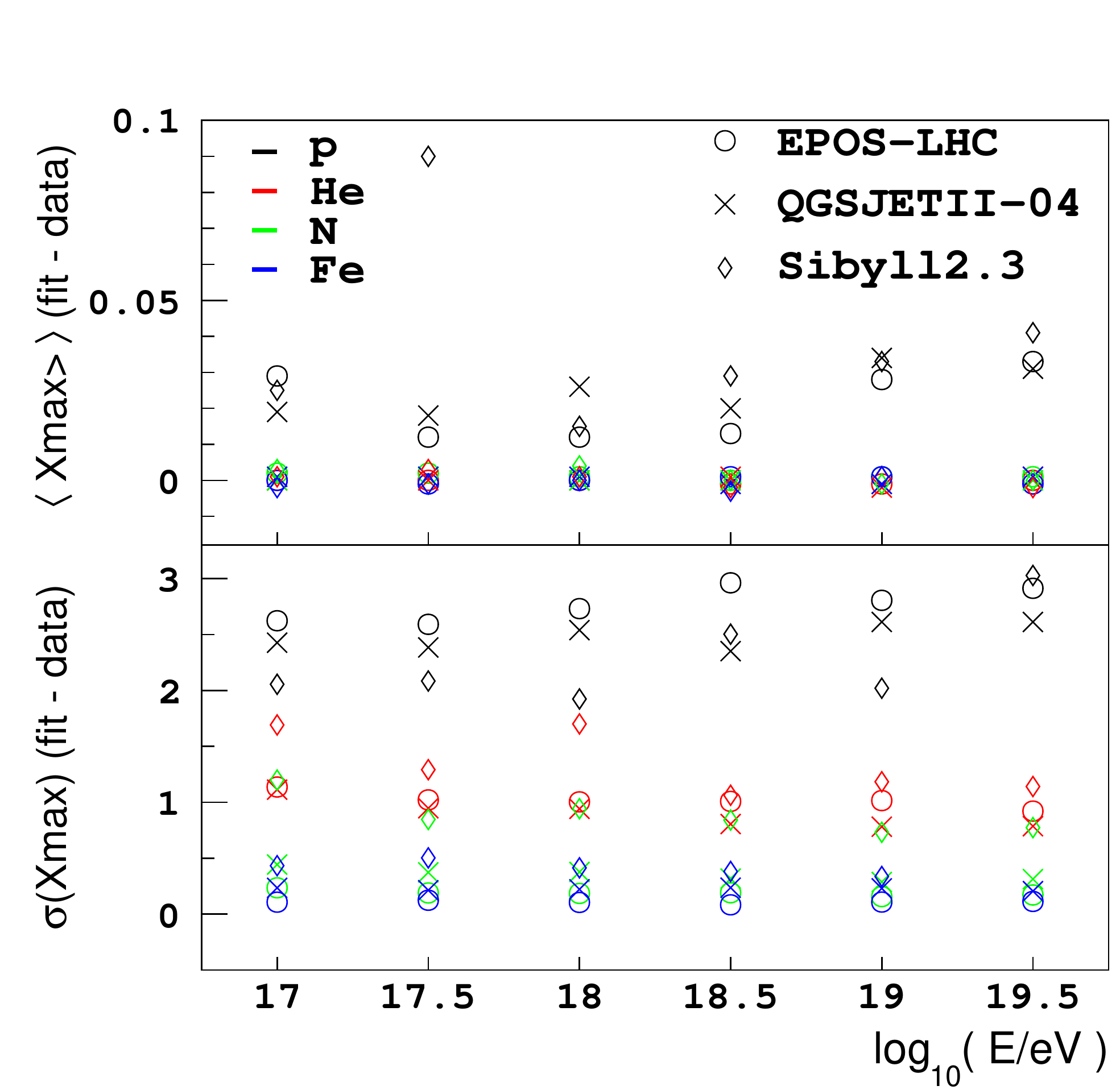}
\caption{Difference in the $\langle\text{X}_{\text{max}}\rangle$ and $\sigma(\text{X}_{\text{max}})$ between the data and the fitted equation.}
\label{fig:EPOS_QGS_diff_mean_sigma_all}
\end{figure*}

The fits of Equation~\eqref{eq:Xmaxbasic} to energy binned \Xmax data are shown in \figsThree{fig:EPOS_all}{fig:QGS_all}{fig:Sib_all}. The differences in the \meanXmax and \sigmaXmax of the data versus the fitted equation are shown in \fig{fig:EPOS_QGS_diff_mean_sigma_all}. For the fitted equation, $\langle\text{X}_{\text{max}}\rangle_{\text{fit}} = t_0 + \lambda$ and  $\sigma(\text{X}_{\text{max}})_{\text{fit}} = \sqrt{\sigma^2 + \lambda^2}$. Although the fitted function (red line) does not always precisely overlap the data (blue line), we see $\langle\text{X}_{\text{max}}\rangle_{\text{fit}}$ is always within $0.1\;\text{g/cm}^{2}$ of $\langle\text{X}_{\text{max}}\rangle_{\text{data}}$. The $\langle\text{X}_{\text{max}}\rangle$ of the distribution is the main property we endeavour to accurately define. $\sigma(\text{X}_{\text{max}})_{\text{fit}}$ is always within $3\; \text{g/cm}^{2}$ of $\sigma(\text{X}_{\text{max}})_{\text{data}}$ which is acceptable.

\subsection{\epos \Xmax distribution fits}
\clearpage
\subsection{\qgs \Xmax distribution fits}
\begin{figure}[!htb]
\centering \includegraphics[width=0.98\linewidth]{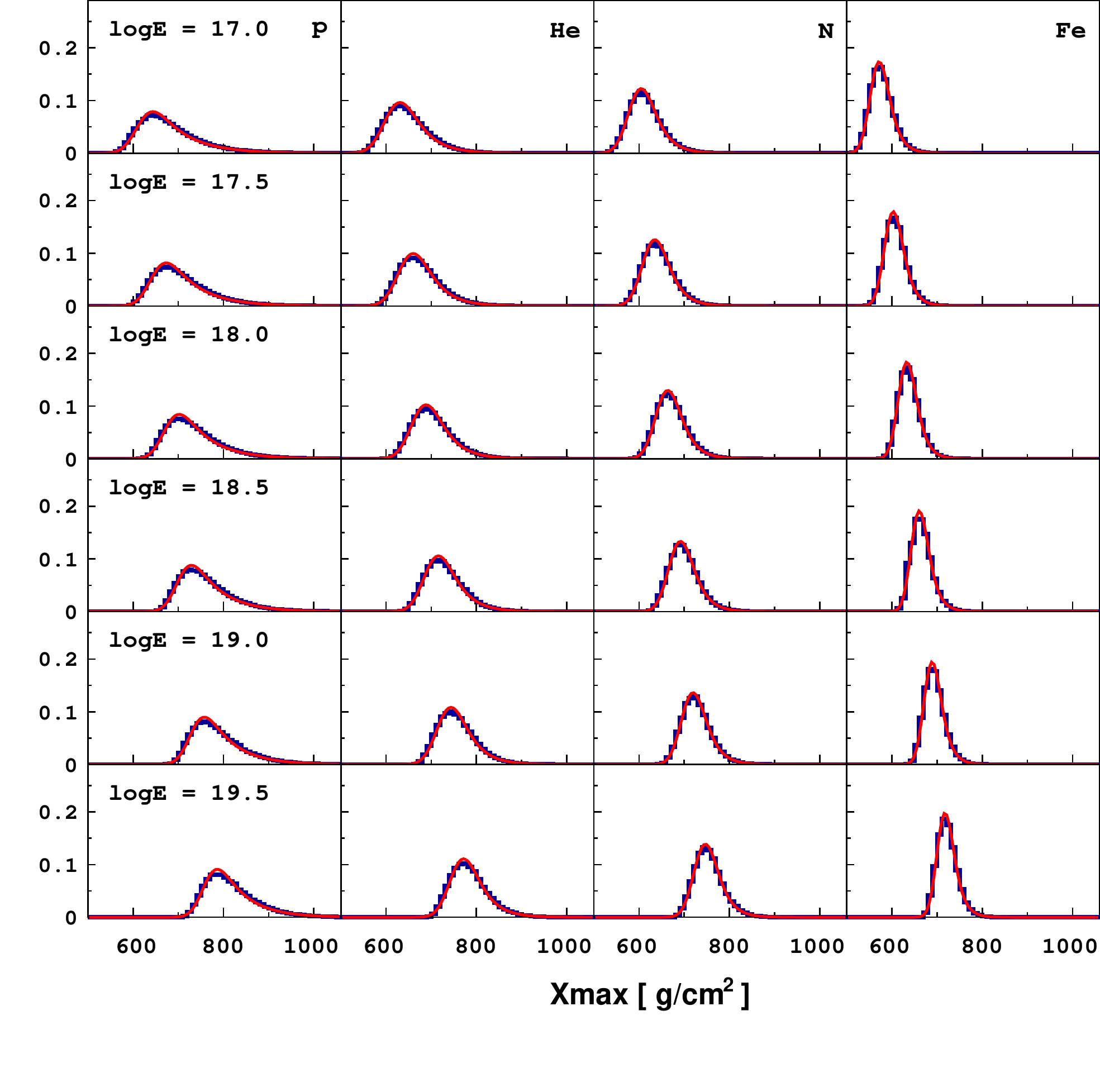}
\caption{Energy binned \qgs \Xmax distributions (blue line) fitted with Equation~\eqref{eq:Xmaxbasic} (red line).}
\label{fig:QGS_all}
\end{figure}
\clearpage
\subsection{\sib \Xmax distribution fits}
\begin{figure}[!htb]
\centering \includegraphics[width=0.98\linewidth]{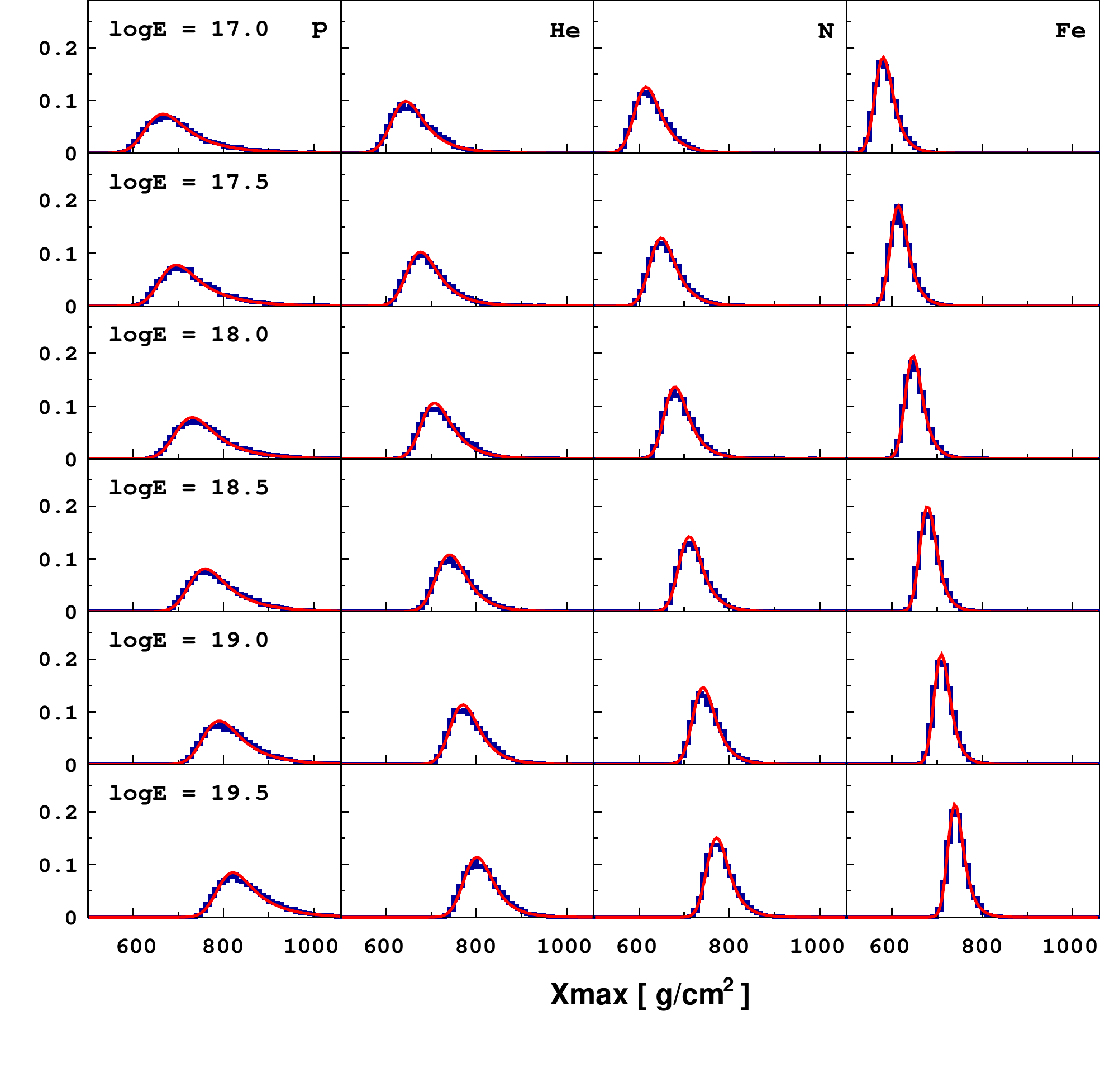}
\caption{Energy binned \sib \Xmax distributions (blue line) fitted with Equation~\eqref{eq:Xmaxbasic} (red line).}
\label{fig:Sib_all}
\end{figure}
\clearpage

\subsection{\Xmax moment comparison between the fitted parameterisation and the data}

\newpage
\section{Mass fraction, \tnorm and \sigmanorm fits of the Auger FD \Xmax data}
\label{AppFDfits_t0sigma}
The \tnorm, \sigmanorm and mass fraction fits of each parameterisation to the Auger FD \Xmax distributions are shown in the following plots. The magenta lines illustrate the measured \Xmax distributions, while the teal lines illustrate the fitted parameterisation. The black, red, green and blue lines are the fitted proton, helium, nitrogen and iron parameterisations respectively.

\begin{figure}[h!]
    	  \centering 
     \includegraphics[width=1\textwidth]{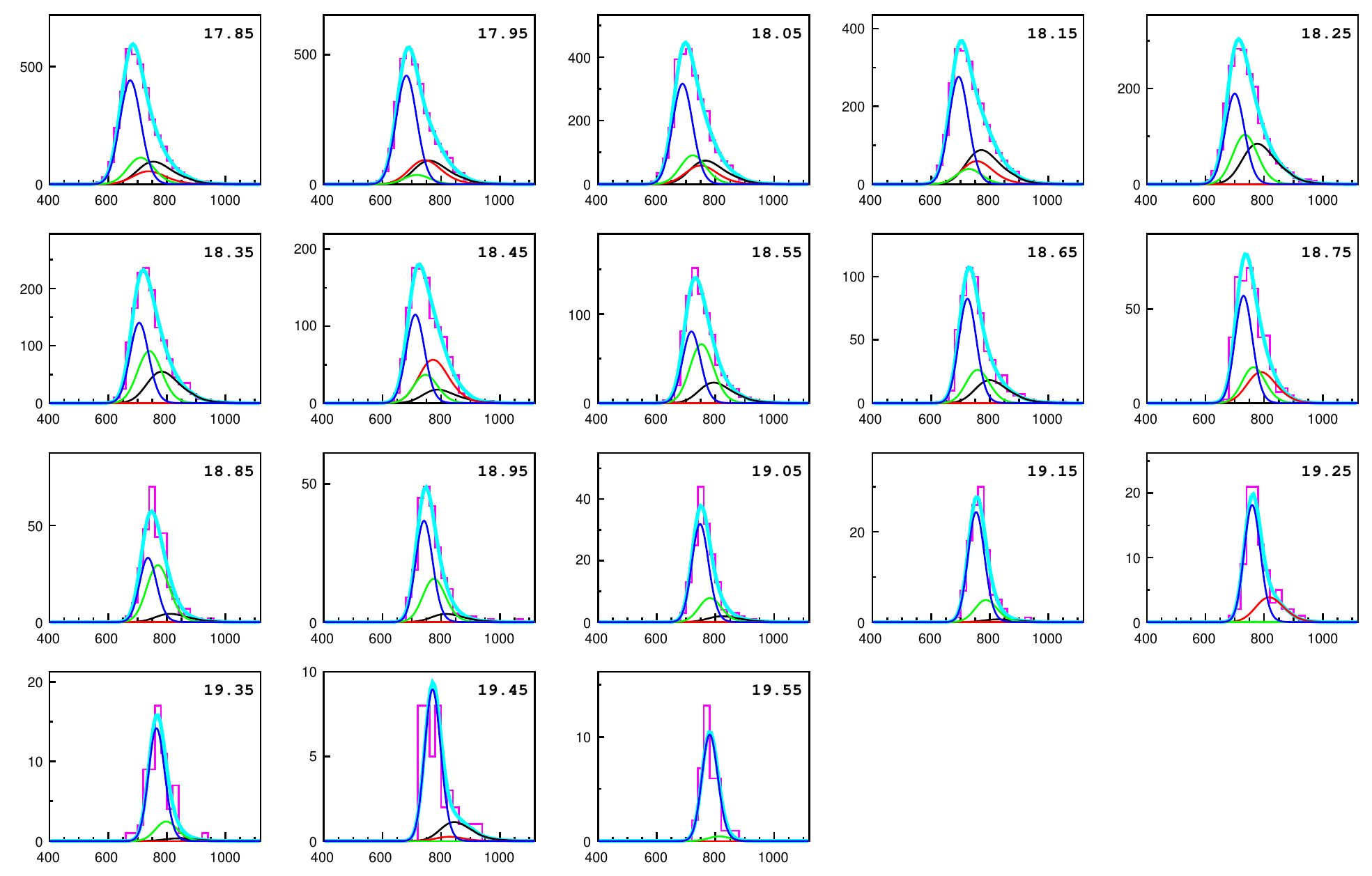}%
   \caption{Fit of the \epos \Xmax parameterisation.}
\label{fig:FDHEAT_t0sigmafit_EPOS}
\end{figure}

\begin{figure}[h!]
    	  \centering 
     \includegraphics[width=1\textwidth]{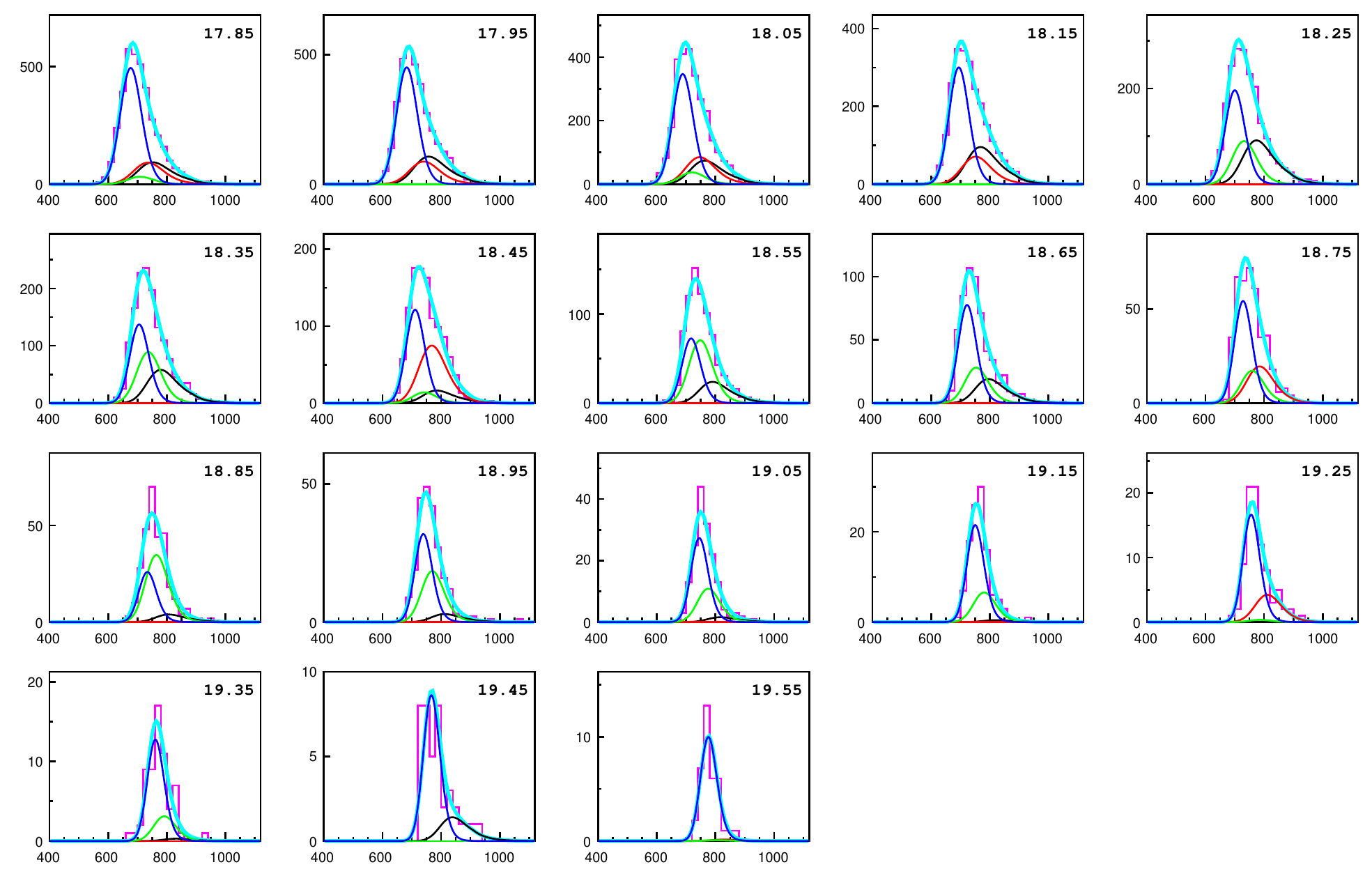}%
   \caption{Fit of the \qgs \Xmax parameterisation.}
\label{fig:FDHEAT_t0sigmafit_QGS}
\end{figure}

\begin{figure}[h!]
    	  \centering 
     \includegraphics[width=1\textwidth]{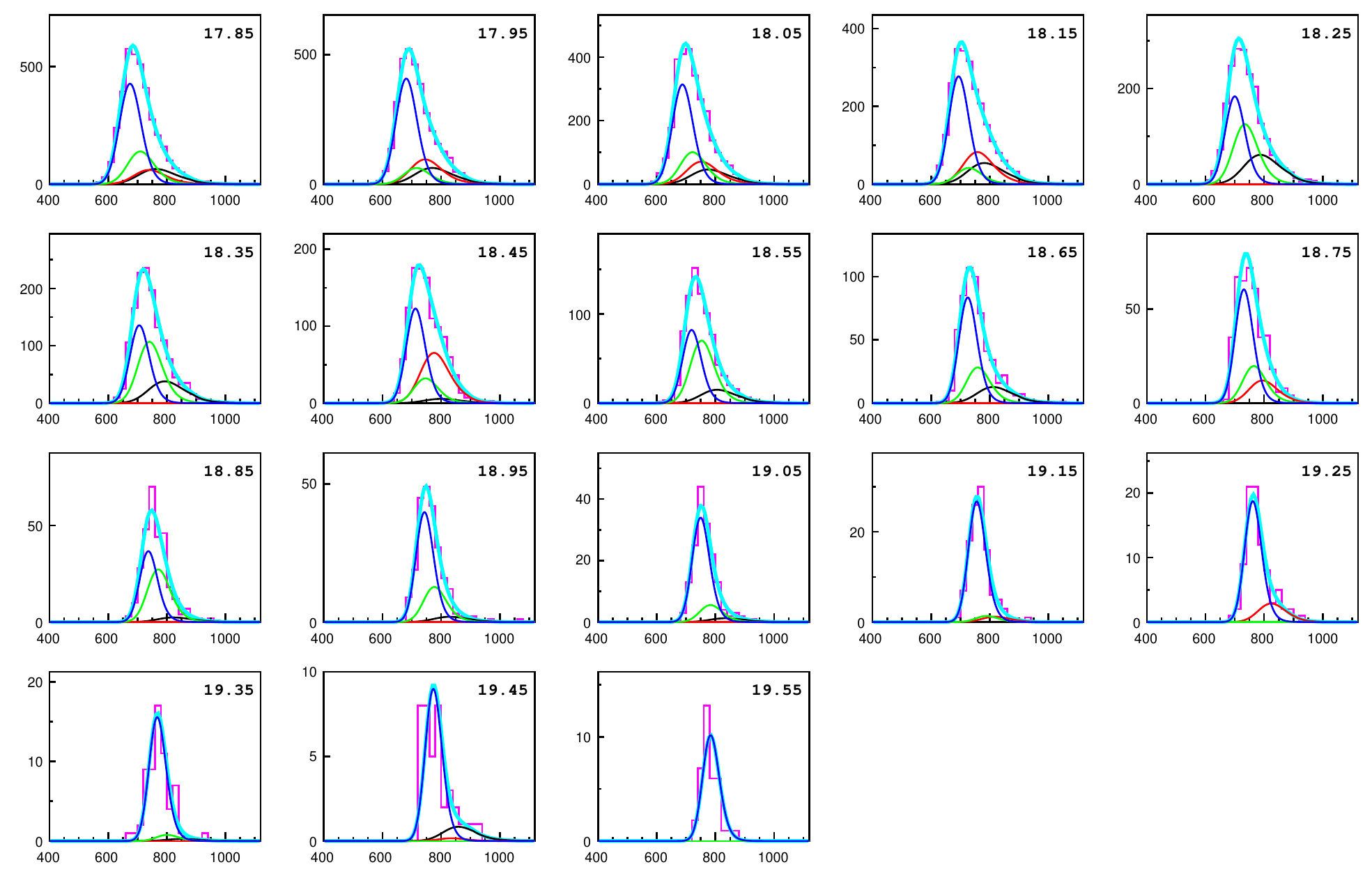}%
   \caption{Fit of the \sib \Xmax parameterisation.}
\label{fig:FDHEAT_t0sigmafit_SIB}
\end{figure}


\clearpage
\section{ Table of coefficients for the \Xmax distribution parameterisations}\label{AppB}
\begin{table}[htb!]
  \renewcommand{\arraystretch}{1.35}
  \setlength\tabcolsep{12pt}
  \centering
  \caption{Coefficients of Equation~\eqref{eq:Xmaxbasicshape} for the \epos,  \qgs and \sib \Xmax distribution predictions, assuming a normalisation energy of $E_0 = \energy{18.24}$. Also in the table, we show the \tnorm and \sigmanorm fitted to the Auger data from the \tnorm, \sigmanorm and mass fraction fit of each of the three models, and the \tnorm fitted to the Auger data from the \tnorm and mass fraction fit of each model.}
  
  \label{tab:comp}  
  \begin{tabular}{| c | c | c | c | c |}
   \hline
    \epos & Proton & Helium & Nitrogen & Iron \\ \hline \hline
    \tnorm & 703  & 697  & 680  & 650 \\ \hline
    $B$ & 2533.29  & 2515.46  & 2548.31  & 2603.31 \\ \hline
    \sigmanorm  & 21.61  & 23.46  &  19.01 & 13.01 \\ \hline
    $C$ & -0.63  & -1.81  & -1.67  & -1.36 \\ \hline  
    $\lambdanorm$ & 59.12  & 34.74  & 20.06  & 13.41 \\ \hline
    $K$ & 5.80  &  -1913.99 & -1828.11  & -1406.72\\ \hline  
    $L$ & -25.93  & 0.063  & 0.035  & 0.027\\ \hline
fitted \tnorm  &  $740\;\text{(stat.)}^{+2}_{-2}$ & 734  & 717  & 688 \\ \hline
              fitted \sigmanorm  &  $37\;\text{(stat.)}^{+2}_{-1}$ & 40 & 32 & 22 \\ \hline  
    fitted \tnorm only &  $731\;\text{(stat.)}^{+1}_{-1}$ & 725  & 708  & 678 \\ \hline
	\hline
       \qgs & Proton & Helium & Nitrogen & Iron \\ \hline \hline
    \tnorm & 688 & 679  &  660 & 635 \\ \hline
    $B$ & 2444.88  & 2410.38  & 2422.37  & 2460.32 \\ \hline
    \sigmanorm  & 24.82 & 26.83  & 23.07  & 16.54 \\ \hline
    $C$ & -1.32  & -1.24  & -0.99  & -0.91 \\ \hline  
    $\lambdanorm$ & 61.29  & 37.5  & 25.84  & 17.46 \\ \hline
    $K$ & 9.35 & 19.32  & -1818.36  & -986.08 \\ \hline  
    $L$ & -17.63  & -6.08  & 0.041  & 0.040 \\ \hline
fitted \tnorm  &  $738\;\text{(stat.)}^{+1}_{-1}$ & 730  & 711  & 685 \\ \hline
              fitted \sigmanorm  &  $32\;\text{(stat.)}^{+1}_{-1}$ & 35 & 30 & 21 \\ \hline  
    fitted \tnorm only &  $729\;\text{(stat.)}^{+1}_{-1}$ & 721  & 702  & 676 \\ \hline
        	\hline
       \sib & Proton & Helium & Nitrogen & Iron \\ \hline \hline
    \tnorm & 715 & 701  &  678 & 650 \\ \hline
    $B$ & 2666.31  & 2705.43  & 2695.22  & 2714.41 \\ \hline
    \sigmanorm  & 28.30  & 24.28  & 19.61  & 14.24 \\ \hline
    $C$ & -1.08  & -0.82  & -1.20  & -0.77 \\ \hline  
    $\lambdanorm$ & 61.52  & 40.31  & 29.48  & 19.20 \\ \hline
    $K$ & 5.81 & 23.70  & -1362.17  & -1349.93 \\ \hline  
    $L$ & -27.47  & -6.84  & 0.083  & 0.044 \\ \hline               
              fitted \tnorm  &  $748\;\text{(stat.)}^{+1}_{-2}$ & 735  & 712  & 684 \\ \hline
              fitted \sigmanorm  &  $42\;\text{(stat.)}^{+1}_{-2}$ & 36 & 29 & 21 \\ \hline  
    fitted \tnorm only &  $741\;\text{(stat.)}^{+1}_{-1}$ & 727  & 704  & 676 \\ \hline
              
  \end{tabular}
\end{table}

\clearpage
\section{\Xmax moments in terms of \lnA moments.}
  \label{AppC}
  The first two \Xmax moments can be parameterised in terms of \lnA as follows \cite{Abreu:2013env}:

\begin{equation}
\meanXmax = X_0 + D\log10\left(\frac{E}{E_0A}\right) + \xi\lnA + \delta\lnA\log10\left(\frac{E}{E_0}\right),
\label{lnA_coeff_1}
\end{equation}
and 
\begin{equation}
\sigma^2(X_\text{max}) = \sigma^2_p \; [1 + a\meanlnA + b\langle(\ln A)^2\rangle],
\label{lnA_coeff_2}
\end{equation}
where
\begin{equation}\begin{split}
\sigma^2_p &= p_0 + p_1\log10\left(\frac{E}{E_0}\right) + p_2\left[\log10\left(\frac{E}{E_0}\right)\right]^2,
\\
a &= a_0 + a_1\log10\left(\frac{E}{E_0}\right).
\end{split}\label{lnA_coeff_3}
\end{equation}

   Using the \tnorm and \sigmanorm fit results of the 2014 FD dataset (see \tab{tab:comp}), the parameters of Equations~\eqref{lnA_coeff_1}, \eqref{lnA_coeff_2} and \eqref{lnA_coeff_3} have been determined, and are displayed in Tables~\ref{tab:lnA_coeff_3} and \ref{tab:lnA_coeff_4}.  The mean and maximum \meanXmax residuals of the fit are $\sim\depth{1}$ and  $\sim\depth{2.5}$ respectively. The mean and maximum \sigmaXmax residuals of the fit are $\sim\depth{1}$ and $\sim\depth{1.5}$ respectively.
  
\par\bigskip
\begin{table}[h]
  \renewcommand{\arraystretch}{1.35}
  \setlength\tabcolsep{12pt}
\begin{center}
  \begin{tabular}{| c | c | c | c |}
    \hline
    parameter & \epos & \qgs & \sib  \\ \hline \hline
    $X_0$ &  842.8 $\pm$ 0.3   &  839.9 $\pm$ 0.3  & 855.7 $\pm$ 0.4 \\ \hline
    $D$ & 54.8 $\pm$ 0.5  &  51.9 $\pm$ 0.4 & 59.1 $\pm$ 0.6 \\ \hline
    $\xi$ &  -0.10 $\pm$ 0.26   &  -1.52 $\pm$ 0.20 &  0.09 $\pm$ 0.33 \\ \hline
    $\delta$ &  0.83 $\pm$ 0.21 & 0.13 $\pm$ 0.16 &  1.20 $\pm$ 0.26 \\ \hline
  \end{tabular}
\end{center}
\caption{Parameters of Equation~\eqref{lnA_coeff_1}, obtained by fitting the predicted \meanXmax from the \tnorm and \sigmanorm fit of the 2014 FD data set. All values are in \gcm.}
\label{tab:lnA_coeff_3}
\end{table}

\begin{table}[h]
  \renewcommand{\arraystretch}{1.35}
  \setlength\tabcolsep{12pt}
\begin{center}
  \begin{tabular}{| c | c | c | c |}
    \hline
    parameter & \epos & \qgs & \sib  \\ \hline \hline
    $p_0 \times \SI{}{g^{-2}cm^4}$ & 4592 $\pm$ 19  & 4402 $\pm$ 32 & 5222 $\pm$ 34  \\ \hline
    $p_1 \times \SI{}{g^{-2}cm^4}$ & -361 $\pm$ 20  & -427 $\pm$ 33 & -413 $\pm$ 35 \\ \hline
    $p_2 \times \SI{}{g^{-2}cm^4}$ & 70 $\pm$ 33 &  71 $\pm$ 54 &  87 $\pm$ 56  \\ \hline
    $a_0$ & -0.377 $\pm$ 0.003 &  -0.372 $\pm$ 0.005 & -0.362 $\pm$ 0.004 \\ \hline
    $a_1$ & -0.0038 $\pm$ 0.0010  &  -0.0004 $\pm$ 0.0017  & -0.0031 $\pm$ 0.0016   \\ \hline
    $b$ & 0.040 $\pm$ 0.001 & 0.041 $\pm$ 0.001  & 0.038 $\pm$ 0.001  \\ \hline
  \end{tabular}
\end{center}
\caption{Parameters of Equation~\eqref{lnA_coeff_2} and Equation~\eqref{lnA_coeff_3}, obtained by fitting the predicted $\sigma^2(X_\text{max})$ from the \tnorm and \sigmanorm fit of the 2014 FD data set. }
\label{tab:lnA_coeff_4}
\end{table}

\newpage
  Using the results from the fit of only \tnorm and the mass fractions to the 2014 FD dataset (see \tab{tab:comp}), the parameters of Equations~\eqref{lnA_coeff_1}, \eqref{lnA_coeff_2} and \eqref{lnA_coeff_3} are displayed in Tables~\ref{tab:lnA_coeff_5} and \ref{tab:lnA_coeff_6}. The \meanXmax and \sigmaXmax residuals of these results are similar to those from the \tnorm, \sigmanorm and mass fraction fit results.

\begin{table}[h]
  \renewcommand{\arraystretch}{1.35}
  \setlength\tabcolsep{12pt}
\begin{center}
  \begin{tabular}{| c | c | c | c |}
    \hline
    parameter & \epos & \qgs & \sib  \\ \hline \hline
    $X_0$ & 833.4 $\pm$ 0.3 & 830.9 $\pm$ 0.3 & 848.2 $\pm$ 0.4 \\ \hline
    $D$ & 54.8 $\pm$ 0.5 & 51.9 $\pm$ 0.4 & 59.1 $\pm$ 0.6 \\ \hline
    $\xi$ & -0.10 $\pm$ 0.26 & -1.52 $\pm$ 0.20 & 0.09 $\pm$ 0.33 \\ \hline
    $\delta$ & 0.83 $\pm$ 0.21 & 0.13 $\pm$ 0.16 & 1.20 $\pm$ 0.26 \\ \hline
  \end{tabular}
\end{center}
\caption{Parameters of Equation~\eqref{lnA_coeff_1}, obtained by fitting the predicted \meanXmax from the \tnorm fit of the 2014 FD data set}
\label{tab:lnA_coeff_5}
\end{table}

\begin{table}[h]
  \renewcommand{\arraystretch}{1.35}
  \setlength\tabcolsep{12pt}
\begin{center}
  \begin{tabular}{| c | c | c | c |}
    \hline
    parameter & \epos & \qgs & \sib \\ \hline \hline
    $p_0 \times \SI{}{g^{-2}cm^4}$ & 3793 $\pm$ 35  & 3990 $\pm$ 44 & 4049 $\pm$ 47  \\ \hline
    $p_1 \times \SI{}{g^{-2}cm^4}$ & -355 $\pm$ 36  & -411 $\pm$ 45 & -392 $\pm$ 49 \\ \hline
    $p_2 \times \SI{}{g^{-2}cm^4}$ & 76 $\pm$ 61 & 74 $\pm$ 76 & 89 $\pm$ 80  \\ \hline
    $a_0$ & -0.459 $\pm$ 0.006 & -0.425 $\pm$ 0.007 & -0.392 $\pm$ 0.008 \\ \hline
    $a_1$ & -0.0022 $\pm$ 0.0021  & -0.0011 $\pm$ 0.0026 &  -0.0041 $\pm$ 0.0027   \\ \hline
    $b$ & 0.059 $\pm$ 0.002  & 0.052 $\pm$ 0.002 & 0.045 $\pm$ 0.002   \\ \hline
  \end{tabular}
\end{center}
\caption{Parameters of Equation~\eqref{lnA_coeff_2} and Equation~\eqref{lnA_coeff_3}, obtained by fitting the predicted $\sigma^2(X_\text{max})$ from the \tnorm fit of the 2014 FD data set }
\label{tab:lnA_coeff_6}
\end{table}

\end{document}